\documentclass[11pt]{article}
\pdfoutput=1 
\usepackage{jheppub}

\usepackage{CJKutf8} 
\usepackage{mathtools,slashed,mathrsfs}
\usepackage[caption=false]{subfig}
\usepackage{dcolumn}
\usepackage{multirow}
\usepackage{tabularx}
\usepackage{booktabs}
\usepackage{orcidlink}
\usepackage{bm}
\usepackage{amsmath}
\usepackage[normalem]{ulem}
\usepackage{comment}
\usepackage{setspace}
\usepackage{xcolor}
\usepackage{cancel}
\usepackage{enumerate}
\usepackage{siunitx}
\usepackage{float}
\usepackage{hyperref}
\usepackage{cancel}
\usepackage{slashed} 
\usepackage{simpler-wick} 

\allowdisplaybreaks

\newcommand{\pwrap}[1]{\mathopen{}#1\mathclose{}\mathord{\vphantom{#1}}} 
\newcommand{\p}[1]{\pwrap{\left(#1\right)}}
\newcommand{\bq}[1]{\pwrap{\left[#1\right]}}
\newcommand{\Bq}[1]{\pwrap{\left\{#1\right\}}}
\newcommand{\bk}[1]{\pwrap{\left\langle#1\right\rangle}}

\newcommand{\ket}[1]{\pwrap{\left|#1\right\rangle}}
\newcommand{\abs}[1]{\left|#1\right|}
\newcommand{\eval}[1]{\left.#1\right|}

\newcommand{\lbar}[1]{\,\overline{\!#1}}

\newcommand{\Cite}[1]{ref.~\cite{#1}}
\newcommand{\fig}[1]{figure~\ref{#1}}
\newcommand{\Fig}[1]{Figure~\ref{#1}}
\newcommand{\tab}[1]{table~\ref{#1}}
\newcommand{\eq}[1]{eq.~(\ref{#1})}
\newcommand{\Eq}[1]{Eq.~(\ref{#1})}
\newcommand{\eqs}[2]{eqs.~(\ref{#1}) and (\ref{#2})}
\newcommand{\Sec}[1]{section~\ref{#1}}

\newcommand{\app}[1]{appendix~\ref{#1}}

\newcommand{\peq}{\phantom{{}={}}}
\newcommand{\n}{\nonumber\\}

\newcommand{\dd}[2][]{\mathop{\mathrm{d}^{#1}#2}\nolimits}

\DeclareSIUnit\parsec{pc}

\definecolor{plotred}{HTML}{DC143C}
\definecolor{plotblue}{HTML}{0018A8}

\newcommand{\Rc}{\mathcal{R}} 
\newcommand{\phipt}{\delta\phi} 
\newcommand{\phicl}{\phi_0} 
\newcommand{\dphicl}{\dot\phi_0}
\newcommand{\chicl}{\chi_0}
\newcommand{\bchicl}{\bar\chi_0}
\newcommand{\chipt}{\delta\chi}
\newcommand{\dmu}{\Delta\mu} 
\newcommand{\phase}{\vartheta}
\newcommand{\Int}{\mathcal{I}} 

\newcommand{\ctime}{\eta} 
\newcommand{\Higgs}{\mathbf{h}} 
\newcommand{\Mpl}{M_{\mathrm{pl}}} 
\newcommand{\Mgut}{M_{\mathrm{U}}} 
\newcommand{\Msusy}{M_{\mathrm{SUSY}}} 
\newcommand{\Mcom}{M_{\mathrm{C}}} 
\newcommand{\fNL}{f_{\mathrm{NL}}} 

\newcommand{\Ps}{\mathcal{P}} 
\newcommand{\tF}{\widetilde{F}} 
\newcommand{\Op}{\mathcal{O}} 

\newcommand{\N}{\mathbb{N}}
\newcommand{\vb}[1]{\mathbf{#1}} 
\newcommand{\order}{\mathcal{O}}
\newcommand{\Res}{\operatorname{Res}} 
\newcommand{\sumint}{\int\mathllap{\sum}}

\newcommand{\arcsinh}{\operatorname{arcsinh}}

\newcommand{\stackon}[2]{\genfrac{}{}{-0.4pt}{}{#1}{#2}}
\newcommand{\hF}[4]{F\p{\stackon{#1,#2}{#3};#4}} 
\newcommand{\hFPQ}[5]{{}_{#1}F_{#2}\p{\stackon{#3}{#4};#5}} 
\newcommand{\mGamma}[2]{\Gamma\p{\stackon{#1}{#2}}} 
\newcommand{\MeijerG}[7] {G^{#1}_{#2}\p{\stackon{#3;#4}{#5;#6};#7}} 

\renewcommand{\Re}{\operatorname{Re}}
\renewcommand{\Im}{\operatorname{Im}}

\newcommand{\cc}{\mathrm{c.c.}} 
\newcommand{\rhs}{\mathrm{RHS}}  
\newcommand{\An}{\mathrm{A}} 
\newcommand{\NAn}{\mathrm{NA}} 
\newcommand{\sq}{\mathrm{sq}} 
\newcommand{\wloop}{\text{1-loop}}
\newcommand{\tree}{\text{tree}}
\newcommand{\sqlimit}[1][\to\infty]{\xrightarrow{p#1}}

\newcommand{\NIST}[1]{\cite[eq. (#1)]{NIST}}

\title{\centering Charged Loops at the Cosmological Collider \\ 
with Chemical Potential
}

\date{\today}
\author[a,b]{Arushi Bodas \orcidlink{0000-0003-4664-4277},}
\author[c]{Edward Broadberry \orcidlink{0000-0003-0652-1862},}
\author[c]{Raman Sundrum \orcidlink{0009-0004-7537-5357},}
\author[c]{Zhaohui Xu \orcidlink{0009-0001-0473-1121} }

\affiliation[a]{Enrico Fermi Institute, University of Chicago, Chicago, IL 60637, USA}
\affiliation[b]{Particle Theory Department, Fermilab, Batavia, Illinois 60510, USA}
\affiliation[c]{Maryland Center for Fundamental Physics, Department of Physics, University of Maryland, College Park, MD 20742, USA}

\emailAdd{arushib@uchicago.edu}
\emailAdd{edbroad@umd.edu}
\emailAdd{raman@umd.edu}
\emailAdd{zhxu1226@umd.edu}

\abstract{
Cosmological collider physics allows the detection of heavy particles at inflationary scales through their imprints on primordial non-Gaussianities.
We study the chemical potential mechanism applied to a pair of charged scalars. We analytically evaluate the resulting one-loop contribution to the bispectrum, using the spectral decomposition. In this way we are able to determine the parametric dependencies for both the signal and the background. We show that a signal strength \(\fNL \sim \order(0.01)\) can be obtained within theoretical control, potentially reachable by \qty{21}{cm} tomography. As an application we consider the colored Higgs bosons in \(\mathrm{SU}(5)\) supersymmetric orbifold grand unification with masses \(M\lesssim\qty{e15}{GeV}\).
}

\begin{document}

\maketitle

\section{Introduction}

Heavy particles produced on-shell during inflation, if they decay into inflatons, could have imprinted themselves on cosmological observables such as the Cosmic Microwave Background (CMB) and Large Scale Structure (LSS) \cite{Chen:2009zp,CCP6,CCP2,CCP4,Arkani-Hamed:2015bza,Lee:2016vti}.
The mechanism, dubbed \emph{cosmological collider physics}, offers an alternative approach to search for new particles.
Compared to terrestrial colliders at the \unit{TeV}-scale, the ``cosmological collider" probes inflationary energy scales, possibly as high as \(\sim\qty{e14}{GeV}\). Therefore, with increasing sensitivity from upcoming LSS experiments \cite{Alvarez:2014vva,SPHEREx:2014bgr,Camera:2014bwa,MoradinezhadDizgah:2017szk,MoradinezhadDizgah:2018ssw,Kogai:2020vzz} and future \qty{21}{cm} tomography \cite{Loeb:2003ya,Munoz:2015eqa,Meerburg:2016zdz}, there is a possibility to detect heavy particles far beyond the reach of upcoming colliders.

In the minimal version of cosmological collider physics particles are created by the expansion of spacetime, characterized by the inflationary Hubble energy scale, \(H\). During inflation, this could be as large as \cite{Planck:2018jri}
\begin{align}\label{eq:mplanck}
    H &< \num{2.5e-5} \Mpl \approx \qty{6e13}{GeV}. 
\end{align}
If the heavy particles can decay into inflatons, they will give the following non-zero contribution to the non-Gaussianity (NG) in the 3-point correlator of the comoving curvature perturbation, \(\Rc\), \cite{Chen:2009zp,Arkani-Hamed:2015bza,CCP2}
\begin{align}\label{eq:na-vanilla}
    \frac{\bk{\Rc_{\vb{k}_1}\Rc_{\vb{k_2}}\Rc_{\vb{k_3}}}'}{\bk{\Rc_{\vb{k}_1} \Rc_{-\vb{k}_1}}' \bk{\Rc_{\vb{k}_3} \Rc_{-\vb{k}_3}}'} &\sim e^{-\pi\mu} \p{\frac{k_3}{k_1}}^{-3/2-i\mu} +\cc,&
    \mu \coloneqq \sqrt{\frac{M^2}{H^2} -\frac{9}{4}},
\end{align}
in the squeezed limit, \(k_1\sim k_2\gg k_3\). Depending on the value of the heavy mass \(M\), this signal can be either oscillatory (\(M>3H/2\)), or a fractional power-law (\(M<3H/2\)). Both behaviors, especially the oscillatory one, cannot be mimicked by local contact interactions of inflatons, and can serve as strong evidence for particle production.

Despite the exciting opportunity in this direction, \eq{eq:na-vanilla} contains an exponential suppression \(e^{-\pi\mu}\) for \(M>3H/2\), which leads to a very narrow observable mass window for oscillatory signals. Fortunately, this suppression can be overcome through an extension of the minimal mechanism, where the kinetic energy of the inflaton background can give rise to a  \emph{chemical potential} for heavy fields through a dimension-5 coupling \cite{Barnaby:2011vw,Chen:2018xck,Adshead:2018oaa,Wang:2019gbi,Wang:2020ioa,Bodas:2020yho,Tong:2022cdz,Bodas:2024hih}:
\begin{equation}\label{eq:chem-potential}
    \mathcal{L} \supset -\frac{1}{\Lambda} \nabla_\mu \phi J^\mu \supset \lambda J^0
\end{equation}
with the chemical potential defined by \(\lambda \coloneqq \dphicl/\Lambda\), and \(J^\mu\) is a non-conserved current made from the heavy fields.
In order to maintain in effective field theory control, \(\Lambda^2\) is required to be larger than \(\dphicl\), i.e. the chemical potential is bounded by \(\lambda < \sqrt{\dphicl}\) where \cite{Planck:2018jri}
\begin{equation}\label{eq:dphicl}
    \dphicl \approx (60H)^2.
\end{equation}
The chemical potential leads to unsuppressed particle production for all fields in \(J\) whose masses, \(M<\lambda\). This can greatly widen the mass window for oscillatory signatures. The simplest signals of this type, both theoretically and phenomenologically,  appear at tree-level, and have been explored in \Cite{Bodas:2020yho,Bodas:2024hih}.

Loop level processes for cosmological correlators are poorly understood, because we usually lack the theoretical control required to calculate them. However, since the inflatons on the external legs are neutral, one-loop processes can be important because particles with nontrivial quantum numbers must be produced and annihilated in pairs. Consequently, including one-loop effects can broaden the range of new physics targets to include charged particles and fermions.

There have been several efforts towards computing loops during inflation:\footnote{This is far from an exhaustive list of papers on loops during inflation, here are a few more papers that focus on similar diagrams to the one we have considered \cite{Cacciatori:2024zrv,Benincasa:2024ptf,Senatore:2009cf,Senatore:2012nq,Pimentel:2012tw}.}
\begin{itemize}
    \item Rough estimates based on ``loop factors \(\times\) propagators \(\times\) vertices" have been accomplished in \cite{Chen:2018xck,Wang:2019gbi,Wang:2020ioa,Bodas:2020yho,Maru:2021ezc,Maru:2022bhr}. There are also approximate one-loop calculations based on a late-time expansion of propagators in \cite{Chen:2016hrz,Wang:2019gbi}. Nonetheless, as mentioned in \cite{Wang:2019gbi}, these results cannot predict the extra parametric dependence on the chemical potential, \(\lambda\), and the particles masses coming from the loop integrals.
    
    \item In \cite{Wang:2021qez}, a numerical calculation of a one-loop process was performed. It is found that the brute-force method is time-consuming because of the high dimensionality and the oscillatory nature of the integrand. Furthermore, the underlying physics is less clear than in an analytic calculation.

    \item There are also analytic results in \cite{Marolf:2010zp,Qin:2022lva,Xianyu:2022jwk,Cohen:2024anu,Qin:2024gtr,Cacciatori:2024zrv,Westerdijk:2025ywh}.
    In particular, \Cite{Marolf:2010zp,Cohen:2024anu} applied the \emph{spectral decomposition} technique to one-loop corrections to the power spectrum, \Cite{Westerdijk:2025ywh} applied this to multi-loops in the special case of conformally coupled scalar, while \Cite{Xianyu:2022jwk} further applied this to a particular one-loop contribution to the bispectrum without chemical potential, reducing it to a one-dimensional integral over the invariant mass or a sum over residues.
    However, most models will also produce triangle diagrams which cannot be captured by the spectral decomposition, a problem which has not so far been addressed. Compared to the numerical approach, this method relies on de Sitter isometry, which is broken down to scale invariance by the inflaton background.  
\end{itemize}

In this paper we evaluate the 3-point correlator in a model where the chemical potential effect is perturbative, such that we can still take advantage of the de Sitter isometry. Furthermore, in this particular model all leading order effects can be computed using the spectral representation.

It turns out that in the limit of heavy particles with a large chemical potential, the signal size from the loop process has a surprisingly simple form: for two scalars with masses \(M_1\), \(M_2\), the amplitude for the 3-point function is related to the amplitude that one would obtain from the tree level exchange of a single scalar by
\begin{align}
    \bk{\Rc_{\vb{k}_1} \Rc_{\vb{k_2}} \Rc_{\vb{k_3}}}'_{\wloop,\text{signal}}
    &\simeq \frac{e^{3\pi i/4}}{2\pi^{3/2}} \sqrt{\frac{M_1M_2}{M_{12}}} \p{\frac{2M_{12}}{\lambda +M_{12}} \frac{2k_3}{k_1}}^{-3/2}\n
    &\peq \times
    \eval{\bk{\Rc_{\vb{k}_1} \Rc_{\vb{k_2}} \Rc_{\vb{k_3}}}'_{\tree}}_{M=M_{12}}
\end{align}
in the squeezed limit, where \(M_{12}\coloneqq M_1+M_2\) and \(\lambda\) is the chemical potential.
A quick derivation of this relation is given in \Sec{subsec:central-result}. In addition to an oscillatory signal, the model predicts the existence of a smooth background. We also evaluate the analytic background, showing that it comes from the resonance limit in the corresponding effective tree-level model. By obtaining the correct parametric dependence on both the chemical potential and the masses, we remove one obstacle for future phenomenological studies in this direction.

As a phenomenological application we consider the prospect of detecting the colored Higgs of supersymmetric \(\mathrm{SU}(5)\) orbifold grand unification. The framework of grand unification is a well motivated scenario for physics beyond the standard model. In GUTs there exist new particles with masses around the scales that cosmological collider physics can probe \cite{GUT1,GUT2,GUT3}. Some of these possibilities have been investigated in \cite{Maru:2021ezc,Maru:2022bhr}. However, the simple models of grand unification are in tension with proton decay bounds, which can be avoided in the higher dimensional ``orbifold GUTs" \cite{Kawamura:1999nj,Kawamura:2000ev,Hall1}. The cosmological collider signatures of orbifold GUTs have been investigated in \cite{Kumar:2018jxz,Bodas:2024hih}, while the idea of detecting supersymmetry in general has been considered in \cite{Baumann:2011nk,CCP18}. In \Cite{Bodas:2024hih}, the authors also considered the signal from an orbifold SUSY GUT with a chemical potential. However, they looked at tree level exchanges, which require the heavy particle to be neutral. This is not possible at tree level with the simple \(\mathrm{SU}(5)\) gauge group, but required more exotic grand unification, e.g. trinification. The fact that we can calculate a loop level process allows us to consider the simpler \(\mathrm{SU}(5)\) gauge group. We find that the one-loop signal could be within the sensitivity of future \qty{21}{cm} experiments.

The rest of the paper is organized as follows. In \Sec{sec:preliminaries}, we introduce our notation and review the necessary background. 
Section~\ref{sec:ChemPot_SpectDecomp} elaborates on the implementation of a chemical potential and the method of spectral representation, which is crucial for evaluating loop diagrams involving a pair of charged scalars. 
We also present an intuitive calculation of the non-analytic part of the bispectrum, which constitutes the cosmological collider signal. 
Sections~\ref{sec:Systematic_tree} and~\ref{sec:Systematic_loop} then provide a more systematic evaluation of both the non-analytic contribution and the smooth analytic background. In \Sec{sec:signal-strength}, we analyze theoretical and observational constraints and estimate the maximum signal strength achievable. 
As a concrete application of our chemical potential model, we assess the detectability of the colored Higgs bosons predicted by the \(\mathrm{SU}(5)\) orbifold SUSY GUT model in \Sec{sec:SUSY_GUT}. 
We conclude with a discussion of future directions in \Sec{sec:discussion}.
The appendices contain supplementary material relevant to our calculations.

\section{Preliminaries}
\label{sec:preliminaries}

In this section, we define our notation and briefly review the in-in formalism for cosmological correlators. 
The mostly plus signature \((-,+,+,+)\) for the metric is used throughout this paper.

\subsection{Primordial non-Gaussianity}

During slow-roll inflation the spacetime is approximately described by the de Sitter metric: 
\begin{align}
    \dd{s}^2 &= -\dd{t}^2 +e^{2Ht} \dd{\vb{x}}^2,\n
    &= \frac{-\dd{\ctime}^2 +\dd{\vb{x}}^2}{H^2\ctime^2},
    \label{eq:dS-metric}
\end{align}
where \(t\) is the proper time, \(H\) is the Hubble parameter during inflation and \(\ctime = -e^{-Ht}/H\) is the conformal time.

Inflation is usually realized by a scalar field \(\phi\) called the ``inflaton". It can be decomposed into a homogeneous classical background, \(\phicl\), that controls the inflation dynamics, plus a quantum fluctuation, \(\phipt\), that is responsible for the primordial fluctuations after inflation. In the spatially flat gauge, \(\phipt\) is related to the gauge-invariant comoving curvature perturbation, \(\Rc\), by \cite{Bardeen:1983qw}
\begin{equation}
    \Rc = -H \frac{\phipt}{\dphicl}
\end{equation}
with the overdot representing the derivatives with respect to the proper time \(t\). Since the inflationary dynamics is invariant under spatial translation, all momentum-space correlators contain a total-momentum \(\delta\)-function:
\begin{equation}
    \bk{\Rc_{\vb{k}_1}\Rc_{\vb{k}_2}\cdots\Rc_{\vb{k}_n}} = (2\pi)^3\delta(\vb{k}_1+\vb{k}_2+\cdots+\vb{k}_n)\cdot \bk{\Rc_{\vb{k}_1}\Rc_{\vb{k}_2}\cdots\Rc_{\vb{k}_n}}'.
\end{equation}

Starting from the 2-point correlation function, current CMB data favors a nearly scale-invariant primordial power spectrum given by
\begin{equation}
    \Ps_\Rc(k) \coloneqq \bk{\Rc_{\vb{k}} \Rc_{-\vb{k}}}' = \frac{H^4}{2\dphicl^2 k^3} \p{\frac{k}{k_*}}^{n_s-1}.
\end{equation}
The scalar spectral index, \(n_s\), has been measured to be around \(0.96\) at the reference comoving momentum scale, \(k_* \approx \qty{0.05}{\mega\parsec^{-1}}\). In this paper, we assume exact scale invariance during inflation as an approximation, i.e. we will treat \(H\) and \(\dphicl\) as constants. From now on we work in units with \(H=1\) in all equations unless stated otherwise. Note that with these two assumptions, the \emph{spacetime scale transformation}:
\begin{equation}
    \ctime,\vb{x} \mapsto \lambda\ctime,\lambda\vb{x}
\end{equation}
remains to be a good symmetry.

While scale invariance forces \(\Ps_\Rc(k)\) to be proportional to \(1/2k^3\), higher-point correlation functions, such as the primordial bispectrum \(\bk{\Rc_{\vb{k}_1}\Rc_{\vb{k}_2}\Rc_{\vb{k}_3}}'\), may contain more intricate momentum dependence.
The bispectrum is conventionally normalized relative to the size of the power spectrum in defining the NG as \cite{Kumar:2017ecc}
\begin{align}\label{eq:3pt-F}
    F(k_1,k_2,k_3) &\coloneqq \frac{5}{6} \frac{\bk{\Rc_{\vb{k}_1}\Rc_{\vb{k}_2}\Rc_{\vb{k}_3}}'}{\Ps_\Rc(k_1) \Ps_\Rc(k_2) +\Ps_\Rc(k_2) \Ps_\Rc(k_3) +\Ps_\Rc(k_3) \Ps_\Rc(k_1)}\n
    &= -\frac{10}{3} \dphicl\cdot \frac{k_1^3 k_2^3 k_3^3}{k_1^3 +k_2^3 +k_3^3} B(k_1,k_2,k_3),
\end{align}
where \(B(k_1,k_2,k_3) = \bk{\phipt_{\vb{k}_1}\phipt_{\vb{k}_2}\phipt_{\vb{k}_3}}'\) is the 3-point correlator of the inflaton fluctuations. The current bound from Planck on the 3-point correlator in the equilateral configuration \cite{Planck:2019kim} is
\begin{equation}
    \fNL \coloneqq F(k,k,k) \lesssim \order(10) ,
\end{equation}
where the detailed value depends on the shape of the NG. Meanwhile, future LSS experiments are expected to be able to probe these signals at \(\fNL\sim\order(1)\) \cite{Alvarez:2014vva}, and more futuristic \qty{21}{cm} tomography can further bring this precision down to \(\fNL\sim\order(0.01)\) \cite{Loeb:2003ya,Munoz:2015eqa,Meerburg:2016zdz}.

\begin{figure}[htbp]
    \centering
    \includegraphics{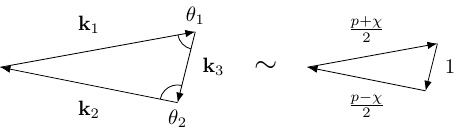}
    \caption{The shape of the momentum triangle and the two parameters. In the squeezed limit \(p\gg1\), the angle parameter \(\chi\simeq\cos\theta_1\simeq-\cos\theta_2\).}
    \label{fig:pchi-param}
\end{figure}

The characteristic cosmological collider signal appears in the squeezed limit \(k_1 \sim k_2 \gg k_3\). We define the squeezing parameter \(p\) and the angle parameter \(\chi\) to characterize the shape of the squeezed triangle up to similarity as follows:
\begin{align}\label{eq:pchi-param}
    p &\coloneqq \frac{k_1 +k_2}{k_3},&
    \chi &\coloneqq \frac{k_1 -k_2}{k_3}.
\end{align}
The corresponding momentum triangle is illustrated in \fig{fig:pchi-param}.
In the limit \(p\gg1\), \(\chi = \cos\theta_1 +\order\p{p^{-1}} = -\cos\theta_2 +\order\p{p^{-1}}\), whereas triangular inequality implies \(p\ge1\) and \(\abs{\chi}\le1\) regardless of whether the squeezed limit is taken.
Under the \((p,\chi)\) parametrization, \eq{eq:3pt-F} becomes
\begin{align}
    F(k_1,k_2,k_3) &= -\frac{10}{3} \dphicl\cdot \frac{4}{p^3 +3\chi^2 p +4} \cdot k_1^3 k_2^3 B\p{\vb{k}_1,\vb{k}_2,\vb{k}_3}\n
    &\sqlimit -\frac{40}{3}\dphicl \cdot p^{-3} \cdot k_1^3 k_2^3 B\p{\vb{k}_1,\vb{k}_2,\vb{k}_3}.
\end{align}
Since the angular dependence does not play a significant role in scalar models, \(\chi=0\) is illustrated in all the plots unless otherwise stated.

\subsection{In-in formalism and mode functions}

In inflationary cosmology, we are mostly interested in the \emph{in-in} correlators, i.e., the expectation values of operators at a specific time (\(t_0\)) in the vacuum state  \cite{Weinberg:2005vy}:
\begin{align}\label{eq:in-in}
    \bk{\Omega_-| \mathcal{O}_H (t_0)|\Omega_-} &= \langle0| \p{\lbar{T}e^{i\int_{-\infty_+}^{t_0} H_{\mathrm{int}}(t)\dd{t}}} \mathcal{O}_I (t_0) \p{Te^{-i\int_{-\infty_-}^{t_0} H_{\mathrm{int}}(t)\dd{t}}} |0\rangle,
\end{align}
where \(\infty_\pm\coloneqq\infty(1\pm i\epsilon)\) contains the \(i\epsilon\)-prescription, which we leave implicit in all expressions henceforth, and \(T\) (\(\lbar{T}\)) represents time (anti-time) ordering. This differs from the usual in-out correlators by having the anti-time ordered exponential.
After expanding the two exponentials in power, one can evaluate in-in correlators using ordinary Feynman rules with two sets of vertices from the time ordered exponential (denoted as \(+\)) and the anti-time ordered exponential (denoted as \(-\)) respectively. Meanwhile, depending on the types of vertices connected in \eq{eq:in-in}, there are four sets of propagators:
\begin{equation}\label{eq:propagator-def}\begin{aligned}
    G_{++}(t_1,t_2) &\coloneqq \bk{0|T\bq{\phi(t_1) \phi(t_2)}|0},\\
    G_{--}(t_1,t_2) &\coloneqq \bk{0|\lbar{T}\bq{\phi(t_1) \phi(t_2)}|0},\\
    G_{-+}(t_1,t_2) &\coloneqq \bk{0|\phi(t_1) \phi(t_2)|0},\\
    G_{+-}(t_1,t_2) &\coloneqq \bk{0|\phi(t_2) \phi(t_1)|0}.
\end{aligned}\end{equation}
The \(G_{++}\) propagator is identical to the Feynman propagator in the usual in-out formalism, while other propagators are either its hermitian conjugate (\(G_{--}\)) or can be obtained from its analytic continuation (\(G_{-+}\) and \(G_{+-}\)).

In de Sitter spacetime it is simpler to work with conformal time, \(\ctime\), and comoving spatial momentum. To calculate the matrix elements in \eq{eq:propagator-def} for scalar fields, we represent the scalar field operator in terms of creation and annihilation operators as
\begin{align}
    \phi(\vb{k},\ctime) &= f_k(\ctime) a_{\vb{k}} +\bar f_k(\ctime) a^\dagger_{-\vb{k}},&
    [a_{\vb{k}'},a^\dagger_{\vb{k}}] &= (2\pi)^3 \delta^3(\vb{k}'-\vb{k}),
\end{align}
where the mode function \(f_k(\ctime)\) is a normalized solution to the free field equation
\begin{equation}\label{eq:mode-function-eq}
    \left\{\begin{aligned}
        \p{\partial_\ctime^2 -\frac{2}{\ctime} \partial_\ctime +k^2 +\frac{M^2}{\ctime^2}} f_k &= 0,\\
        -i\p{f_k \partial_\ctime \bar{f}_k - \bar{f}_k \partial_\ctime f_k} &= \ctime^2.
    \end{aligned}\right.
\end{equation}
With this representation, the four propagators in \eq{eq:propagator-def} are given by
\begin{equation}\begin{aligned}\label{eq:general-propagator}
    G_{++}(k,\ctime_1,\ctime_2) &\coloneqq f_k(\ctime_1) \bar{f}_k(\ctime_2) \theta(\ctime_1-\ctime_2) +\bar{f}_k(\ctime_1) f_k(\ctime_2) \theta(\ctime_2-\ctime_1),\\
    G_{--}(k,\ctime_1,\ctime_2) &\coloneqq \bar{f}_k(\ctime_1) f_k(\ctime_2) \theta(\ctime_1-\ctime_2) +f_k(\ctime_1) \bar{f}_k(\ctime_2) \theta(\ctime_2-\ctime_1),\\
    G_{-+}(k,\ctime_1,\ctime_2) &\coloneqq f_k(\ctime_1) \bar{f}_k(\ctime_2),\\
    G_{+-}(k,\ctime_1,\ctime_2) &\coloneqq \bar{f}_k(\ctime_1) f_k(\ctime_2),
\end{aligned}\end{equation}
where \(\theta\) is the Heaviside step function.
To fix the solution, we must further impose the \emph{Bunch-Davies} initial condition at \(\ctime\to-\infty\):
\begin{equation}
    f_k(\ctime) \sim e^{-ik\ctime},\qquad \ctime\to-\infty.
\end{equation}
This choice ensures that the spacetime contains no particles in the infinite past. The solution to \eq{eq:mode-function-eq} is then given by
\begin{equation}\label{eq:heavy-mode-function}
    f_k(\ctime) = \sqrt{\frac{\pi}{4}} e^{\pi i/4 -\pi \mu/2} (-\ctime)^{3/2} H_{i\mu}^{(1)} (-k\ctime),\qquad
\end{equation}
where \(H^{(1)}\) is the Hankel function of first kind and \(\mu\) is the de Sitter mass parameter defined by
\begin{equation}
    i\mu \coloneqq \begin{dcases}
        i\sqrt{M^2 -\frac{9}{4}}, & M>3/2,\\
        \sqrt{\frac{9}{4} -M^2}, & M<3/2.
    \end{dcases}.
\end{equation}
In the special case of the inflaton, we have \(M=0\) and
\begin{equation}
    f_k(\ctime) = \frac{i}{\sqrt{2k^3}} \p{1 +ik\ctime} e^{-ik\ctime}.
\end{equation}
The primordial fluctuations are measured at the end of inflation, i.e., \(\ctime\to0\), so we also need the bulk-to-boundary propagators for the inflaton:
\begin{equation}\label{eq:inf-propagator}
    D_\pm(k,\ctime) \coloneqq G_{+\pm}(k,0,\ctime) = \frac{1}{2k^3} (1 \mp ik\ctime) e^{\pm ik\ctime}.
\end{equation}
In particular, the leading-order inflaton power spectrum is given by
\begin{equation}\label{eq:inf-power-spectrum}
    \Ps_\phi(k) \coloneqq \bk{\phipt_{\vb{k}} \phipt_{-\vb{k}}}' = D_+(k,0) = \frac{1}{2k^3}.
\end{equation}

\section{Charged-pair Chemical Potential Model and Spectral Decomposition}
\label{sec:ChemPot_SpectDecomp}

In this section we construct a model for a pair of complex scalars (\(\chi_1\) and \(\chi_2\)) with a chemical potential, and opposite charges under all symmetry groups. We then derive the Källén-Lehmann representation for the 3-point correlation function, and discuss some features of the spectral density.

\subsection{Chemical potential for a pair of charged scalars}

For two complex scalar fields, \(\chi_1\), and, \(\chi_2\), a chemical potential-like coupling is achieved by coupling their \(\mathrm{U}(1)\) currents to the derivative of the inflaton field, \(\phi\), as follows \cite{Bodas:2020yho}:
\begin{align}\label{eq:chem-potential-u(1)}
    \mathcal{L} &\supset -\sum_{I=1,2} \frac{1}{2\Lambda} \nabla_\mu\phi J^\mu_I,&
    J^\mu_I &\coloneqq i\p{\bar\chi_I \nabla^\mu \chi_I -\chi_I \nabla^\mu \bar\chi_I}.
\end{align}
After plugging in the classical background, \(\phicl\), the Lagrangian density contains a term \(\frac{\lambda}{2} J_0\) for each field, giving a chemical potential difference of
\begin{equation}\label{eq:lambda}
    \lambda \coloneqq \frac{\dphicl}{\Lambda},
\end{equation}
between \(\chi_1\), \(\chi_2\) and their conjugates \(\bar\chi_1\), \(\bar\chi_2\). An alternative way to write the chemical potential of \eq{eq:chem-potential-u(1)} is through the following Lagrangian:
\begin{align}\label{eq:chem-potential-u(1)-L}
    \mathcal{L}_\chi &\supset -\sum_{I=1,2} \bq{ \p{\nabla^\mu +i\frac{\nabla^\mu\phi}{2\Lambda}} \bar\chi_I \p{\nabla_\mu -i\frac{\nabla_\mu\phi}{2\Lambda}} \chi_I +M_I^2 \abs{\chi_I}^2}\n
    &= -\sum_{I=1,2} \biggl[ \abs{\nabla\chi_I}^2 +M_I^2 \abs{\chi_I}^2 +\frac{1}{2\Lambda} \nabla_\mu\phi J_I^\mu +\frac{1}{4\Lambda^2} \p{\nabla\phi}^2 \abs{\chi_I}^2 \biggr],
\end{align}
which gives the same dimension-5 interaction up to a dimension-6 coupling.

In the form of \eq{eq:chem-potential-u(1)-L}, it is clear that the chemical potential has no nontrivial effect if the full theory is invariant under the following \(\mathrm{U}(1)_A\) symmetry:
\begin{equation}
    \mathrm{U}(1)_A:\quad \chi_1\to \chi_1 e^{i\theta},\qquad \chi_2\to \chi_2 e^{i\theta}.
\end{equation}
In this case a field redefinition \(\chi_I\to \chi_I e^{i\phi/2\Lambda}\) could eliminate the chemical potential term, while leaving any non-derivative couplings of \(\chi_I\) invariant. Absent any further symmetry considerations, the lowest dimensional symmetry-breaking term would be a linear term in \(\chi_I\), leading to the model discussed in \Cite{Bodas:2020yho}, where a heavy particle is exchanged at tree-level. In this paper we will focus on charged scalars, where we assume that \(\chi_1\) and \(\chi_2\) have opposite charges under all other symmetries.
For example if the scalars are also charged under some other \(\mathrm{U}(1)_V\), they should transform under its action as\footnote{One can also build a model with \(\chi_1=\chi_2=\chi\) as a single complex scalar. In this case the same argument applies if \(\chi\) is odd under a \(\mathbb{Z}_2\) symmetry, or it transforms under a nontrivial real representation of some symmetry group. Equivalently, one can start with two real scalars with the same set of symmetries, in which the \(\mathrm{U}(1)_A\) breaking term \(\alpha\chi^2\) naturally comes from the mass splitting between the two real scalars.}
\begin{equation}
    \mathrm{U}(1)_V:\quad \chi_1\to \chi_1 e^{i\theta},\qquad \chi_2\to \chi_2 e^{-i\theta}.
\end{equation}
In this case, the lowest dimensional term that breaks \(\mathrm{U}(1)_A\) is given by \(\chi_1\chi_2\) and we arrive at the following Lagrangian:
\begin{align}\label{eq:1-loop-chem-basis}
    \mathcal{L}_\chi &= -\sum_{I=1,2} \biggl[ \abs{\nabla\chi_I}^2 +M_I^2 \abs{\chi_I}^2 +\kern-1.8ex\underbrace{\frac{1}{2\Lambda} \nabla_\mu\phi J_I^\mu}_{\text{chemical potential}}\kern-1.8ex{} +\frac{1}{4\Lambda^2} \p{\nabla\phi}^2 \abs{\chi_I}^2 \biggr]
    -\underbrace{\alpha \chi_1\chi_2 -\bar{\alpha} \bar\chi_1\bar\chi_2}_{\text{explicit }\cancel{\mathrm{U}(1)}_A}.
\end{align}
The flatness of the inflaton potential is preserved
due to the derivative \(\phi\) couplings.

The role of the explicit \(\mathrm{U}(1)_A\) breaking term is more apparent after the field redefinition \(\chi_I \mapsto \chi_I e^{i\phi/2\Lambda}\), where the Lagrangian in \eq{eq:1-loop-chem-basis} becomes \footnote{One could have started with \(\chi_I\) having different charges. This does not make any difference if only \(\chi_1\chi_2\) term is considered. When one consider the \(\chi_1\bar\chi_2\) term together with uneven charges, one can write down a model with two sources injecting different energies.}
\begin{align}\label{eq:1-loop-exp-basis}
    \mathcal{L}_\chi &= -\sum_{I=1,2} \p{ \abs{\nabla\chi_I}^2 +M_I^2 \abs{\chi_I}^2 } -\alpha \chi_1\chi_2 e^{i\phi/\Lambda} -\bar{\alpha} \bar\chi_1\bar\chi_2 e^{-i\phi/\Lambda},\n
    &=  -\sum_{I=1,2} \p{ \abs{\nabla\chi_I}^2 +M_I^2 \abs{\chi_I}^2 } -\alpha \chi_1\chi_2 e^{i\lambda t} e^{i\phipt/\Lambda} -\bar{\alpha} \bar\chi_1\bar\chi_2 e^{-i\lambda t} e^{-i\phipt/\Lambda}.
\end{align}
The \(\mathrm{U}(1)_A\) breaking term effectively acts as an external source that injects (extracts) energy of order \(\lambda\) into (from) the system and creates (annihilates) \(\chi_1\chi_2\) pairs.
The Lagrangian in this basis still contains a shift symmetry for the inflaton:
\begin{equation}\label{eq:loop-exp-shift}
    \phi \to \phi +c,\qquad \chi_I\to \chi_I e^{ic/2\Lambda}.
\end{equation}
This symmetry plays the same role as the original shift symmetry in preserving the flatness of the inflaton potential.

\subsection{One-loop diagram under spectral representation}

\begin{figure}[htbp]
    \centering
    \begin{tabular}{cccc}
        \includegraphics[scale=0.9]{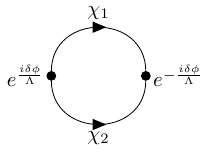} &
        \includegraphics[scale=0.9]{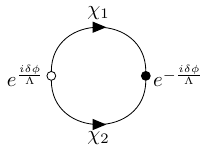} &
        \includegraphics[scale=0.9]{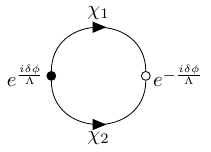} &
        \includegraphics[scale=0.9]{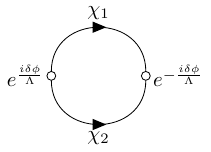}
    \end{tabular}
    \caption{The 1PI one-loop subdiagrams contributing at \(\order\p{\alpha^2}\). The solid dots, \(\bullet\), represents \((+)\)-type vertices, while the empty dots, \(\circ\), represent \((-)\)-type vertices.}
    \label{fig:1-loop-feyn}
\end{figure}

\begin{figure}[htbp]
    \centering
    \begin{tabular}{c@{\hspace{4em}}c}
        \includegraphics[scale=0.9]{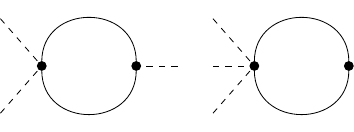} &
        \includegraphics[scale=0.9]{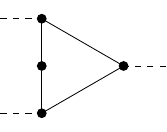}\\
        \(\order\p{\alpha^2}\) & \(\order\p{\alpha^4}\)
    \end{tabular}
    \caption{(\emph{Left}) At \(\order\p{\alpha^2}\), there are only two types of topologies contributing to the 3-point function and both of them admit the spectral density representation. (\emph{Right}) Triangle diagram contributions first appear at \(\order\p{\alpha^4}\). These contributions are subdominant though cannot be represented by a single spectral integral.}
    \label{fig:1-loop-feyn-2}
\end{figure}

The time-dependent coupling in \eq{eq:1-loop-exp-basis} explicitly breaks de Sitter covariance down to scale invariance. However, the breaking effects will remain in control if the mixing, \(\alpha\), can be treated perturbatively.
Since \(\chi_1\chi_2\) are created and annihilated in pairs, the leading-order contribution to the bispectrum is \(\order\p{\alpha^2}\) at one-loop, as illustrated in \fig{fig:1-loop-feyn}. In the in-in formalism, it is given by
\begin{align}
    \bk{\phipt_{\vb{k}_1} \phipt_{\vb{k}_2} \phipt_{\vb{k}_3}} &= \abs{\alpha}^2 \sum_{s_1,s_2=\pm} (-is_1)(-is_2) \int \sqrt{-g_1} \dd[4]{x_1} \int \sqrt{-g_2} \dd[4]{x_2} (-\ctime_1)^{-i\lambda} (-\ctime_2)^{i\lambda}\n
    &\peq \times \bk{\phipt_{\vb{k}_1}\phipt_{\vb{k}_2}\phipt_{\vb{k}_3}\cdot e^{i\phipt(x_1)/\Lambda} e^{-i\phipt(x_2)/\Lambda}}_{s_1s_2}
    \bk{\Op(x_1) \lbar\Op(x_2)}_{s_1s_2},
\end{align}
where \(\Op = \chi_1\chi_2\) and the two signs \(s_1,s_2=+/-\) indicate whether the vertex comes from the time or anti-time ordered exponential. At leading order, the two correlators of \(\phipt\) and \(\chi_I\) can be evaluated as in the free field theory, where the second correlator gives rise to a loop in momentum space:
\begin{equation}\label{eq:loop-integral}
    \bk{\Op(\vb{k},\ctime_1) \lbar\Op(\vb{k},\ctime_2)}'_{s_1s_2} = \int \frac{\dd[3]{\vb{q}}}{(2\pi)^3} G_{s_1s_2}(\abs{\vb{q}},\ctime_1,\ctime_2;\mu_1) G_{s_1s_2}(\abs{\vb{k}-\vb{q}},\ctime_1,\ctime_2;\mu_2),
\end{equation}
with \(G_{s_1s_2}\) defined in \eqs{eq:general-propagator}{eq:heavy-mode-function}.
Note that at \(\order\p{\alpha^2}\), there is only two insertions of the exponentials \(e^{\pm i\phipt(x_1)/\Lambda}\) with which \(\phipt_{\vb{k}}\) can contract, as shown in \fig{fig:1-loop-feyn-2}. In particular, triangle diagrams do not appear at \(\order\p{\alpha^2}\) and are thus higher order in the \(\alpha\)-expansion.

\begin{figure}[htbp]
    \centering
    \includegraphics{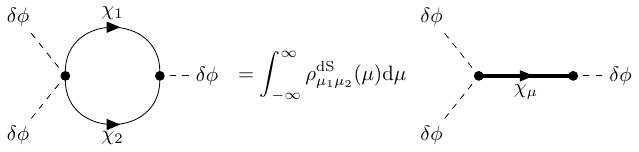}
    \caption{The spectral representation of the 3-point correlator at one-loop. This diagram only shows one of the ways \(\phipt\) can contract with \(e^{\pm i\phipt/\Lambda}\).}
    \label{fig:1-loop-kallen}
\end{figure}

Loop integrations such as the one in \eq{eq:loop-integral} are, in general, hard to evaluate either analytically or numerically. Nevertheless, since the propagators appearing in \eq{eq:loop-integral} are free propagators respecting the de Sitter isometries, it can be simplified through the \emph{spectral representation}. The idea is similar to the K\"all\'en–Lehmann representation in Minkowski spacetime: to calculate the matrix element \(\bk{\Op\lbar\Op}_{-+} =\bk{0|\Op\lbar\Op|0}\), we first insert a complete set of states to get
\begin{equation}\label{eq:kallen-1}
    \bk{\Op(\vb{k},\ctime_1) \lbar\Op(\vb{k},\ctime_2)}_{-+}
    = \sumint \dd{n} \int\frac{\dd[3]{\vb{k}'}}{(2\pi)^3} \bk{0|\Op(\vb{k},\ctime_1)|\vb{k}',\mu_n} 
    \overline{\bk{0|\Op(\vb{k},\ctime_2)|\vb{k}',\mu_n}},
\end{equation}
where \(\ket{\vb{k},\mu_n}\) is a state with momentum, \(\vb{k}\), and a mass parameter, \(\mu_n\), that can either be a single-particle state or a multi-particle state with a definite invariant mass. The de Sitter isometry then implies that the matrix element has a unique form up to a multiplicative constant:\footnote{The easiest way to derive this is to consider the de Sitter Casimir operator \(C_1\), which has a constant value \(\mu^2+9/4\) over irreducible representations, while acting on scalar operators as the d'Alembert operator:
\[
    [C_1,\Op(x)] = \Box \Op(x) = \ctime^2 \p{ -\partial_\ctime^2 +\frac{2}{\ctime} \partial_\ctime +\nabla^2} \Op(x).
\]
This implies that the matrix element \(\bk{0|\Op(\vb{k}',\ctime)|\vb{k},\mu_n}\) satisfies the differential equation \eq{eq:mode-function-eq} with \(M^2=\mu_n^2+9/4\).}
\begin{equation}
    \bk{0|\Op(\vb{k}',\ctime)|\vb{k},\mu_n} = \bk{0|\Op|\mu_n} \cdot f_k(\ctime;\mu_n)\cdot (2\pi)^3\delta^3(\vb{k}'-\vb{k}),
\end{equation}
where \(\bk{0|\mathcal{O}|\mu_n}\) is the reduced matrix element and \(f_k(\ctime;\mu)\) is defined by \eq{eq:heavy-mode-function}. Plugging this into \eq{eq:kallen-1} gives
\begin{align}
    \bk{\Op(\vb{k},\ctime_1) \lbar\Op(\vb{k},\ctime_2)}_{-+}' &= \sumint \abs{\bk{0|\Op|\mu_n}}^2 G_{-+}(k,\ctime_1,\ctime_2;\mu_n) \dd{n}\n
    &\eqqcolon \int_0^\infty \rho^{\text{dS}}_{\Op} (\mu)  G_{-+}(k,\ctime_1,\ctime_2;\mu) \dd{\mu},
\end{align}
where
\begin{equation}
    \rho^{\text{dS}}_{\Op} (\mu) \coloneqq \sumint \abs{\bk{0|\Op|\mu_n}}^2 \delta(\mu_n-\mu) \dd{n}
\end{equation}
is the de Sitter spectral density of \(\Op\). Equivalently:\footnote{This form of spectral decomposition is only correct when either at least one scalar has \(M_I\ge3/2\), or if both scalars have \(M_I<3/2\) then they must satisfy
\[
    \sqrt{\frac{9}{4}-M_1^2} +\sqrt{\frac{9}{4}-M_2^2}\le\frac{3}{2}.
\]
For these values of \(M_I\), the multi-particle states all have \(M\ge3/2\).}
\begin{equation}\label{eq:kallen}
    \bk{\Op(x_1) \lbar\Op(x_2)}_{s_1s_2} = \int_0^\infty \rho^{\text{dS}}_{\Op} (\mu) \bk{\chi_\mu(x_1) \bar\chi_\mu(x_2)}_{s_1s_2} \dd{\mu},
\end{equation}
where \(\chi_\mu\) is a fictitious scalar with mass parameter \(\mu\). We then arrive at the following relationship between the one-loop correlator and a fictitious tree-level correlator:
\begin{equation}\label{eq:kallen-bispectrum}
    B_{\wloop}(k_1,k_2,k_3) = \int_0^\infty \rho^{\text{dS}}_{\mu_1\mu_2} (\mu) B_{\tree}(k_1,k_2,k_3;\mu) \dd{\mu},
\end{equation}
This is shown diagrammatically as in \fig{fig:1-loop-kallen}.

\begin{figure}[htbp]
    \centering
    \includegraphics[scale=0.8]{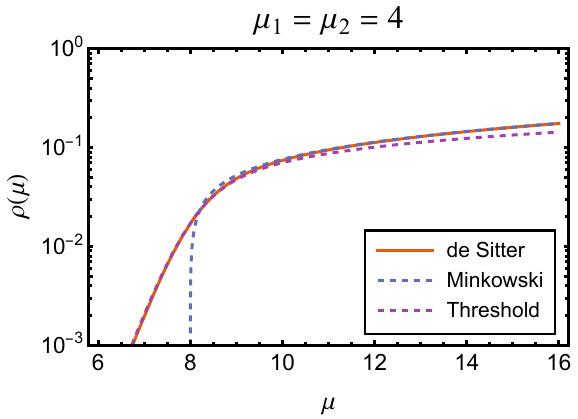}
    \caption{The de Sitter spectral density \eq{eq:ds-spectral} compared with the Minkowski spectral density \eq{eq:flat-spectral}, and the near-threshold approximation \eq{eq:ds-spectral-threshold}, for \(\mu_1=\mu_2=4\).}
    \label{fig:ds-spectral}
\end{figure}

The corresponding de Sitter spectral density for \(\Op=\chi_1\chi_2\) has been evaluated and has a surprisingly simple form \cite{Bros:2009bz, Loparco:2023rug, Hogervorst:2021uvp}:\footnote{We are also aware of another expression for the spectral density in terms of the hypergeometric function \({}_7F_6\) in \cite{Marolf:2010zp,Xianyu:2022jwk}. It can be numerically verified that our spectral density is related to the spectral density in \cite{Xianyu:2022jwk}, denoted as \(\mathrm{P}^{\text{dS}}_{\mu_0}(\mu)\), through the following equation:
\begin{equation*}
    \rho_{\mu_0\mu_0}^{\mathrm{dS}}(\mu) = \frac{2\mu}{\pi} \Im \mathrm{P}^{\text{dS}}_{\mu_0}(\mu).
\end{equation*}}
\begin{equation}\label{eq:ds-spectral}
    \rho^{\text{dS}}_{\mu_1\mu_2} (\mu) = \frac{1}{8\pi^4} \frac{\mu\sinh\pi\mu}{\pi} \frac{\prod_{\epsilon,\epsilon_1,\epsilon_2=\pm} \Gamma\bq{\frac{3}{4} +\frac{i}{2} \epsilon\p{\mu +\epsilon_1\mu_1 +\epsilon_2\mu_2}}}{\Gamma\p{\frac{3}{2}-i\mu} \Gamma\p{\frac{3}{2}+i\mu}}.
\end{equation}
In this paper we will not go through the derivation of this expression, but only discuss some of its features:

\begin{itemize}
    \item At small distances the spacetime curvature is unimportant and we should expect that \eq{eq:ds-spectral} recovers its Minkowski form. This is achieved by taking the limit \(\mu,\mu_1,\mu_2\gg1\), where all gamma functions in \eq{eq:ds-spectral} can be replaced by their Stirling approximation. One obtains
    \begin{align}
        \rho^{\text{dS}}_{\mu_1\mu_2} (\mu)
        &\simeq \frac{1}{8\pi^2 \mu}  \prod_{\epsilon_1,\epsilon_2=\pm} \sqrt{\abs{\mu +\epsilon_1\mu_1 +\epsilon_2\mu_2}}\n
        &\peq \times \begin{dcases}
            1, & \mu>\mu_{12},\\
            e^{-\pi(\mu_{12} -\mu)}, & \abs{\mu_1-\mu_2} < \mu < \mu_{12},\\
            e^{-2\pi(\max\Bq{\mu_1,\mu_2} -\mu)}, & \mu < \abs{\mu_1-\mu_2},
        \end{dcases}
    \end{align}
    where \(\mu_{12}\coloneqq\mu_1+\mu_2\). For comparison, we quote the Minkowski spectral density for \(\Op=\chi_1\chi_2\):
    \begin{align}\label{eq:flat-spectral}
        \rho^{\text{Mink}}_{M_1M_2} (M)
        &= \frac{1}{8\pi^2 M}  \prod_{\epsilon_1,\epsilon_2=\pm} \sqrt{M +\epsilon_1M_1 +\epsilon_2M_2} \cdot \theta(M -M_1 -M_2).
    \end{align}
    The comparison between the two spectral densities is presented in \fig{fig:ds-spectral}. For \(\mu_1,\mu_2\gg1\), the two spectral densities are close for \(\mu-\mu_1-\mu_2\gg1\), while they have major difference for \(\mu<\mu_1+\mu_2\). In particular, the Minkowski spectral density \eq{eq:flat-spectral} has a sharp cutoff at \(\mu=\mu_1+\mu_2\), whereas the de Sitter spectral density decays exponentially for \(\mu<\mu_1+\mu_2\).

    \item As will be shown in the next section, the oscillatory signal is determined by the \emph{threshold limit} of the spectral density, i.e. \(\mu=\mu_{12}+\varepsilon\) with \(\varepsilon=\order(1)\ll \mu_1,\mu_2\). In this limit all gamma functions in \eq{eq:ds-spectral} can be approximated except
    \begin{equation}
        \Gamma\bq{\frac{3}{4} \pm \frac{i}{2}\p{\mu -\mu_{12}}} = \Gamma\p{\frac{3}{4} \pm\frac{i\varepsilon}{2}},
    \end{equation}
    giving
    \begin{equation}\label{eq:ds-spectral-threshold}
        \rho^{\text{dS}}_{\mu_1\mu_2} (\mu_{12}+\varepsilon) \simeq \frac{1}{4\pi^3} \sqrt{\frac{\mu_1\mu_2}{\mu_{12}}} \Gamma\p{\frac{3}{4} +\frac{i\varepsilon}{2}} \Gamma\p{\frac{3}{4} -\frac{i\varepsilon}{2}} e^{\pi\varepsilon/2}.
    \end{equation}
    This form of the spectral density is plotted in \fig{fig:ds-spectral}, showing good agreement with the general form for small values of \(\mu\). It also differs from the flat space case \eq{eq:flat-spectral}, where one has
    \begin{equation}\label{eq:flat-spectral-threshold}
        \rho^{\text{Mink}}_{\mu_1\mu_2} (M_{12}+E) \simeq \frac{1}{4\pi^2} \sqrt{\frac{M_1 M_2}{M_{12}}} \sqrt{2E} \cdot \theta(E).
    \end{equation}
    This difference is important in our chemical potential model because we will show later that the proper time elapsed from the creation to the annihilation of the heavy particle(s) is larger than \(H^{-1}\) in the squeezed limit. Consequently, the particle(s) can see the curvature in the time direction and distinguish de Sitter from flat spacetime. \Eq{eq:ds-spectral-threshold} can be better understood by deriving it from the non-relativistic limit of de Sitter spacetime.
    Since this deviates from the main purpose of this paper, we leave the discussion in \app{app:NRdS}.

    \item As a function of complex \(\mu\), \(\rho^{\text{dS}}_{\mu_1\mu_2} (\mu)\) is meromorphic, and only has contains simple poles at
    \begin{equation}
        \mu = \pm\mu_1 \pm\mu_2  \pm\p{\frac{3}{2} +2\N}i.
    \end{equation}
    Meanwhile, the late-time behavior of the mode function \eq{eq:heavy-mode-function} is given by
    \begin{align}
        f_k(\ctime;\mu) &= \sum_{k=0}^\infty \bq{ A_k (-\ctime)^{\Delta^+ +2k} +B_k (-\ctime)^{\Delta^- +2k} },&
        \begin{dcases}
            \Delta^+ \coloneqq \frac{3}{2}+i\mu,\\
            \Delta^- \coloneqq \frac{3}{2}-i\mu.
        \end{dcases}.
    \end{align}
    Consequently, poles in the upper half plane correspond to the \(\mu\) values when \(\Delta^+ = \Delta_1^\pm +\Delta_2^\pm\). From the dS/CFT perspective, these poles can further be attributed to different late-time operators appearing in the operator product expansion of \(\chi_1\chi_2\) \cite{Hogervorst:2021uvp}.
\end{itemize}

\subsection{Central result for the signal}

\label{subsec:central-result}

A systematic calculation of all contributions to the one-loop bispectrum will be carried out in the next section. However, in this section we give a quick derivation of the non-analytic piece in the squeezed limit. This piece is the oscillatory signal in cosmological collider physics. Our starting point will be the tree-level amplitude in \eqref{eq:kallen-bispectrum}, which is given by
\begin{align}\label{eq:tree-ampl}
    &B_{\tree}\p{k_1,k_2,k_3;\mu}= \abs{\alpha}^2 \sum_{s_1,s_2} (is_1)(is_2) \int \sqrt{-g_1} \dd[4]{x_1} \int \sqrt{-g_2} \dd[4]{x_2} (-\ctime_1)^{-i\lambda} (-\ctime_2)^{i\lambda}\n
    &\peq\hspace{2.3cm}\times \bk{\phipt_{\vb{k}_1}\phipt_{\vb{k}_2}\phipt_{\vb{k}_3}\cdot e^{i\phipt(x_1)/\Lambda} e^{-i\phipt(x_2)/\Lambda}}_{s_1s_2}
    \bk{\chi_\mu(x_1) \bar\chi_\mu(x_2)}_{s_1s_2}.
\end{align}
The amplitude for the fictitious scalar has been computed in \Cite{Bodas:2020yho} as part of a tree-level model. The dominant contribution comes from the \((-+)\) diagram. 

\begin{figure}[htbp]
    \centering
    \begin{tabular}{cc}
        \includegraphics[scale=0.75]{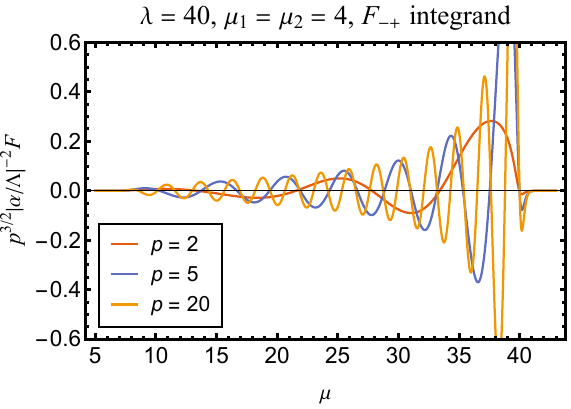} &
        \includegraphics[scale=0.75]{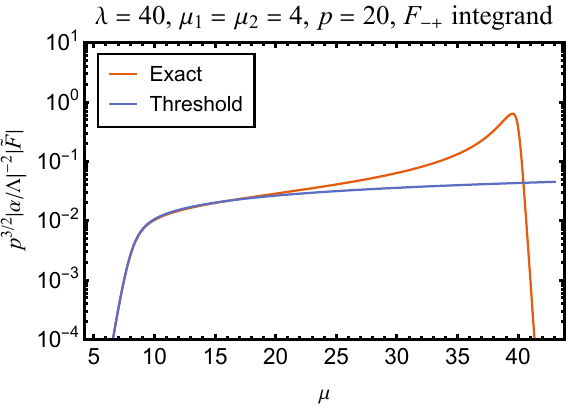}
    \end{tabular}
    \caption{(\emph{Left}) The integrand in \eq{eq:kallen-bispectrum} for \(\lambda=40\), \(\mu_1=\mu_2=4\) and various \(p\). The normalized bispectrum \eq{eq:3pt-F} is used instead as the \(y\)-axis when making the plot. (\emph{Right}) The magnitude of the integrand for \(p=20\). The orange line represents the exact integrand while the blue line represents the modified integrand to extract the near-threshold approximation in \eq{eq:1-loop-spa-integral-2}.}
    \label{fig:(-+)-integrand}
\end{figure}

In the left panel of figure \ref{fig:(-+)-integrand} we plot the integrand in \eq{eq:kallen-bispectrum} for the \((-+)\) diagram.
The integrand has the form of an oscillatory function multiplied by a smooth function. In particular, as we will see shortly, the phase does not have a stationary point. 
However, the smooth envelope changes rapidly around \(\mu=\lambda\), and \(\mu=\mu_{12}\). Near \(\mu=\lambda\), the tree-level amplitude \(B_\tree\) peaks and then decays exponentially once \(\mu>\lambda\). Near \(\mu=\mu_{12}\), the spectral density also decays exponentially for \(\mu<\mu_{12}\). The integral is therefore dominated by the two regions where \(\mu\simeq\mu_{12},\lambda\).

For values of the mass parameter, \(\mu \lesssim \lambda\), the squeezed limit tree level amplitude can be approximately calculated in the \emph{Stationary Phase Approximation} (SPA), and it is given by \cite{Bodas:2020yho}
\begin{align}\label{eq:tree-MP-spa}
    B_{\tree,\text{SPA}}(k_1,k_2,k_3;\mu) &= \frac{\pi}{16k_1^3 k_2^3} \frac{\abs{\alpha}^2}{\Lambda^3} \frac{\lambda^{1/2} \mu^{-1/2}}{\lambda^2 -\mu^2} p^{3/2 -i(\lambda-\mu)} e^{i\phase_*(\lambda,\mu)} +\cc, 
\end{align}
where the phase \(\phase_*\) is given by\footnote{Ref. \cite{Bodas:2020yho} does not present an expression for this phase as it is unimportant for their computation. We will derive it in \Sec{sec:tree-SPA}.}
\begin{align}\label{eq:tree-phase-sq}
    \phase_*(\lambda,\mu) &\coloneqq \lambda \log2\lambda +\mu \log2\mu -(\lambda+\mu) \log(\lambda+\mu).
\end{align}
Plugging this into \eq{eq:kallen-bispectrum}, we obtain
\begin{align}\label{eq:1-loop-spa-integral}
    B_{\wloop}(k_1,k_2,k_3) &\sim \int_0^\infty \rho^{\text{dS}}_{\mu_1\mu_2} (\mu) B_{\tree}^{(\text{SPA})}(k_1,k_2,k_3;\mu)\n
    &\sim \frac{\pi}{16k_1^3 k_2^3} \frac{\abs{\alpha}^2}{\Lambda^3} \int_0^\lambda  \frac{\lambda^{1/2} \mu^{-1/2}}{\lambda^2 -\mu^2} p^{3/2 -i(\lambda-\mu)} e^{i\phase_*(\lambda,\mu)} \rho^{\text{dS}}_{\mu_1\mu_2} (\mu) \dd{\mu}.
\end{align}
Note that the superficial divergence when \(\mu\to\lambda\) is an artifact of SPA and does not exist in the exact expression.

The full phase\footnote{Even for moderate \(p\), \(\phase\) is given by \eq{eq:tree-phase} and its derivative is given by
\[
    \frac{\partial\phase}{\partial\mu} = \frac{1}{2}\log\frac{\lambda +\mu-px_*}{\lambda -\mu-px_*}-\log\frac{\lambda+\mu}{\lambda-\mu} > 0,
\]
where \(x_*=x_*^-\) is given by \eq{eq:stationary-phase}. The latter is clearly positive-definite.}
\begin{align}\label{eq:tree-phase-sq2}
    \phase(\lambda,\mu,p) &\coloneqq -(\lambda-\mu)\log p +\phase_*(\lambda,\mu),&
    \frac{\partial\phase}{\partial\mu} &= \log\frac{2\mu p}{\lambda+\mu} > 0
\end{align}
does not, as we mentioned before, have a stationary point on its own. Inspection of \eq{eq:1-loop-spa-integral} tells us that the non-analyticity is proportional to \(p^{-i(\lambda -\mu)}\). Thus, only the \(\mu\to\mu_{12}\) limit contribution has oscillating behavior as a function of \(p\). 

To isolate this limit, we consider a modified integral:
\begin{align}\label{eq:1-loop-spa-integral-2}
    B_{\wloop}^{(\mu_{12})}(k_1,k_2,k_3) &\simeq B_{\tree}^{(\text{SPA})}(k_1,k_2,k_3;\mu_{12}) \int_{-\infty}^{\infty_+} \rho^{\text{dS}}_{\mu_1\mu_2} (\mu_{12} +\varepsilon) \beta^{i\varepsilon} \dd{\varepsilon},\\
    \beta &\coloneqq \eval{ \exp\p{\frac{\partial\phase}{\partial\mu}} }_{\mu=\mu_{12}} = \frac{2\mu_{12} p}{\lambda +\mu_{12}}.
\end{align}
As shown in the right panel of \fig{fig:(-+)-integrand}, this new integral leaves the envelope of the integrand near \(\mu\simeq\mu_{12}\) unmodified, but also lets it continues to oscillate beyond \(\mu=\lambda\), eliminating the contribution near that limit. To simplify the calculation, we have also expanded the full phase \eq{eq:tree-phase-sq2} to linear order around \(\mu=\mu_{12}\).

We can now further approximate \(\rho^{\text{dS}}_{\mu_1\mu_2} (\mu)\) by \eq{eq:ds-spectral-threshold} and obtain an approximate form for the signal
\begin{align}\label{eq:1-loop-na}
    B_{\wloop,\sq}^{(\mu_{12})} (k_1,k_2,k_3) &\simeq \frac{1}{2\pi^{3/2}} \sqrt{\frac{\mu_1\mu_2}{\mu_{12}}} \p{\frac{2\mu_{12} p}{\lambda +\mu_{12}}}^{-3/2} e^{3\pi i/4} B_{\tree}^{\rm(SPA)}(k_1,k_2,k_3;\mu_{12})\n
    &= \frac{1}{32\sqrt{\pi} k_1^3 k_2^3} \frac{\abs{\alpha}^2}{\Lambda^3}  \frac{\lambda^{1/2} \mu_{12}^{-1/2}}{\lambda^2 -\mu_{12}^2} \sqrt{\frac{\mu_1\mu_2}{\mu_{12}}}\n
    &\peq \times\p{\frac{2\mu_{12} }{\lambda +\mu_{12}}}^{-3/2} p^{-i(\lambda-\mu_{12})} e^{i\phase_*(\lambda,\mu_{12}) +3\pi i/4}.
\end{align}
Here we have used the Gamma function integral identity in \eq{eq:GammaIntegral} to perform the integral over \(\epsilon\). Matching to the bispectrum parametrized by \eq{eq:3pt-F}, the non-analytic piece of the signal is given by
\begin{align}
    F_{\wloop,\sq}^{(\mu_{12})} (k_1,k_2,k_3) &= \widetilde{F}_{\wloop,\sq}^{(\mu_{12})} (k_1,k_2,k_3) +\cc,\\\label{eq:1-loop-na-F}
    \widetilde{F}_{\wloop,\sq}^{(\mu_{12})} (k_1,k_2,k_3) 
    &\simeq f_{\text{oscil}}(\lambda,\mu_1,\mu_2)p^{-3-i(\lambda -\mu_{12})},
\end{align}
where the function, \(f_{\text{oscil}}\), is given by
\begin{align}
    f_{\text{oscil}}&= -\frac{5}{12\sqrt{\pi}} \frac{\abs{\alpha}^2} {\Lambda^2} \frac{\lambda^{3/2} \mu_{12}^{-1/2}}{\lambda^2 -\mu_{12}^2} \sqrt{\frac{\mu_1\mu_2}{\mu_{12}}}
    \p{\frac{2\mu_{12} }{\lambda +\mu_{12}}}^{-3/2}  e^{i\phase_*(\lambda,\mu_{12}) +3\pi i/4}.
\end{align}

Before going through a more systematic derivation, we list some important features of the signal here:

\begin{itemize}
    \item The \(\beta\)-factor defined in \eq{eq:1-loop-spa-integral-2} has a clear physical meaning: it measures the proper time lapse between the creation and annihilation of the heavy particle(s) in its own reference frame. This can be seen by noting that the \(\mu\)-dependence in the phase \(\phase(\lambda,\mu,p)\) comes from the time evolution of the heavy particle(s), thus
    \begin{align}\label{eq:time-lapse}
        \log \beta &= \frac{\partial}{\partial\mu_{12}} \int_{\ctime_{1*}}^{\ctime_{2*}} E(\ctime) \frac{\dd{\ctime}}{(-\ctime)} = \int_{\ctime_{1*}}^{\ctime_{2*}} \frac{\mu_{12}}{E(\ctime)} \frac{\dd{\ctime}}{(-\ctime)} = \Delta T,
    \end{align}
    where
    \begin{align}
        E(\ctime) &\coloneqq \sqrt{k_3^2 \ctime^2 +\mu_{12}^2}
    \end{align}
    is the local energy of the heavy particle(s). This shows that \(e^{\Delta T}\) is proportional to the squeeze parameter \(p\) in the squeezed limit.

    \item Compared to the \(p^{-3/2}\) dilution in tree-level exchanges, the one-loop signal decays faster as \(p^{-3}\). This is because \(p^{-3/2}\) is  a result of the matter-like dilution of the heavy particle between its production and decay. Due to the presence of two heavy particles in the loop, the cost of dilution is doubled.

    \item At large \(p\) the oscillation frequency of the one-loop amplitude is the same as would be obtained from a tree-level exchange, except for a phase shift \(3\pi i/4\). 
    As has been discussed in \Cite{Qin:2022lva}, this phase can be used to separately determine \(\lambda\) and \(\mu\) from the data, when combined with the oscillating frequency. In \Cite{Bodas:2020yho,Bodas:2024hih}, it is also suggested that such separation can be achieved by measuring the frequency of the non-analytic component as a function of \(p\) for small to moderate values of \(p\). For this purpose, we will also derive the SPA approximation for the signal applicable to moderate \(p\) in \Sec{subsec:signal-2}. 
\end{itemize}

In this section we have approximately calculated the non-analytic piece of the bispectrum, but we did note that the integral in \eq{eq:kallen-bispectrum} has a large contribution from the region \(\mu\rightarrow \lambda\). In this region the contribution is neither oscillatory, nor suppressed. Therefore, this model is a good case study where the exchange of the heavy scalars provides both the signal and background.

The following two sections will be dedicated to a systematic calculation of all one-loop contributions to the bispectrum, including the backgrounds.

\section{Systematic Computation: Tree-level Model Revisited}
\label{sec:Systematic_tree}

In this section we will evaluate the tree-level amplitude \eq{eq:tree-ampl} systematically. The amplitude \eq{eq:tree-ampl}, when evaluated by series expanding \(e^{\pm i\phipt/\Lambda}\), contains late-time divergence in individual contributions. We will perform a change of basis in the effective tree-level Lagrangian to eliminate this divergence explicitly. Besides the results in \Cite{Bodas:2020yho}, we will also derive some extra quantities necessary for the one-loop calculation, such as the full form of the phase in \eq{eq:tree-phase}, the improved approximated form in \eq{eq:tree-MP-spa2}, and the contributions from the \((++)\)-diagrams in \Sec{subsec:tree-subdom}.

\subsection{Perturbative expansion under late-time-finite basis}

\begin{figure}[htbp]
    \centering
    \begin{tabular}{cc}
        \includegraphics{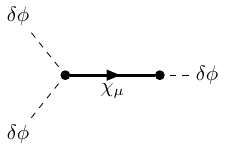} &
        \includegraphics{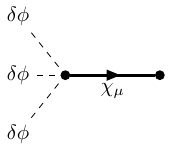}
    \end{tabular}
    \caption{The effective tree-level diagrams from the spectral decomposition. There are another two diagrams with \(\chi_\mu\) flowing in the opposite direction. In the new basis described by \eq{eq:tree-int-basis}, only the first diagram is present.}
    \label{fig:tree-exp}
\end{figure}

In \Cite{Bodas:2020yho}, the amplitude \eq{eq:tree-ampl} is evaluated by series expanding \(e^{\pm i\phipt/\Lambda}\). This gives rise to two diagrams, shown in \fig{fig:tree-exp}. 
It turns out that both diagrams contain divergence in the late-time region. 
To understand its origin note that, at leading order in \(\alpha\), the tree-level amplitude can be derived from the following Lagrangian:
\begin{align}\label{eq:tree-exp-basis}
    \mathcal{L}_\chi &= -\abs{\nabla\chi_\mu}^2 -M^2 \abs{\chi_\mu}^2 -\alpha_\mu \chi_\mu e^{i\phi/\Lambda} -\bar{\alpha}_\mu \bar\chi_\mu e^{-i\phi/\Lambda},\n
    &=  -\abs{\nabla\chi_\mu}^2 -M^2 \abs{\chi_\mu}^2 -\alpha_\mu (-\ctime)^{-i\lambda} \chi_\mu e^{i\phipt/\Lambda} -\bar{\alpha}_\mu (-\ctime)^{i\lambda} \bar\chi_\mu e^{-i\phipt/\Lambda},
\end{align}
where \(t=-\log(-\ctime)\) and \(M^2=\mu^2+9/4\). Note that the \(\mu\)-dependent coupling \(\alpha_\mu\) should account for the variation in spectral density: by matching to the tree-level result, we should have
\begin{equation}
    \abs{\alpha_\mu}^2 = \rho^{\text{dS}}_{\mu_1\mu_2} (\mu) \abs{\alpha}^2.
\end{equation}
In the rest of this section, we will drop the subscript \(\mu\) on \(\chi_\mu\) and \(\alpha_\mu\) for simplicity.

The Lagrangian in \eq{eq:tree-exp-basis} is invariant under
\begin{equation}
    \phi \to \phi +c,\qquad \chi\to \chi e^{ic/\Lambda},
\end{equation}
following from \eq{eq:loop-exp-shift}, which ensures no late-time divergence. However, the shift symmetry is not manifest if one expands \(e^{\pm i\phipt/\Lambda}\) in a power series for perturbative calculations. As a result, late-time divergences are present in individual diagrams, and only cancel after summing over all diagrams at the same order.

We are free to perform field redefinitions of \(\chi\) in the Lagrangian \eq{eq:tree-exp-basis} because the contribution to the amplitude in \eq{eq:tree-ampl} will remain unchanged.
To avoid complications with late-time divergences, consider the field redefinition \(\chi \mapsto \chi e^{-i\phi/\Lambda}\) that brings the Lagrangian back to its chemical potential form:
\begin{align}\label{eq:tree-chem-basis}
    \mathcal{L}_\chi &= -\p{\nabla^\mu +i\frac{\nabla^\mu\phi}{\Lambda}} \bar\chi \p{\nabla_\mu -i\frac{\nabla_\mu\phi}{\Lambda}}\chi -M^2 \abs{\chi}^2 -\alpha \chi -\bar{\alpha} \bar\chi\n
    &= -\abs{\nabla\chi}^2 -M^2 \abs{\chi}^2 -\alpha \chi -\bar{\alpha} \bar\chi +\frac{1}{\Lambda} \nabla_\mu\phi J^\mu -\frac{1}{\Lambda^2} \p{\nabla\phi}^2 \abs{\chi}^2.
\end{align}
Treating the last two terms as interactions, the equation of motion of \(\chi\) at free field order is given by
\begin{equation}
    \bq{-\p{\nabla_\mu -i\frac{\nabla_\mu\phicl}{\Lambda}}^2 +M^2}\chi = -\bar\alpha,\qquad
    \phicl=\dphicl t=-\dphicl\log(-\ctime).
\end{equation}
The inhomogeneous term on the r.h.s. gives rise to a homogeneous classical background for \(\chi\):
\begin{equation}\label{eq:chi-vev}
    \chicl = \frac{\bar\alpha}{\lambda^2 +3i\lambda -M^2} = -\frac{\bar\alpha}{\p{\frac{3}{2} -i\lambda}^2 +\mu^2}.
\end{equation}
Now, consider the following field redefinition:
\begin{equation}
    \chi \mapsto \chicl +\chipt e^{i\phi/\Lambda}.
\end{equation}
Substituting this into \eq{eq:tree-chem-basis} gives
\begin{align}\label{eq:tree-int-basis}
    \mathcal{L}_{\chipt} &= -\abs{\nabla\chipt} -M^2\abs{\chipt}^2 -\frac{\abs{\chicl}^2}{\Lambda^2}\p{\nabla\phi}^2 +\mathcal{L}_{\chipt}^{(\text{int})}\n
    &\supset -\abs{\nabla\chipt} -M^2\abs{\chipt}^2 -\frac{\abs{\chicl}^2}{\Lambda^2}\p{\nabla\phipt}^2 +\mathcal{L}_{\chipt}^{(\text{int})},\\[1em]
    \mathcal{L}_{\chipt}^{(\text{int})} &= \bchicl (-\ctime)^{-i\lambda} \bq{\frac{i}{\Lambda} \Box\phipt +2(-\ctime)\frac{\lambda}{\Lambda} \phipt' -\frac{1}{\Lambda^2} (\nabla\phipt)^2} e^{i\phipt/\Lambda} \chipt +\cc\n
    &= \bchicl (-\ctime)^{-i\lambda} \bq{\frac{i}{\Lambda} \Box\phipt +2(-\ctime)\frac{\lambda}{\Lambda} \phipt' -\frac{1}{\Lambda^2} (\nabla\phipt)^2} \p{1 +\frac{i}{\Lambda} \phipt +\cdots} \chipt +\cc,
\end{align}
where we have dropped all constant terms and tadpoles in \(\phipt\) at the second line of \(\mathcal{L}_{\chipt}\). 
This new basis is advantageous over the exponential basis \eq{eq:tree-exp-basis} as at least one of the \(\phipt\) in each interaction has a derivative acting on it. This improves the late-time behavior and leads to finite results for each individual diagram.

\begin{figure}[htbp]
    \centering
    \newcommand{\scale}{0.75}
    \begin{tabular}{cccc}
        \includegraphics[scale=\scale]{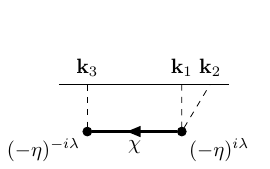} &
        \includegraphics[scale=\scale]{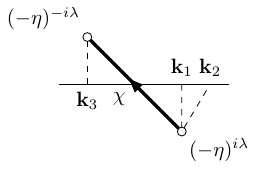} &
        \includegraphics[scale=\scale]{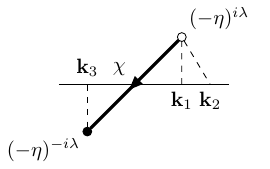} &
        \includegraphics[scale=\scale]{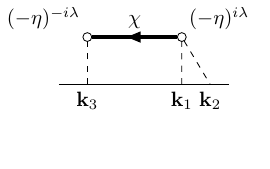}\\
        \(B_{++}^{(-)}\) & \(B_{-+}^{(-)}\) & \(B_{+-}^{(-)}\) & \(B_{--}^{(-)}\)\\[1.5em]
        \includegraphics[scale=\scale]{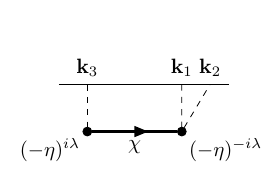} &
        \includegraphics[scale=\scale]{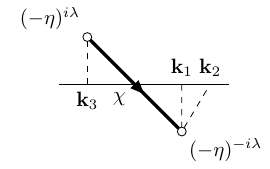} &
        \includegraphics[scale=\scale]{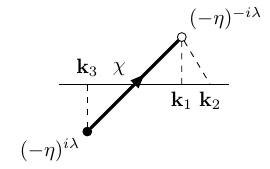} &
        \includegraphics[scale=\scale]{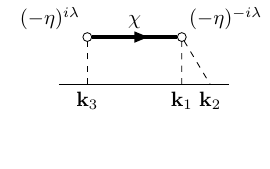}\\
        \(B_{++}^{(+)}\) & \(B_{-+}^{(+)}\) & \(B_{+-}^{(+)}\) & \(B_{--}^{(+)}\)
    \end{tabular}
    \caption{The various effective tree diagrams contributing to the bispectrum. The solid dots \(\bullet\) represent \((+)\)-type vertices, while the empty dots \(\circ\) represent \((-)\)-type vertices. For each diagram shown, there are also another two diagrams where \(\vb{k}_3\) is permuted with \(\vb{k}_1\) or \(\vb{k_2}\). Here we also use the upper/lower half plane notation for the two types of vertices to indicate the direction of time: \((+)\)-type vertices are in the lower half plane with time direction pointing upward, while \((-)\)-type vertices are in the upper half plane with time direction pointing downward. The signs in parentheses correspond to the direction of \(\chi\) number flow, indicated in the diagram by the arrow.}
    \label{fig:tree-feyn}
\end{figure}

In the new basis \eq{eq:tree-int-basis}  onlythe first diagram in \fig{fig:tree-exp} contributes to the bispectrum. In the in-in formalism this breaks up into 8 Feynmann diagrams as illustrated in \fig{fig:tree-feyn}.
The total contribution to the bispectrum at this order is thus given by
\begin{align}
    B(k_1,k_2,k_3) = \sum_{s_1,s_2,s_3=\pm} B^{(s_3)}_{s_1s_2}(k_1,k_2,k_3) +(k_3\to k_1,k_2),
\end{align}
where \(s_1\) gives the index of the 2-point vertex, \(s_2\) gives the index of the 3-point vertex, and \(s_3\) indicates the direction of \(\chi\) number flow.
One can derive the following two relations between the 8 contributions:
\begin{align}\label{eq:relation}
\begin{aligned}
    B^{(-s_3)}_{-s_1,-s_2}(k_1,k_2,k_3) &= B^{(s_3)*}_{s_1s_2}(k_1,k_2,k_3),\\
    B^{(-s_3)}_{s_1s_2}(k_1,k_2,k_3) &= -\eval{B^{(s_3)}_{s_1s_2}(k_1,k_2,k_3)}_{\lambda\to-\lambda}.
\end{aligned}
\end{align}
Therefore, only \(B_{++}^{(+)}\) and \(B_{-+}^{(-)}\) need to be evaluated explicitly. Nonetheless, since the approximated form relies on the sign of \(\lambda\), \(B_{++}^{(\pm)}\) and \(B_{-+}^{(\pm)}\) will be treated separately in approximate calculations. In both cases, we have
\begin{align}\label{eq:tree-MP}
    B_{-+}^{(\pm)} &= \frac{\lambda\abs{\chicl}^2}{4k_1^3k_2^3\Lambda^3} \int^{\infty}_0 \dd{x_2} \int^{\infty}_0 \dd{x_1} \Int_{1,-+}^{(\pm)}(\lambda;x_1) \Int_{2,-+}^{(\pm)}(\lambda,p,\chi;x_2),\\\label{eq:tree-PP}
    B_{++}^{(\pm)} &= -\frac{\lambda\abs{\chicl}^2}{4k_1^3k_2^3\Lambda^3} \biggl[ \int^{\infty}_0 \dd{x_2} \int^{x_2}_0 \dd{x_1} \Int_{1,++}^{(\pm)}(\lambda;x_1)  \Int_{2,-+}^{(\pm)}(\lambda,p,\chi;x_2)\n
    &\peq \phantom{-\frac{\lambda\abs{\chicl}^2}{4k_1^3k_2^3\Lambda^3} \biggl[}
    +\int^{\infty}_0 \dd{x_2} \int^{\infty}_{x_2} \dd{x_1} \Int_{1,-+}^{(\mp)*} (\lambda;x_1) \Int_{2,++}^{(\mp)*}(\lambda,p,\chi;x_2) \biggr],
\end{align}
where we have made the substitution \(x_{1,2}\coloneqq -k_3\ctime_{1,2}\). The functions, \(\Int_{1,2}\), are given by 
\begin{align}\label{eq:int}
    \begin{aligned}
        \Int^{(s_3)}_{1,s_1+}(\lambda;x_1) &\coloneqq e^{-\pi\mu/2} \sqrt{\frac{\pi}{2}} x_1^{-1/2+is_3\lambda} e^{-is_1x_1} H_{i\mu}^{(1)}(x_1),\\
        \Int^{(s_3)}_{2,s_1+}(\lambda,p,\chi;x_2) &\coloneqq e^{-\pi\mu/2} \sqrt{\frac{\pi}{2}} \sum_{n=0}^2 (-s_1i)^n\cdot c_n^{s_3}\cdot x_2^{-1/2+n-is_3\lambda} e^{is_1px_2} H_{i\mu}^{(1)*}(x_2).
    \end{aligned}
\end{align}
The coefficients inside \(\mathcal{I}_2\) are given by
\begin{align}\label{eq:c-coef}
   \begin{aligned}
        c_0^\pm &\coloneqq -\frac{\vb{k}_1\cdot\vb{k}_2}{k_3^2} \pm i\lambda \frac{k_1^2 +k_2^2}{k_3^2}
        = \frac{1}{2} \bq{\p{1 \pm 2i\lambda} \p{p^2 -2Q} -1},\\
        c_1^\pm &\coloneqq -\frac{\vb{k}_1\cdot\vb{k}_2 (k_1 +k_2)}{k_3^3} \pm \frac{i\lambda k_1k_2 (k_1 +k_2)}{k_3^3} = \frac{1}{2} p \bq{p^2 -2(1 \mp i\lambda) Q -1},\\
        c_2^\pm &\coloneqq -\frac{\p{\vb{k}_1\cdot\vb{k}_2} k_1k_2}{k_3^4} +\frac{k_1^2k_2^2}{k_3^4}
        = \frac{1}{2} Q (p^2 -1),
    \end{aligned}
\end{align}
where \(Q\coloneqq k_1k_2/k_3^2 = \frac{1}{4} \p{p^2 -\chi^2}\). Notice in \eq{eq:int} that as \(x_i\to 0\), the integrands go as \(x_i^{-1/2+n\pm i\lambda}\) with \(n\ge0\). The amplitude is now expressed in a form that is manifestly free of late-time divergences.

\subsection{Contributions from \texorpdfstring{\(B_{-+}\)}{B-+}}

\label{sec:tree-SPA}

For \(\lambda,\mu,\lambda-\mu\gg1\), the SPA can be used to calculate \eqs{eq:tree-MP}{eq:tree-PP}. Starting from the Hankel functions \(H_{i\mu}^{(1)}(x)\) with \(\mu\gg1\), we have Debye's asymptotic representation of the Hankel function:
\begin{align}\begin{aligned}
    H_{i\mu}^{(1)}(x) &\simeq \sqrt{\frac{2}{\pi}} e^{\pi\mu/2} \cdot\frac{e^{ih(x,\mu)}}{\p{x^2 +\mu^2}^{1/4}} \Bq{1 +\order\bq{\frac{\mu^2}{\p{x^2 +\mu^2}^{3/2}}}},\\
    h(x,\mu) &= -\frac{\pi}{4} +\sqrt{x^2 +\mu^2} -\mu \arcsinh\frac{\mu}{x},\hspace{4em}
    \frac{\partial h}{\partial x} = \sqrt{1 +\frac{\mu^2}{x^2}},
\end{aligned}\end{align}
which follows directly from the WKB method. Substituting into \eq{eq:int}, the two integrands factorize into slowly-varing parts and highly oscillatory parts:
\begin{align}\label{eq:int-spa}
    \begin{aligned}
        \Int^{(s_3)}_{1,s_1+}(\lambda;x_1) &\simeq \frac{1}{x_1^{1/2} \p{x_1^2 +\mu^2}^{1/4}} e^{ig_{1,s_1}^{(s_3)}(x_1,\lambda,\mu)},\\
        \Int^{(s_3)}_{2,s_1+}(\lambda,p,\chi;x_2) &\simeq \frac{\sum_{n=0}^2 (-s_1ix_2)^n c_n^{s_3}}{x_2^{1/2} \p{x_2^2 +\mu^2}^{1/4}} e^{ig_{2,s_1}^{(s_3)}(x_2,\lambda,\mu,p)},
    \end{aligned}
\end{align}
where the phases are given by
\begin{align}\begin{aligned}
    g_{1,s_1}^{(s_3)}(x_1,\lambda,\mu) &\coloneqq -s_1x_1 +s_3\lambda\log x_1 +h(x_1,\mu),&
    \frac{\partial g_{1,s_1}^{(s_3)}}{\partial x_1} &= -s_1 +\frac{s_3\lambda}{x_1} +\sqrt{1 +\frac{\mu^2}{x_1^2}},\\
    g_{2,s_1}^{(s_3)}(x_2,\lambda,\mu,p) &\coloneqq s_1px_2 -s_3\lambda\log x_2 -h(x_2,\mu),&
    \frac{\partial g_{2,s_1}^{(s_3)}}{\partial x_2} &= s_1p -\frac{s_3\lambda}{x_2} -\sqrt{1 +\frac{\mu^2}{x_2^2}}.
\end{aligned}\end{align}
The dominant contributions to \eqs{eq:tree-MP}{eq:tree-PP} then come from those diagrams for which a stationary phase, \(g'(x_*)=0\), with \(x_* > 0\), exists in both integrals. The stationary points, along with the condition for their existence, are tabulated in \tab{tab:stationary-phases}. For the \(x_2\) integrals the stationary point is given by
\begin{equation}\label{eq:stationary-phase}
    x_*^\pm \coloneqq \frac{\lambda p \pm\sqrt{\lambda^2 +\mu^2\p{p^2-1}}}{p^2-1}  \sqlimit[\gg\lambda/\mu] \frac{\lambda \pm \mu}{p}
\end{equation}
in the squeezed limit. For ease of notation we give only the squeezed limit expression in \tab{tab:stationary-phases}.
\begin{table}[htbp]
    \centering
    \begin{tabular}{lll}\hline
        Integrand &  Stationary phase point \(x_{1,2}\) & Conditions\\\hline
        \(\Int_{1,-+}^{(s_3)}\) & \(\dfrac{\lambda^2-\mu^2}{2\lambda}\) & \(s_3=-,\quad \lambda>\mu\)\rule{0pt}{4.3ex}\\[2ex]
        \(\Int_{1,++}^{(s_3)}\) & \(\dfrac{\mu^2-\lambda^2}{2\lambda}\) & \(s_3=-,\quad \lambda<\mu\)\\[2ex]
        \(\Int_{2,-+}^{(s_3)}\) & 
        \(\dfrac{\lambda -\mu}{p} \) & \(s_3=-,\quad \lambda>\mu\)\\[2ex]
         \multirow{2}{*}[-0.4em]{\(\Int_{2,++}^{(s_3)}\)} & \( -\dfrac{\lambda-\mu}{p} \) & \(s_3=-,\quad \lambda<\mu\)\\[2ex]
        & \( \dfrac{\lambda +\mu}{p} \) & \(s_3=+\)\\[1.7ex]\hline
    \end{tabular}
    \caption{The stationary phase points for the integrands in \eq{eq:int-spa} in the squeezed limit. The third column states the conditions for a stationary phase point to exist on the positive real axis.}
    \label{tab:stationary-phases}
\end{table}

By comparing \tab{tab:stationary-phases} with \eqs{eq:tree-MP}{eq:tree-PP}, it is manifest that only \(B_{-+}^{(-)}\) has stationary points for both the \(x_1\)- and the \(x_2\)-integral, and is the dominant contribution to the whole \(B_{\tree}\). Using SPA
\begin{align}\label{eq:tree-MP-spa2}
     B_{\tree}(k_1,k_2,k_3) &\simeq B_{-+}^{(-)}(k_1,k_2,k_3)\n
     %
     %
     &\simeq \frac{\pi}{4k_1^3k_2^3} \frac{\abs{\alpha}^2}{\Lambda^3} \frac{\lambda^{1/2} e^{i\phase(\lambda,\mu,p)}}{\p{\lambda^2-\mu^2} \bq{\lambda^2+\mu^2(p^2-1)}^{1/4}} \sum_{n=0}^2 d_n (ix_*)^{n-2},
\end{align}
where \(x_*\coloneqq x_*^-\) is defined by \eqref{eq:stationary-phase},
\begin{equation}\label{eq:d-coef}
    d_0 \coloneqq 1,\qquad
    d_1 \coloneqq p,\qquad
    d_2 \coloneqq \frac{1}{4}\p{p^2-\chi^2}.
\end{equation}
The phase, \(\phase\), is given by
\begin{align}\label{eq:tree-phase}
    \phase(\lambda,\mu,p) &= \lambda\log\frac{2\lambda x_*}{\lambda^2-\mu^2} -\mu \log\frac{\lambda +\mu}{\lambda -\mu} +\mu \arcsinh \frac{\mu}{x_*}\n
    &\simeq \begin{dcases}
        -(\lambda-\mu)\log p +\phase_*(\lambda,\mu), & p\gg\lambda/\mu,\\
        -\lambda \log\frac{p+1}{2}
        , & p\ll\lambda/\mu,\ \lambda\gg\mu.
    \end{dcases}
\end{align}
In the squeezed limit, this reduces to \eq{eq:tree-MP-spa}.

The amplitude, \(B_{-+}^{(-)}\), can also be calculated exactly by noting that the two time integrals in \eq{eq:tree-MP} factorize and each of them can be reduced to a standard integral given in \eq{eq:hankel-int}
\begin{align}\label{eq:tree-MP-exact}
    B_{-+}^{(-)}(k_1,k_2,k_3) &= \frac{1}{4k_1^3k_2^3} \frac{\abs{\alpha}^2}{\Lambda^3} \frac{\lambda e^{\pi\lambda}}{\p{\lambda^2 -\mu^2 -\frac{9}{4}}^2 +9\lambda^2} \mathbf{F}_{i\mu}^{\frac{1}{2}-i\lambda} (1) \sum_{n=0}^2 \frac{c_n^-}{2^n} \mathbf{F}_{i\mu}^{\frac{1}{2}+n+i\lambda} (p)\n
    &= -\frac{1}{2k_1^3k_2^3} \frac{\abs{\alpha}^2}{\Lambda^3} \frac{\lambda e^{\pi\lambda}}{\p{\frac{3}{2}+i\lambda}^2 +\mu^2} \mathbf{F}_{i\mu}^{\frac{1}{2}-i\lambda} (1) \sum_{n=0}^2 \frac{d_n}{2^n} \mathbf{F}_{i\mu}^{-\frac{3}{2}+n+i\lambda} (p),
\end{align}
where\footnote{We use the following compact notation for products of gamma functions: 
\[\mGamma{a_1,a_2,\cdots,a_m}{b_1,b_2,\cdots,b_n} \coloneqq \frac{\Gamma(a_1)\Gamma(a_2)\cdots \Gamma(a_m)}{\Gamma(b_1)\Gamma(b_2)\cdots
\Gamma(b_n)}.\]\label{foot:defs}}
\begin{equation}\label{eq:f-func}
    \mathbf{F}_{i\mu}^{\rho}(p) \coloneqq \mGamma{\rho+i\mu,\rho-i\mu}{\frac{1}{2}+\rho} \hF{\rho+i\mu}{\rho-i\mu}{\frac{1}{2}+\rho}{\frac{1-p}{2}},
\end{equation}
and \(\hF{a}{b}{c}{z} = {}_2F_1(a,b,c;z)\) is the hypergeometric function.
In the second line in \eq{eq:tree-MP-exact} the recurrence relation given in \eq{eq:f-func-recur} for \(\mathbf{F}_{i\mu}^{\rho}(p)\) has been used, after which our result is equivalent to the one in \Cite{Bodas:2020yho} evaluated in the basis in \eq{eq:tree-exp-basis}. Our calculation is manifestly finite at every step compared to the one in the exponential basis.

The exact expression in \eq{eq:tree-MP-exact} reduces back to the stationary phase expression in \eq{eq:tree-MP-spa2} for \(\lambda,\mu,\lambda-\mu\gg1\), as shown in \fig{fig:tree-magnitude-breakdown}. Meanwhile, the naive singularity at \(\mu\to\lambda\) in \eqs{eq:tree-MP-spa}{eq:tree-MP-spa2} becomes a finite resonance peak in the exact expression.
The other contribution, \(B_{-+}^{(+)}\), can be obtained from \(B_{-+}^{(-)}\) using \eq{eq:relation}:
\begin{equation}
    B_{-+}^{(+)}(k_1,k_2,k_3) = -\eval{B_{-+}^{(-)}(k_1,k_2,k_3)}_{\lambda\to-\lambda}
    \sim e^{-2\pi\lambda} B_{-+}^{(-)}(k_1,k_2,k_3),
\end{equation}
i.e. it is always exponentially suppressed for \(\lambda\gg1\).

\subsection{Contributions from \texorpdfstring{\(B_{++}\)}{B++}}
\label{subsec:tree-subdom}
To derive the full tree-level amplitude, we must also evaluate \(B_{++}^{(\pm)}\) as well as all permuted diagrams (\(k_3\rightarrow k_1,k_2\)). We defer the discussion of the permuted diagrams to \Sec{subsec:1-loop-perm}.

Now we focus on the contribution from \(B_{++}^{(\pm)}\), given by the integral \eq{eq:tree-PP}. We start by breaking this integral into two parts for calculation convenience:
\begin{align}\label{eq:tree-PP-split}
    B_{++}^{(\pm)}(k_1,k_2,k_3) &= B_{++,\NAn}^{(\pm)}(k_1,k_2,k_3) +B_{++,\An}^{(\pm)}(k_1,k_2,k_3),\\[1em]
    \label{eq:tree-PP-NA}
    B_{++,\NAn}^{(\pm)}(k_1,k_2,k_3) &\coloneqq -\frac{\lambda\abs{\chicl}^2}{4k_1^3k_2^3\Lambda^3} \int^{\infty}_0 \dd{x_2} \int^{\infty}_0 \dd{x_1} \Int_{1,-+}^{(\mp)*} (\lambda;x_1) \Int_{2,++}^{(\mp)*}(\lambda,p,\chi;x_2),\\
    \label{eq:tree-PP-A}
    B_{++,\An}^{(\pm)}(k_1,k_2,k_3) &\coloneqq -\frac{\lambda\abs{\chicl}^2}{4k_1^3k_2^3\Lambda^3} \biggl[ \int^{\infty}_0 \dd{x_2} \int^{x_2}_0 \dd{x_1} \Int_{1,++}^{(\pm)}(\lambda;x_1)  \Int_{2,-+}^{(\pm)}(\lambda,p,\chi;x_2)\n
    &\peq \phantom{-\frac{\lambda\abs{\chicl}^2}{4k_1^3k_2^3\Lambda^3}}
    -\int^{\infty}_0 \dd{x_2} \int_0^{x_2} \dd{x_1} \Int_{1,-+}^{(\mp)*} (\lambda;x_1) \Int_{2,++}^{(\mp)*}(\lambda,p,\chi;x_2) \biggr].
\end{align}
We will soon see that in the squeezed limit they correspond to the non-analytic and analytic contributions respectively. 
Since \(B_{++,\NAn}^{(\pm)}(k_1,k_2,k_3)\) also factorizes into two time integrals, it can be evaluated using \eq{eq:hankel-int}:
\begin{align}\label{eq:tree-PP-NA-exact}
    B_{++,\NAn}^{(\pm)}(k_1,k_2,k_3) &= \frac{1}{2k_1^3k_2^3} \frac{\abs{\alpha}^2}{\Lambda^3} \frac{\lambda e^{\pm \pi\lambda}}{\p{\frac{3}{2} \mp i\lambda}^2 +\mu^2} \mathbf{F}_{i\mu}^{\frac{1}{2}\pm i\lambda} (1) \sum_{n=0}^2 (-1)^n \frac{d_n}{2^n} \mathbf{F}_{i\mu}^{-\frac{3}{2}+n\mp i\lambda} (-p+i\epsilon).
\end{align}
As shown in \Cite{Bodas:2020yho}, \(B_{++,\NAn}^{(+)}\) is suppressed by \(e^{-\pi\abs{\lambda-\mu}}\) and only becomes comparable to \(B_{-+}^{(-)}\) when \(\mu\simeq\lambda\), while one can similarly show that \(B_{++,\NAn}^{(-)}\) is suppressed by \(e^{-\pi\max\Bq{\lambda,\mu}}\) and can be always neglected.

The second part of \eq{eq:tree-PP} defined by \eq{eq:tree-PP-A} does not factorize between \(x_1\) and \(x_2\), and we will leave its exact evaluation to \app{app:exact-B++}. 
Here we will derive its approximate form in a different approach.
As implied by \tab{tab:stationary-phases}, the two time integrals in \eq{eq:tree-PP-A} never have stationary phase points simultaneously. 
Therefore, it is expected that the heavy propagator is far off-shell, so that the amplitude is dominated by the \(\ctime_1\to\ctime_2\) limit and can be described by an effective local Lagrangian of the inflaton:\footnote{For \(M\gg\lambda\), \(\chi\) decouples from the theory and this effective Lagrangian has the usual interpretation under effective field theory. For \(\lambda>M\), such interpretation is obscured by the fact that \(\chi\) are massively produced on-shell.}
\begin{align}\label{eq:tree-local-L}
    \mathcal{L}_{\phi} &= \abs{\alpha}^2 e^{i\phi/\Lambda} \frac{1}{-\Box +M^2} e^{-i\phi/\Lambda}\n
    &= \abs{\alpha}^2 \bq{-\p{\nabla^\mu -\frac{i\nabla^\mu\phi}{\Lambda}} \p{\nabla_\mu -\frac{i\nabla_\mu\phi}{\Lambda}} +M^2}^{-1} \cdot 1\n
    &= -\abs{\alpha}^2 \sum_{k=0}^\infty \frac{\bq{-\Box +\frac{2i\nabla^\mu\phipt \nabla_\mu +i\Box\phipt}{\Lambda} +\frac{(\nabla\phipt)^2}{\Lambda^2}
    -2\lambda \p{i\ctime\partial_\ctime -\frac{3i}{2} +\frac{\ctime\phipt'}{\Lambda}}}^k}{\p{\lambda^2 -M^2}^{k+1}} \cdot 1.
\end{align}
For \(\abs{\lambda^2-M^2} \gtrsim\order\p{\lambda^2}\), this becomes an expansion in \(\lambda^{-1}\). The leading-order contribution to the 3-point function is given by
\begin{align}
    \mathcal{L}^{\text{(loc)}}_{\phipt^3} &= \frac{4\lambda \abs{\alpha}^2}{\Lambda^3 \p{\lambda^2 -M^2}^3} (-\ctime) \phipt' \p{\nabla\phipt}^2 +\frac{8\lambda^3 \abs{\alpha}^2}{\Lambda^3 \p{\lambda^2 -M^2}^4} (-\ctime)^3 \phipt'^3 +\order\p{\lambda^{-7}}.
\end{align}
From this local Lagrangian we arrive at the following estimation for the analytic contribution:
\begin{align}
    B_{++} \simeq B_{++,\mathrm{A}} &\simeq B_{\text{loc}}(k_1,k_2,k_3) \coloneqq B_{+,\text{loc}} (k_1,k_2,k_3)  +(k_3\to k_1,k_2) +\cc,\\[1em] \label{eq:tree-PP-loc}
    B_{+,\text{loc}}(k_1,k_2,k_3) &= \frac{1}{8k_1^3k_2^3k_3^3} \int_{-\infty_-}^0 \frac{\dd{\ctime}}{\ctime^4} e^{i(k_1+k_2+k_3)\ctime}  \Biggl\{ \frac{8i\lambda \abs{\alpha}^2}{\Lambda^3 \p{\lambda^2 -M^2}^3} (-\ctime) \p{k_3^2\ctime}\n
    &\peq\peq \times\ctime^2 \bq{ -k_1^2 k_2^2 \ctime^2 -\p{\vb{k}_1\cdot\vb{k}_2} (1-ik_1\ctime)(1-ik_2\ctime)}\n
    &\peq +\frac{1}{3} \frac{48i\lambda^3 \abs{\alpha}^2}{\Lambda^3 \p{\lambda^2 -M^2}^4} (-\ctime)^3 \p{k_1^2k_2^2k_3^2 \ctime^3} \Biggr\}\n
    &= -\frac{1}{k_1^3k_2^3} \frac{\abs{\alpha}^2}{\Lambda^3} \frac{\lambda}{\p{\lambda^2 -M^2}^3} 
    \bq{ \frac{p^2}{p+1} -\frac{Q(p+2)}{(p+1)^2} -\frac{1}{2} -\frac{4\lambda^2}{\lambda^2-M^2} \frac{Q^2}{(p+1)^3} },
\end{align}
where \(Q\coloneqq k_1k_2/k_3^2 = \frac{1}{4} \p{p^2 -\chi^2}\). At large \(p\), \eq{eq:tree-PP-loc} becomes
\begin{align}
    B_{+,\text{loc}}(k_1,k_2,k_3) &\sqlimit[\gg1] -\frac{3}{4k_1^3k_2^3} \frac{\abs{\alpha}^2}{\Lambda^3} \frac{\lambda}{\p{\lambda^2-M^2}^3} \p{1 -\frac{1}{3} \frac{\lambda^2}{\lambda^2-M^2} } p,\\
    F_{+,\text{loc}}(k_1,k_2,k_3) &\sqlimit[\gg1] 10 \abs{\frac{\alpha}{\Lambda}}^2 \frac{\lambda^2}{\p{\lambda^2-M^2}^3} \p{1 -\frac{1}{3} \frac{\lambda^2}{\lambda^2-M^2} } p^{-2}.
\end{align}
Compared with \eq{eq:tree-MP-spa}, we find that \(B_{++}(k_1,k_2,k_3)\) is suppressed by \(\lambda^{-3}\) for \(\lambda,\mu\gg 1\), except near \(\mu\simeq\lambda\). It is also worth noting that the \(p^{-2}\) behavior in the squeezed limit of the normalized bispectrum is a feature applied to all models of single-field inflation \cite{Creminelli:2004yq}.

\begin{figure}[htbp]
    \centering
    \includegraphics[scale=0.8]{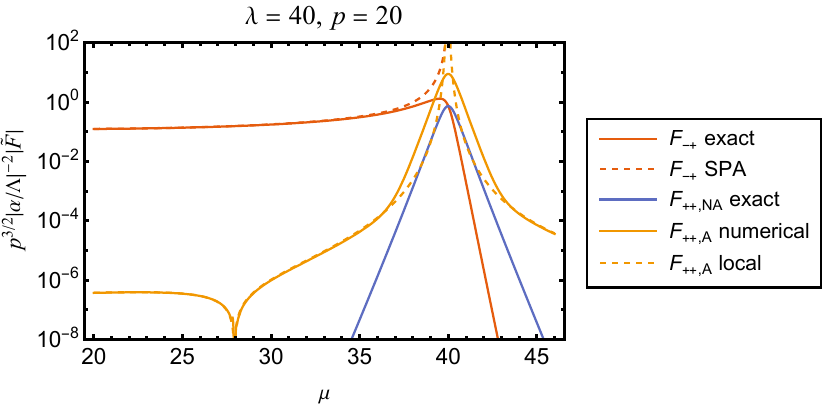}
    \caption{The magnitudes of contributions from \(F_{-+}^{(-)}\), \(F_{++,\NAn}^{(+)}\) and \(F_{++,\An}\) versus \(\mu\) for \(\lambda=40\), and \(p=20\). For \(F_{++,\mathrm{A}}\) we take the average over permuted diagrams to make the numerical result and local estimation \eq{eq:tree-local-L} comparable. From this plot we can see all important contributions near \(\mu=\lambda\), where most approximations break down. Note that there are other contributions that are exponentially suppressed by at least \(e^{-\pi\lambda}\) and thus invisible on the plot.}
    \label{fig:tree-magnitude-breakdown}
\end{figure}

\Fig{fig:tree-magnitude-breakdown} shows the sizes of \(B_{-+}^{(-)}\), \(B_{++,\NAn}^{(+)}\) and \(B_{++,\An}^{(\pm)}\) as functions of \(\mu\) for \(\lambda=40\) and \(p=20\). It is obvious that \(B_{++,\NAn}^{(+)}\) and \(B_{++,\An}^{(\pm)}\) are both negligible in magnitude compared to \(B_{-+}^{(-)}\), except near \(\mu\simeq\lambda\) where they become comparable. Since the one-loop amplitude involves integrating over, \(\mu\) as in \eq{eq:kallen-bispectrum}, \(B_{++,\NAn}^{(+)}\) and \(B_{++,\An}^{(\pm)}\) cannot be dropped in the full one-loop evaluation.

\section{Systematic Computation: One-loop Model}
\label{sec:Systematic_loop}

In this section we will compute the one-loop amplitude by performing the integral over the spectral representation in \eq{eq:kallen-bispectrum}, using the tree-level results of the previous section.
We will first derive the signal and the background for general values of the squeezing parameter, \(p\), using the same approximation as in \Sec{subsec:central-result}. We then demonstrate how the full amplitude at \(\order\p{\alpha^2}\) can be evaluated, with the full results derived in \app{app:full-results}. Finally, we will include the combinatorial backgrounds from permuted diagrams, and discuss higher order corrections in the last subsection.

\subsection{Signal from the threshold limit}

\label{subsec:signal-2}

For \(\lambda,\mu_1,\mu_2,\lambda-\mu_1-\mu_2\gg1\), we have argued in \Sec{subsec:central-result} that the \(B_{-+}\) amplitude is dominated by the region around \(\mu=\mu_{12}\) and \(\mu=\lambda\). Including both contribution gives the following approximation
\begin{align}\label{eq:1-loop-spa}
    B_{\wloop}(k_1,k_2,k_3) &= \int_0^\infty \rho^{\text{dS}}_{\mu_1\mu_2} (\mu) B_{\tree}(k_1,k_2,k_3;\mu)\n
    &\simeq B_{\wloop}^{(\mu_{12})}(k_1,k_2,k_3) +B_{\wloop}^{(\lambda)}(k_1,k_2,k_3).
\end{align}
To extract the \(\mu\to\mu_{12}\) limit we use a slightly modified form of \eq{eq:1-loop-spa-integral-2}
\begin{equation}
    \label{eq:signal-spa}
    B_{\wloop}^{(\mu_{12})}(k_1,k_2,k_3) \simeq B_{\tree,\text{SPA}}\p{k_1,k_2,k_3;\mu_{12}} \int_{-\infty}^{\infty_+} \rho^{\text{dS}}_{\mu_1\mu_2} (\mu_{12} +\varepsilon) \beta^{i\varepsilon} e^{\frac{i}{2} \gamma\varepsilon^2} \dd{\varepsilon},
\end{equation}
where
\begin{gather}
    \label{eq:beta-phase}%
   \begin{aligned}
        \beta &\coloneqq \eval{\exp\p{\frac{\partial\phase}{\partial\mu}}}_{\mu=\mu_{12}} = \frac{\lambda-\mu_{12}}{\lambda+\mu_{12}} \sqrt{\frac{\lambda +\mu_{12}-px_*}{\lambda -\mu_{12}-px_*}}
        \sqlimit[\gg\lambda/\mu] \frac{2\mu_{12} p}{\lambda +\mu_{12}},\\
        \gamma &\coloneqq \eval{\frac{\partial^2\phase}{\partial\mu^2}}_{\mu=\mu_{12}} = \frac{\lambda}{\lambda^2-\mu_{12}^2} \bq{\frac{\lambda p}{\sqrt{\lambda^2+\mu_{12}^2(p^2-1)}} -1}
        \sqlimit[\gg\lambda/\mu] \frac{\lambda}{\mu_{12}(\lambda+\mu_{12})}.
    \end{aligned}
\end{gather}
Note that compared to \eq{eq:1-loop-spa-integral-2}, we have included the second order expansion of the phase \(\phase\) near \(\varepsilon\) in \eq{eq:signal-spa} to improve the estimation for moderate \(p\).
Similarly, we extract the \(\mu \to \lambda\) background with
\begin{align}\label{eq:bkg-spa}
    B_{\wloop}^{(\lambda)}(k_1,k_2,k_3) &\simeq \rho_{\mu_1\mu_2}^{\text{Mink}}(\lambda) \int_{-\infty}^\infty B_{\tree}\p{k_1,k_2,k_3;\lambda+\varepsilon} \dd{\varepsilon}.
\end{align}
In other words, we can use the SPA form of \(B_{\tree}\) near \(\mu\sim\mu_{12}\) and the Minkowski limit of \(\rho_{\mu_1\mu_2}^{\text{dS}}(\mu)\) near \(\mu\sim\lambda\) respectively.

The fact that the integrand \(\rho_{\mu_1\mu_2}^{\text{dS}}(\mu) B_{\tree}(\mu)\) drops rapidly to zero near \(\mu\lesssim\mu_{12}\) and has a peak at \(\mu\simeq\lambda\) has clear physical meanings: the former corresponds to the \emph{threshold limit} where the mass \(\mu\) of the multi-particle state is close to or drops below the total rest mass \(\mu_{12}\) of the two components; the latter corresponds to the \emph{resonance limit}, which can be traced back to the resonance behavior in the homogeneous background of the \(\chi\) field given by \eq{eq:chi-vev} as \(\mu\to\lambda\). As has been argued in \Sec{subsec:central-result}, the threshold limit gives the non-analytic signal, while the resonance limit contributes to the analytic background.

\begin{figure}[htbp]
    \centering
    \begin{tabular}{cc}
        \includegraphics[scale=0.75]{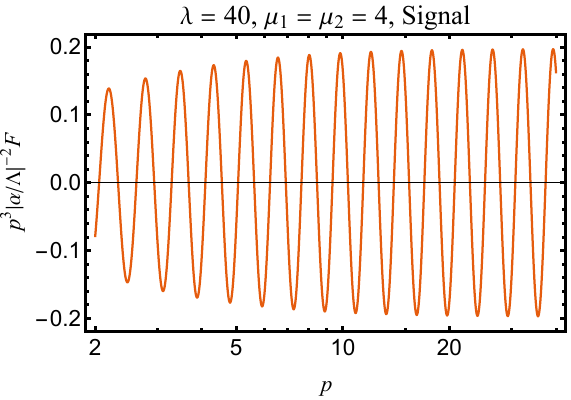} &
        \includegraphics[scale=0.75]{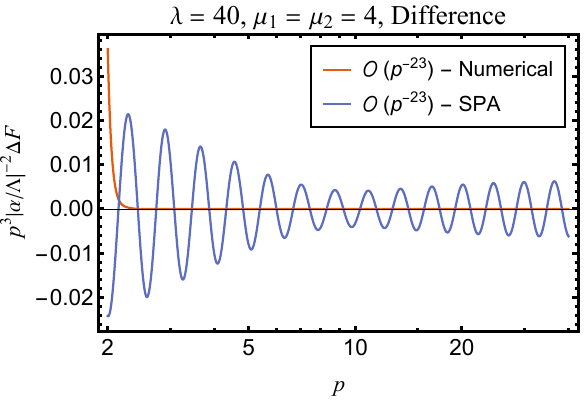}
    \end{tabular}
    \caption{(\emph{Left}) The one-loop oscillatory signal coming from the threshold limit for \(\lambda=40\) and \(\mu=4\), rescaled by \(p^3\).
    (\emph{Right}) The difference between two approximate calculations and the numerical evaluation of \eq{eq:kallen-bispectrum}: (i) the analytic sum over poles up to \(\order\p{p^{-23}}\) (see \Sec{subsec:1-loop-exact} and \app{app:full-results}); (ii) the SPA result \eq{eq:1-loop-na2}. In calculating the difference between the analytic sum and the numerical result, the difference in the full \(F_{-+}^{(-)}\) amplitude is computed, which includes both the signal and the analytic background in \(F_{-+}^{(-)}\). For \(p\) away from \(2\), the numerical result agrees with the analytic sum, while the error for the improved SPA result is within \(5\%\).} 
    \label{fig:loop-signal}
\end{figure}

\begin{figure}[htbp]
    \centering
    \includegraphics[scale=0.8]{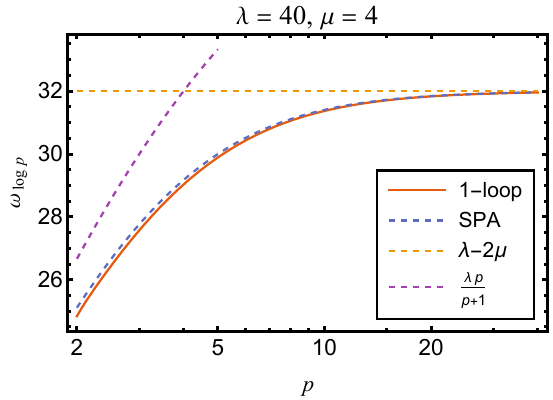}
    \caption{The oscillation frequency of the signal for \(\lambda=40\) and \(\mu_1=\mu_2=4\). The frequency of the signal is defined as \(\omega_{\log p}\coloneqq -\Im \frac{\partial\log\widetilde F}{\partial\log p}\). The SPA curve is given by \(-\frac{\partial\phase}{\partial\log p}\) with \(\phase\) defined by \eq{eq:tree-phase}. The other two lines are derived from the small \(p\) and the large \(p\) limits in \eq{eq:tree-phase}, showing that we are able to extract both the chemical potential and total mass independently from data, as emphasized in \cite{Bodas:2020yho,Bodas:2024hih}.}
    \label{fig:1-loop-freq}
\end{figure}

For the signal, we can combine \eq{eq:ds-spectral-threshold} with \eq{eq:signal-spa} to get a more accurate stationary phase estimation:
\begin{align}\label{eq:1-loop-na2}
    B_{\wloop}^{(\mu_{12})} (k_1,k_2,k_3) &\simeq \frac{1}{2\pi^{3/2}} \sqrt{\frac{\mu_1\mu_2}{\mu_{12}}} B_{\tree}(k_1,k_2,k_3;\mu_{12})\n
    &\peq \times e^{3\pi i/4}\sum_{k=0}^\infty \frac{(3/2)_k}{k!} \beta^{-3/2-2k} e^{-\frac{i}{2} \gamma\p{3/2+2k}^2}. & \p{\log\beta \gg \gamma}
\end{align}
where \((a)_m \coloneqq a(a+1)\cdots(a+m-1)\) is the Pochhammer symbol.
Note that we have approximated the integral as a sum over residues of the poles in \eq{eq:ds-spectral-threshold}, which is only valid up to corrections of order
\begin{equation}
    \eval{\rho^{\text{dS}}_{\mu_1\mu_2} (\mu_{12} +\varepsilon)}_{\gamma\varepsilon +\log\beta=0} \sim e^{-\pi \log\beta/\gamma}.
\end{equation}
\Eq{eq:1-loop-na2} reduces to \eq{eq:1-loop-na} in the squeezed limit with \(\gamma\ll1\).
\Fig{fig:loop-signal} shows the shape of the non-analytic signal in terms of the bispectrum, \(F\), and its comparison with the numerical result and the analytic form we will derive later in \Sec{subsec:1-loop-exact}. We also calculate the oscillating frequency at different \(p\)-value and the result is presented in \fig{fig:1-loop-freq}. As argued in \cite{Bodas:2020yho,Bodas:2024hih}, the frequency at small \(p\) and large \(p\) can be used together to infer both the total mass \(\mu_{12}\), and the chemical potential \(\lambda\).

\subsection{Resonance backgrounds}

The one-loop amplitude also differs from the tree-level amplitude by having an analytic background coming from the resonance production. 
Such a background cannot be neglected as the amplitude of the one-loop signal decays as \(\order\p{p^{-3}}\), which is faster than a typical analytic contribution, decreasing as \(\order\p{p^{-2}}\) in the large-\(p\) limit. 
In this subsection, we will approximately evaluate this background coming from the resonance limit of \(B_{-+}\), given by \eq{eq:bkg-spa}, as well as a similar one from \(B_{++}\). 
Starting from \eq{eq:bkg-spa}, we need to substitute terms singular near \(\mu\simeq\lambda\) in the SPA form of \eq{eq:tree-MP-spa2} by their exact expression in \eq{eq:tree-MP-exact}:
\begin{align}
    \frac{(ix_*)^{n-2}}{\lambda^2 -\mu^2} &\to -\frac{1}{2\pi} \frac{\Gamma\p{\frac{1}{2} -i\lambda +i\mu} \Gamma\p{-\frac{3}{2}+n+i\lambda-i\mu}}{\p{\frac{3}{2} +i\lambda}^2 +\mu^2} \p{\frac{x_*}{\lambda-\mu}}^{n-2} e^{\pi\p{\lambda-\mu}}.
\end{align}
Note that \(x_*/(\lambda-\mu)\to 1/p\) as \(\mu\to\lambda\). For \(\lambda-\mu\gg1\) one can apply the Stirling approximation on the RHS and obtain the LHS, while this does not apply when \(\mu\simeq\lambda\). Besides these terms, we can approximate all other slow-varying terms in \eq{eq:tree-MP-spa2} by their values at \(\mu=\lambda\),
\begin{align}
    B_{\tree,-+}^{(-)} (k_1,k_2,k_3; \lambda+\varepsilon) &\simeq \frac{\pi}{8k_1^3k_2^3} \frac{\abs{\alpha}^2}{\Lambda^3} \frac{p^{3/2 +i\varepsilon}}{2i\lambda \p{\frac{3}{2} -i\varepsilon}} \p{\frac{1}{\frac{3}{2} +i\varepsilon} -\frac{p^2 -\chi^2}{4p^2}} \frac{e^{-\pi\varepsilon}}{\cosh \pi\varepsilon}.
\end{align}
One can now evaluate the \(\mu\)-integral analytically, closing the contour in the upper half plane and applying
using the residue theorem:
\begin{align}
    &\peq \int_{-\infty}^{\infty} \frac{p^{i\varepsilon}}{\frac{3}{2}-i\varepsilon} \p{\frac{1}{\frac{3}{2} +i\varepsilon} -\frac{p^2 -\chi^2}{4p^2}} \frac{e^{-\pi\varepsilon}}{\cosh \pi\varepsilon} \dd{\varepsilon}\n
    &= -\frac{2i}{3} p^{-3/2} \p{p +\log p +\pi i -\sum_{k=0}^\infty \frac{p^{-1-k}}{k+1} +\frac{p^2+3\chi^2}{4} \sum_{k=0}^\infty \frac{p^{-1-k}}{k+2}},\n
    &= -\frac{2i}{3} p^{-3/2} \bq{p +\log\p{p-1} +\pi i -\frac{p^2+3\chi^2}{4} \p{ 1 +p\log\frac{p-1}{p} } },
\end{align}
giving
\begin{align}
    B_{\wloop,-+}^{(\lambda)}(k_1,k_2,k_3) 
    %
    &\simeq -\frac{1}{192\pi k_1^3k_2^3} \frac{\abs{\alpha}^2}{\Lambda^3}
    \frac{\prod_{\epsilon_1,\epsilon_2=\pm} \sqrt{\lambda +\epsilon_1\mu_1 +\epsilon_2\mu_2}}{\lambda^2}\n
    &\peq \times \bq{ p +\log(p-1) +\pi i -\frac{p^2+3\chi^2}{4} \p{1 +p \log\frac{p-1}{p}} }.
    %
\end{align}
In terms of the properly normalized bispectrum:
\begin{align}\label{eq:1-loop-a-MP}
    &F_{\wloop,-+}^{(\lambda)} \simeq \frac{5}{36\pi} \frac{\abs{\alpha}^2}{\Lambda^2} \frac{\prod_{\epsilon_i} \sqrt{\lambda +\epsilon_1\mu_1 +\epsilon_2\mu_2}}{\lambda \p{p^3 +3\chi^2 p +4}}
    \bq{ p +\log(p-1) -\frac{p^2+3\chi^2}{4} \p{1 +p \log\frac{p-1}{p}} },\n
    &\hspace{1cm}\sqlimit[\gg1] \frac{5}{32\pi} \frac{\abs{\alpha}^2}{\Lambda^2} \frac{\prod_{\epsilon_i} \sqrt{\lambda +\epsilon_1\mu_1 +\epsilon_2\mu_2}}{\lambda p^2} \p{1 +\frac{2 +24\log p}{27p} -\frac{5 +16\chi^2}{6p^2} +\cdots},
\end{align}
where we also include the contribution from the complex conjugate.

By combining \eq{eq:1-loop-na2} with \eq{eq:1-loop-a-MP} we have approximately \(B_{-+}\). This contribution contains an oscillatory signal on top of a smooth background. If this was the tree level exchange of \cite{Bodas:2020yho} or \cite{Bodas:2024hih} then we would have successfully calculated the dominant contribution to the bispectrum. However, recall from \Sec{subsec:tree-subdom} that we \(B_{++}\) is not suppressed in the region \(\mu\simeq\lambda\), which we must include in the loop integration. Therefore, we must take into account a further smooth background from the \(B_{++}\) diagram.

The tree level amplitude for \(B_{++}\) has been split into \(B_{++,\NAn}^{(\pm)}\) and \(B_{++,\An}^{(\pm)}\) in \eq{eq:tree-PP-split}. According to \Sec{subsec:tree-subdom}, \(B_{++,\NAn}^{(-)}\) is always exponentially suppressed and can be neglected. For \(B_{++,\NAn}^{(+)}\), we use the following hypergeometric function identity \NIST{15.8.4}:
\begin{align}
    \mathbf{F}_{i\mu}^{\rho} (-p+i\epsilon) &= \mGamma{\rho+i\mu,\rho-i\mu}{\frac{1}{2}+\rho} \biggl[ \mGamma{\frac{1}{2}+\rho,\frac{1}{2}-\rho}{\frac{1}{2}+i\mu,\frac{1}{2}-i\mu} \hF{\rho+i\mu}{\rho-i\mu}{\frac{1}{2}+\rho}{\frac{1-p}{2}}\n
    &\peq +\mGamma{\frac{1}{2}+\rho,-\frac{1}{2}+\rho}{\rho+i\mu,\rho-i\mu} \p{\frac{1-p}{2}+i\epsilon}^{1/2-\rho} \hF{\frac{1}{2}+i\mu}{\frac{1}{2}-i\mu}{\frac{3}{2}-\rho}{\frac{1-p}{2}} \biggr].
\end{align}
For \(\rho=-3/2+n-i\lambda\), the Stirling approximation implies that
\begin{equation}
    \begin{aligned}
        &\mGamma{\frac{1}{2}+\rho,\frac{1}{2}-\rho}{\frac{1}{2}+i\mu,\frac{1}{2}-i\mu} \simeq -i (-1)^n e^{-\pi(\lambda-\mu)},\\
        &\abs{ \mGamma{\frac{1}{2}+\rho,-\frac{1}{2}+\rho}{\frac{1}{2}+i\mu,\frac{1}{2}-i\mu} \p{\frac{1-p}{2}+i\epsilon}^{1/2-\rho} } \sim e^{-\pi(2\lambda-\mu)}.
    \end{aligned}
\end{equation}
Consequently, the second term is suppressed by \(e^{-\pi\lambda}\) compared to the first term, and we have
\begin{align}
    (-1)^n \mathbf{F}_{i\mu}^{-\frac{3}{2}+n-i\lambda} (-p+i\epsilon) &\simeq -ie^{-\pi(\lambda-\mu)} \mathbf{F}_{i\mu}^{-\frac{3}{2}+n-i\lambda}(p)\\
    B_{\tree,++,\NAn}^{(+)} (k_1,k_2,k_3; \lambda+\varepsilon)
    &\simeq ie^{\pi\varepsilon} \lbar{B}_{\tree,-+}^{(-)} (k_1,k_2,k_3; \lambda+\varepsilon),
\end{align}
where we changed variables to \(\mu = \lambda + \varepsilon\), and in the last line we used \eqs{eq:tree-MP-exact}{eq:tree-PP-NA-exact} to relate the two amplitudes. The integral over \(\varepsilon\) can now be evaluated explicitly in a similar manner to give
\begin{align}\label{eq:1-loop-a-PP-NA}
    &F_{\wloop,++,\NAn}^{(\lambda)} \simeq \frac{5}{36\pi} \frac{\abs{\alpha}^2}{\Lambda^2} \frac{\prod_{\epsilon_i} \sqrt{\lambda +\epsilon_1\mu_1 +\epsilon_2\mu_2}}{\lambda \p{p^3 +3\chi^2 p +4}}\bq{ p -\log(p+1) +\frac{p^2+3\chi^2}{4} \p{1 -p \log\frac{p+1}{p}} }\n
    &\hspace{1cm}\sqlimit[\gg1] \frac{5}{32\pi} \frac{\abs{\alpha}^2}{\Lambda^2} \frac{\prod_{\epsilon_i} \sqrt{\lambda +\epsilon_1\mu_1 +\epsilon_2\mu_2}}{\lambda p^2} \p{1 -\frac{2 +24\log p}{27p} -\frac{5 +16\chi^2}{6p^2}}.
\end{align}

\begin{figure}[htbp]
    \centering
    \begin{tabular}{cc}
        \includegraphics[scale=0.75]{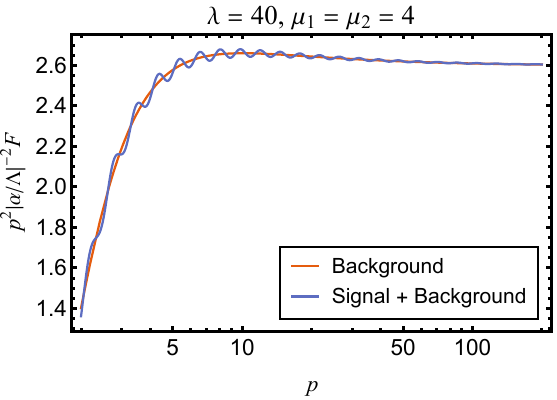} &
        \includegraphics[scale=0.75]{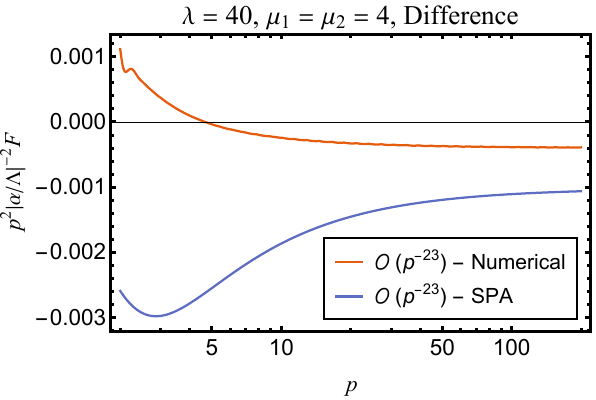}
    \end{tabular}
    \caption[The one-loop smooth background coming from the resonance limit]
    {(\emph{Left}) The one-loop smooth background coming from the resonance limit of \(F_{-+}\), \(F_{++,\NAn}\) and \(F_{++,\An}\) for \(\lambda=40\) and \(\mu_1=\mu_2=4\), rescaled by \(p^2\).
    (\emph{Right}) The difference between three different calculations: (i) the analytic sum over poles up to \(\order\p{p^{-23}}\) (see \Sec{subsec:1-loop-exact} and \app{app:full-results}); (ii) numerical evaluation of \eq{eq:kallen-bispectrum}; (iii) the SPA result \eq{eq:1-loop-a}. For \(p\) away from \(2\), both the numerical result and the improved SPA result agree with the analytic sum within \(0.1\%\). The deviation in the orange line at large \(p\) is due to the \(\tF_{++,\An,\wloop}^{(s_1,s_2\mu_{12},k)}\) series which, in principle, requires a full resummation even at \(\order\p{p^{-2}}\). In making the plot, we include terms up to \(k=40\). The details are discussed at the end of \app{app:full-results}.}
    \label{fig:loop-bkg}
\end{figure}

Finally, the tree level contribution to \(B_{++,\An}^{(\pm)}\) has been calculated exactly in \app{app:exact-B++}. To obtain an approximate expression we expand \eq{eq:tree-PP-A-exact}, with a useful identity derived in \eq{eq:h-func-spa}, around \(\mu=\lambda+\varepsilon\). Assuming \(\mu,\lambda\gg1\) we obtain
\begin{align}
    B_{++,\An}^{(\pm)}(k_1,k_2,k_3) &\simeq
    \frac{1}{16k_1^3k_2^3} \frac{\abs{\alpha}^2}{\Lambda^3} \frac{p^{-1}}{\lambda \p{\frac{9}{4} +\varepsilon^2} \p{\frac{1}{2} \mp i\varepsilon}} \sum_{n=0}^1 \hat c_n \hF{1+n}{\frac{1}{2}\mp i\varepsilon}{\frac{3}{2}\mp i\varepsilon}{-\frac{1}{p}},
\end{align}
with \(\hat c_0\coloneqq p^2-2Q\), \(\hat c_1\coloneqq Q\), and \(Q\coloneqq \frac{1}{4} \p{p^2-\chi^2}\). It turns out that the function
\begin{equation}
     \frac{1}{\frac{1}{2} \mp i\varepsilon} \hF{1+n}{\frac{1}{2}\mp i\varepsilon}{\frac{3}{2}\mp i\varepsilon}{-\frac{1}{p}}
\end{equation}
is meromorphic in \(\varepsilon\) with poles at \(1/2\pm i\varepsilon+n=0\) (\(n=0,1,2\cdots\)), and decays as \(\order\p{\varepsilon^{-1}}\) as \(\varepsilon\to\infty\). 
The integration over \(\varepsilon\) can thus be closed on either the upper or the lower half plane:
\begin{align}
    \int_{-\infty}^\infty \frac{\dd{\varepsilon}}{\p{\frac{9}{4} +\varepsilon^2} \p{\frac{1}{2} \mp i\varepsilon}} \hF{1+n}{\frac{1}{2}\mp i\varepsilon}{\frac{3}{2}\mp i\varepsilon}{-\frac{1}{p}} &= \frac{\pi}{3} \hF{1+n}{2}{3}{-\frac{1}{p}}\n
    &= \frac{2\pi}{3} \begin{dcases}
        1 -p\log\frac{p+1}{p}, & n=0,\\
        -\frac{p^2}{p+1} +p^2\log\frac{p+1}{p}, & n=1.
    \end{dcases}
\end{align}
We thus have
\begin{align}\label{eq:1-loop-a-PP-A}
    F_{\wloop,++,\An}^{(\lambda)} &\simeq -\frac{5}{72\pi} \frac{\abs{\alpha}^2}{\Lambda^2} \frac{\prod_{\epsilon_i} \sqrt{\lambda +\epsilon_1\mu_1 +\epsilon_2\mu_2}}{\lambda \p{p^3 +3\chi^2 p +4}}
    \bq{ \p{p^2+3\chi^2} \p{1 -p \log\frac{p+1}{p}} +\frac{p^2-\chi^2}{p+1} }\n
    &\sqlimit[\gg1] -\frac{5}{48\pi} \frac{\abs{\alpha}^2}{\Lambda^2}
    \frac{\prod_{\epsilon_i} \sqrt{\lambda +\epsilon_1\mu_1 +\epsilon_2\mu_2}}{\lambda p^2} \p{ 1 -\frac{8}{9p} +\frac{5-16\chi^2}{6p^2} +\cdots }.
\end{align}
Combining eqs.~\eqref{eq:1-loop-a-MP}, \eqref{eq:1-loop-a-PP-NA} and \eqref{eq:1-loop-a-PP-A} gives an approximate expression for the smooth background
\begin{align}\label{eq:1-loop-a}
    F_{\wloop,\mathrm{bkg}} &\simeq \frac{5}{18\pi} \frac{\abs{\alpha}^2}{\Lambda^2} \frac{\prod_{\epsilon_1,\epsilon_2=\pm} \sqrt{\lambda +\epsilon_1\mu_1 +\epsilon_2\mu_2}}{\lambda \p{p^3 +3\chi^2 p +4}}\n
    &\peq \times \bq{ p -\frac{1}{2}\log\frac{p+1}{p-1} -\frac{p^2+3\chi^2}{4} \p{1 -\frac{p}{2} \log\frac{p+1}{p-1}} -\frac{p^2-\chi^2}{4(p+1)} }\n
    &\sqlimit[\gg1] \frac{5}{24\pi} \frac{\abs{\alpha}^2}{\Lambda^2}
    \frac{\prod_{\epsilon_1,\epsilon_2=\pm} \sqrt{\lambda +\epsilon_1\mu_1 +\epsilon_2\mu_2}}{\lambda} p^{-2} \p{ 1 +\frac{4}{9p} -\frac{5+8\chi^2}{3p^2} +\cdots }.
\end{align}

\begin{figure}[htbp]
    \centering
    \begin{tabular}{cc}
        \includegraphics[scale=0.75]{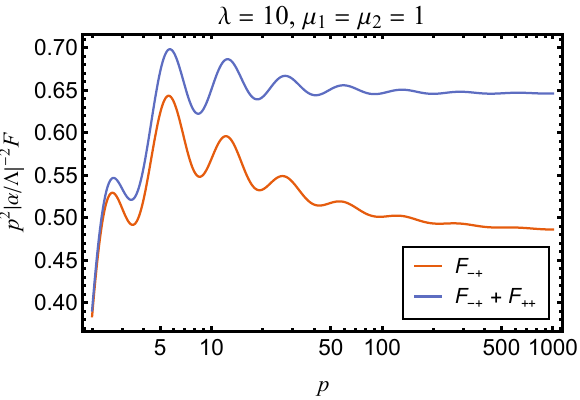} &
        \includegraphics[scale=0.75]{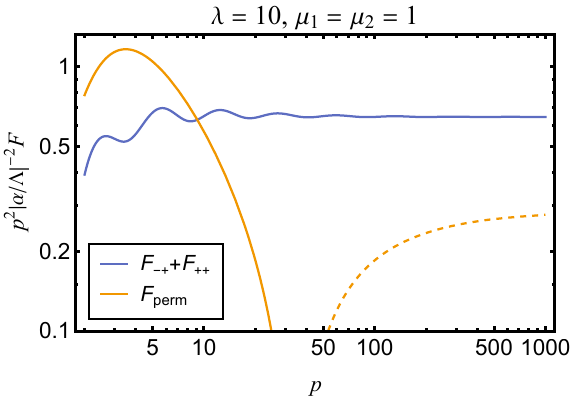}\\
        \includegraphics[scale=0.75]{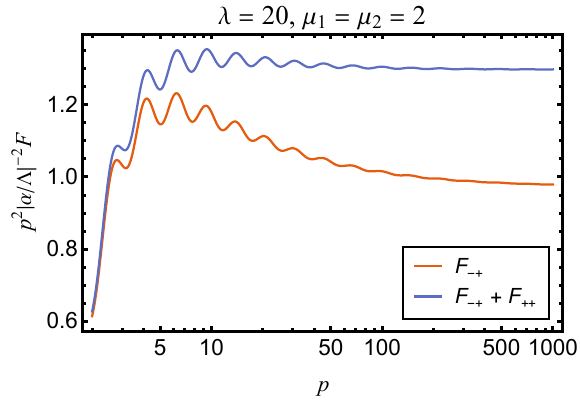} &
        \includegraphics[scale=0.75]{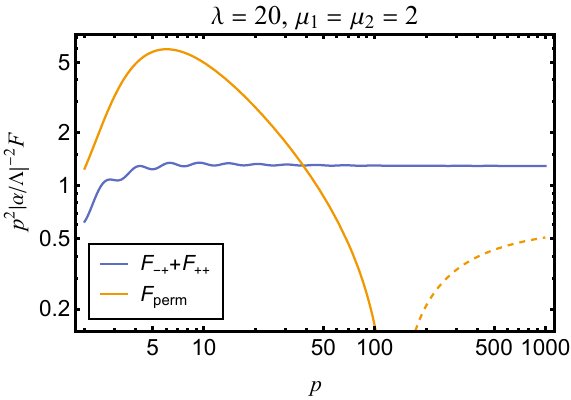}\\
        \includegraphics[scale=0.75]{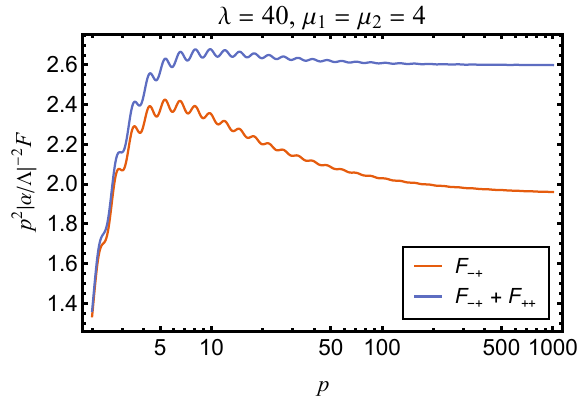} &
        \includegraphics[scale=0.75]{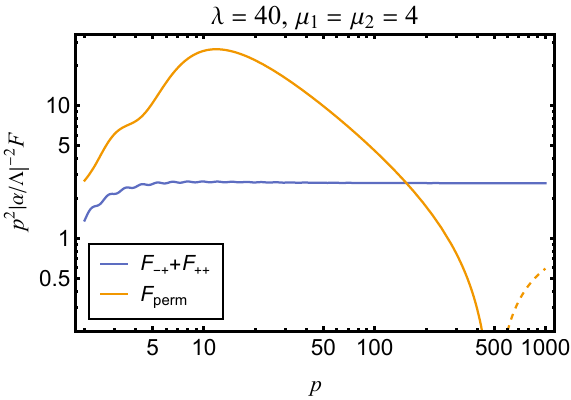}\\
   \end{tabular}
    \caption{Different contributions to the one-loop correlators for various \(\lambda\) and \(\mu_1,\mu_2\), rescaled by \(p^2\). (\emph{Left}) The contribution from \(F_{-+}\) and \(F_{++}\). (\emph{Right}) The contribution from the two permuted diagrams where the soft momentum \(k_3\) is contracted with the 3-point vertex in \fig{fig:tree-feyn}. Negative values are shown in dashed lines.}
    \label{fig:1-loop-breakdown}
\end{figure}

The shape of the background is shown in \fig{fig:loop-bkg} and the left panel of \fig{fig:1-loop-breakdown}. 
One can see that the resonance background is an order of magnitude larger than the signal for \(\lambda=40\) and \(\mu_1=\mu_2=4\). It will thus be detected before the signal in the one-loop dominated scenario. 
Fortunately, since the background is a smooth function of \(p\), the non-analytic signal can be separated from it as long as the experimental error is smaller than the amplitude of the oscillations. This detectability issue will be further discussed in \Sec{subsec:1-loop-perm}.

\subsection{Beyond stationary phase approximation}
\label{subsec:1-loop-exact}

Besides the SPA result \eqs{eq:1-loop-na}{eq:1-loop-na2}, the spectral representation also allow us to expand the exact result for the one-loop amplitude as an infinite sum over analytic terms, which is effectively an expansion in \(1/p\). To see how this works, we start with the integration for \(B_{-+}\) given by
\begin{equation}\label{eq:kallen-bispectrum-MP}
    B_{-+,\wloop}^{(-)}(k_1,k_2,k_3) = \int_0^\infty \rho^{\text{dS}}_{\mu_1\mu_2} (\mu) B^{(-)}_{-+,\tree}\p{k_1,k_2,k_3;\mu} \dd{\mu},
\end{equation}
where \(B^{(-)}_{-+,\tree}\) is given by \eq{eq:tree-MP-exact}. 

There is one difficulty in applying the residue theorem to \eq{eq:kallen-bispectrum-MP}: in the complex plane the integrand blows up as \(\Im \mu \rightarrow \pm \infty\). This is analogous to the situation in flat space where the integrand is a trigonometric function, i.e \(\cos(x)\sim e^{ix}+e^{-ix}\). The integrand blows up in both the upper and lower half plane, but once we split the integrand up into complex exponentials the contour can be closed in both terms individually.

To achieve such separation for the hypergeometric function we rewrite \(\vb{F}^\rho_{i\mu}(p)\) in \eq{eq:tree-MP-exact} using the following identity \NIST{15.8.2}:
\begin{align}
\label{eq:Fseparation}
    \mathbf{F}_{i\mu}^{\rho}(p) &= \frac{1}{2\sqrt{\pi}} \bq{ 2^{2i\mu} \Gamma(i\mu) \Gamma(\rho -i\mu) \mathbf{\widetilde F}_{i\mu}^{\rho-i\mu}(p) +(\mu\to -\mu)},
\end{align}
where
\begin{equation}
    \mathbf{\widetilde F}_{i\mu}^s(p) \coloneqq \p{\frac{p-1}{2}}^{-s} \hF{s}{\frac{1}{2}-i\mu}{1-2i\mu}{\frac{2}{1-p}}. \label{eq:tilde-f-func}
\end{equation}
Since the integrand in \eq{eq:kallen-bispectrum-MP} is an even function in \(\mu\), we then have
\begin{align}\label{eq:kallen-bispectrum-MP-contour}
    B_{-+,\wloop}^{(-)}(k_1,k_2,k_3) &= \int_{-\infty}^\infty \rho^{\text{dS}}_{\mu_1\mu_2} (\mu) \widetilde{B}^{(-)}_{-+,\tree}\p{k_1,k_2,k_3;\mu} \dd{\mu},\\[1em]
    \widetilde{B}^{(-)}_{-+,\tree}\p{k_1,k_2,k_3;\mu} &\coloneqq -\frac{1}{4\sqrt{\pi} k_1^3k_2^3} \frac{\abs{\alpha}^2}{\Lambda^3} \frac{2^{2i\mu} \lambda e^{\pi\lambda}}{\p{\frac{3}{2}+i\lambda}^2 +\mu^2} \sum_{n=0}^2 \frac{d_n}{2^n} \mathbf{\widetilde F}_{i\mu}^{-\frac{3}{2}+n+i\lambda-i\mu} (p)\n
    &\peq \times \mGamma{i\mu,-\frac{3}{2}+n+i\lambda -i\mu,\frac{1}{2}-i\lambda+i\mu,\frac{1}{2}-i\lambda-i\mu}{1-i\lambda}.
\end{align}

The integrand in \eq{eq:kallen-bispectrum-MP-contour} has improved asymptotic behavior compared to the one in \eq{eq:kallen-bispectrum-MP}. In particular, for \(p>1\), \(\Re\mu=\mu_*\) and \(\Im\mu=M\), we have the following asymptotic bound as \(M\to+\infty\):\footnote{For the gamma functions, the asymptotic property \(\Gamma(a-i\mu)\sim (-i\mu)^a \Gamma(-i\mu)\) and the identity \(\Gamma(a+i\mu) \Gamma(1-a-i\mu) = \pi \csc \pi(a+i\mu)\) are used. For \(\mathbf{\widetilde F}_{i\mu}^{\rho -i\mu}\), the asymptotic expansion is given by \eq{eq:tilde-F-asymp-3}, where we leave its derivation to \app{app:asymp-bounds}.}
\begin{align}\label{eq:asymp-bound}
    \rho^{\text{dS}}_{\mu_1\mu_2} (\mu) \widetilde{B}^{(-)}_{-+,\tree}\p{k_1,k_2,k_3;\mu} &\sim \mathcal{P}\p{\lambda,\mu,\mu_1,\mu_2}  M^{-\frac{3}{2}-i\lambda} \p{p +\sqrt{p^2-1}}^{-M},
\end{align}
where
\begin{equation}
    \mathcal{P}\p{\lambda,\mu,\mu_1,\mu_2} = \frac{\sin \pi\p{\frac{3}{2}+i\mu}}{\sin\pi\p{\frac{1}{2}-i\lambda+i\mu} \prod_{\epsilon_1,\epsilon_2=\pm} \sin\pi\bq{\frac{3}{4}+\frac{i}{2} \p{\mu +\epsilon_1\mu_1 +\epsilon_2\mu_2}}}
\end{equation}
is bounded if \(\mu\) does not coincide with the zeros of any sine functions in the denominator. Therefore, we can close the contour in the upper \(\mu\)-plane.

\begin{figure}[htbp]
    \centering
    \includegraphics{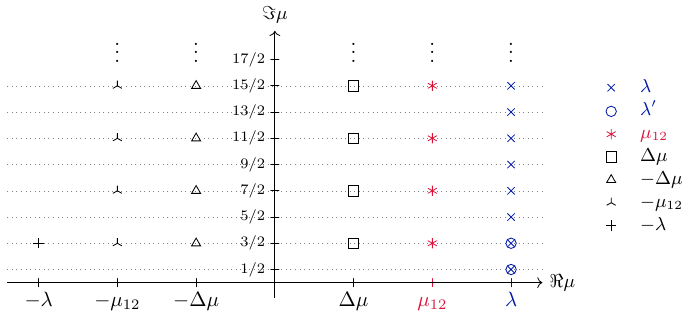}
    \caption{The analytic structure of the integrand in \eq{eq:kallen-bispectrum-MP-contour} in the upper half plane. The two colored columns are poles with residues that are not exponentially suppressed. Given the bound \eq{eq:asymp-bound}, the integral can be represented as a sum over the residues of all these poles for \(p>1\).}
    \label{fig:pole-structure-MP}
\end{figure}

\begin{table}[htbp]
    \centering
    \begin{tabular}{lllll}\hline
         Series (\(\Re\mu\)) &  source & & \(\mu\) & residues\\\hline
         \(\lambda\) & \(\widetilde{B}^{(-)}_{-+,\tree},\) & \(\Gamma\p{\frac{1}{2}-i\lambda +i\mu}\) & \color{plotblue}\(\lambda +\p{\frac{1}{2}+k}i\) & \color{plotblue}\(\order(1)\)\\
         \(\lambda'\) & \(\widetilde{B}^{(-)}_{-+,\tree},\) & \(\Gamma\p{-\frac{3}{2}+n+i\lambda-i\mu}\) & \color{plotblue}\(\lambda +\frac{1}{2}i,\ \lambda +\frac{3}{2}i\) & \color{plotblue}\(\order(1)\)\\
         \(-\lambda\) & \(\widetilde{B}^{(-)}_{-+,\tree},\) & \(\frac{1}{(3/2+i\lambda)^2 +\mu^2}\) & \(-\lambda +\frac{3}{2}i\) & \(\order\p{e^{-\pi\lambda}}\)\\
         \(\mu_{12}\) & \(\rho^{\text{dS}}_{\mu_1\mu_2},\) & \(\Gamma\bq{\frac{3}{4} +\frac{i}{2} \p{\mu -\mu_{12}}}\) & \color{plotred}\(\mu_{12} +\p{\frac{3}{2}+2k}i\) & \color{plotred}\(\order(1)\)\\
         \(-\mu_{12}\) & \(\rho^{\text{dS}}_{\mu_1\mu_2},\) & \(\Gamma\bq{\frac{3}{4} +\frac{i}{2} \p{\mu +\mu_{12}}}\) & \(-\mu_{12} +\p{\frac{3}{2}+2k}i\) & \(\order\p{e^{-\pi\mu_{12}}}\)\\
         \(\dmu\) & \(\rho^{\text{dS}}_{\mu_1\mu_2},\) & \(\Gamma\bq{\frac{3}{4} +\frac{i}{2} \p{\mu -\dmu}}\) & \(\dmu +\p{\frac{3}{2}+2k}i\) & \(\order\p{e^{-2\pi\mu_2}}\)\\
         \(-\dmu\) & \(\rho^{\text{dS}}_{\mu_1\mu_2},\) & \(\Gamma\bq{\frac{3}{4} +\frac{i}{2} \p{\mu +\dmu}}\) & \(-\dmu +\p{\frac{3}{2}+2k}i\) & \(\order\p{e^{-\pi\mu_{12}}}\)\\\hline
    \end{tabular}
    \caption{The position of poles of the integrand in \eq{eq:kallen-bispectrum-MP-contour} in the upper half plane and their sources (\(n=0,1,2\) and \(k\in\N\)). The residues at these poles are estimated logarithmically for \(\lambda,\mu_1,\mu_2,\lambda-\mu_{12}\gg1\), where \(\mu_1>\mu_2\) is assumed w.l.o.g. For the logarithmic estimation of all ranges of \(\lambda\), see \tab{tab:1-loop-order-estimation}.}
    \label{tab:pole-structure-MP}
\end{table}

The analytic structure of the integrand in \eq{eq:kallen-bispectrum-MP-contour} is illustrated in \fig{fig:pole-structure-MP} and the position of the poles are listed in \tab{tab:pole-structure-MP}, where the simplified notation \(\dmu \coloneqq \abs{\mu_1 -\mu_2}\) is used. The \(p\)-dependence of each pole for large \(p\) can be read off from \eqref{eq:tilde-f-func} directly:
\begin{equation}
    \Res_{\mu_* +i\delta} \widetilde{B}^{(-)}_{-+,\tree} \sqlimit p^{3/2 -\delta -i(\lambda -\mu_*)},
\end{equation}
where the imaginary part of the exponent is given by \(\lambda-\mu_*\), while poles with larger imaginary part, \(\delta\), are more suppressed in the squeezed limit.

For \(\lambda,\mu_1,\mu_2,\lambda-\mu_{12}\gg1\), one can use the Stirling approximation on all the \(\Gamma\) functions in \eq{eq:kallen-bispectrum-MP-contour}.
By doing so, we find that all but three series of poles are exponentially suppressed, where the unsuppressed poles have either \(\mu_* = \lambda\) or \(\mu_*=\mu_{12}\). The former case corresponds to the resonance limit, and contributes to the bispectrum as\footnote{The \(\log p\) at \(p^{-3}\) comes from the overlapping pole at \(\mu = \lambda +\frac{3}{2} i\), while the other apparent overlapping pole at \(\mu = \lambda +\frac{1}{2}i\) does not contain \(\log p\). See \app{app:full-results} for more details.}
\begin{equation}
    F_{-+,\wloop}^{(-,\lambda)} \sqlimit p^{-2},\ \ p^{-3},\ \ p^{-3} \log p,\ \ p^{-4},\ \ \cdots,
\end{equation}
while the latter case corresponds to the threshold limit, and contributes to the bispectrum as
\begin{equation}
    F_{-+,\wloop}^{(-,\mu_{12})} \sqlimit p^{-3-i(\lambda-\mu_{12})},\ \ p^{-5-i(\lambda-\mu_{12})},\ \ p^{-7-i(\lambda-\mu_{12})},\ \ \cdots.
\end{equation}
In particular, when the contribution from one pole is considered, the rest of \eq{eq:kallen-bispectrum-MP-contour} can be approximated as in \eq{eq:1-loop-spa-integral}. This further justifies the use of \eq{eq:signal-spa} as a leading-order approximation.

The same technique can also be used to evaluate \(B_{++,\NAn}^{(\pm)}\) and \(B_{++,\An}^{(\pm)}\) for the one-loop model. For \(B_{++,\NAn}^{(\pm)}\), the following identity is used to obtain an integrand with an improved asymptotic behavior:
\begin{align}
    \mathbf{F}_{i\mu}^{\rho}(-p +i\epsilon) &= \frac{e^{-\rho\pi i}}{2\sqrt{\pi}} \bq{ 2^{2i\mu} e^{-\pi\mu} \Gamma(i\mu) \Gamma(\rho -i\mu) \mathbf{\widetilde F}_{i\mu}^{\rho-i\mu}(p) +(\mu\to -\mu)},
\end{align}
while for \(B_{++,\An}^{(\pm)}\), \eq{eq:tree-PP-A-exact} is already in the desired form. Eventually:
\begin{align}\label{eq:kallen-bispectrum-PP-contour}
    B_{++,\NAn/\An,\wloop}^{(\pm)}(k_1,k_2,k_3) &= \int_{-\infty}^\infty \rho^{\text{dS}}_{\mu_1\mu_2} (\mu) \widetilde{B}^{(\pm)}_{++,\NAn/\An,\tree}\p{k_1,k_2,k_3;\mu} \dd{\mu},\\[1em]
    \widetilde{B}^{(\pm)}_{++,\NAn,\tree} \p{k_1,k_2,k_3;\mu} &\coloneqq -\frac{i}{4\sqrt{\pi} k_1^3k_2^3} \frac{\abs{\alpha}^2}{\Lambda^3} \frac{2^{2i\mu} \lambda e^{-\pi\mu}}{\p{\frac{3}{2} \mp i\lambda}^2 +\mu^2} \sum_{n=0}^2 \frac{d_n}{2^n} \mathbf{\widetilde F}_{i\mu}^{-\frac{3}{2}+n\mp i\lambda-i\mu}(p)\n
    &\peq \times \mGamma{i\mu, -\frac{3}{2}+n\mp i\lambda -i\mu, \frac{1}{2} \pm i\lambda+i\mu,\frac{1}{2} \pm i\lambda-i\mu}{1 \pm i\lambda}, \\[1em]
    \widetilde{B}^{(\pm)}_{++,\An,\tree} \p{k_1,k_2,k_3;\mu} &\coloneqq \frac{i}{8k_1^3k_2^3} \frac{\abs{\alpha}^2}{\Lambda^3} \frac{\lambda}{\mu \bq{\p{\lambda^2 -\mu^2 -\frac{9}{4}}^2 +9\lambda^2}} \sum_{n=0}^2 \frac{c_n^\pm}{2^n} \mathbf{H}_{n,i\mu}^{\pm i\lambda} (p)\n
    &= \frac{i}{8k_1^3k_2^3} \frac{\abs{\alpha}^2}{\Lambda^3} \frac{\lambda}{\mu \bq{\p{\lambda^2 -\mu^2 -\frac{9}{4}}^2 +9\lambda^2}} \sum_{n=0}^2 \frac{c_n^\pm}{2^n}\n
    &\peq \times\sum_{m=0}^\infty \frac{(-)^m (m+1)_n \p{\frac{1}{2}+i\mu}_m}{\p{\frac{1}{2} +m \pm i\lambda +i\mu} (1+2i\mu)_m} \mathbf{\widetilde{F}}_{i\mu}^{1+n+m} (p). &(p>3)
\end{align}
The full results for both the signal and all the backgrounds have an intricate form and can be found in \app{app:full-results}.

\begin{figure}[htbp]
    \centering
    \begin{tabular}{cc}
        \includegraphics[scale=0.75]{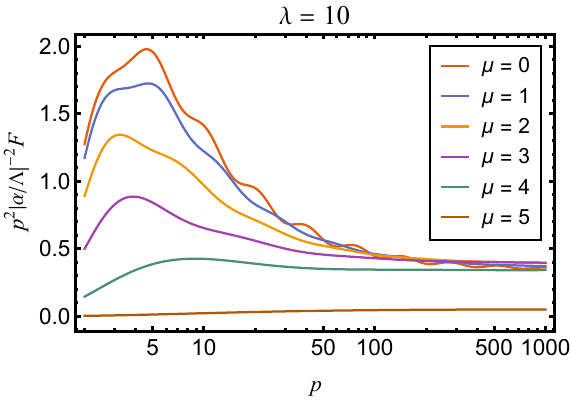} &
        \includegraphics[scale=0.75]{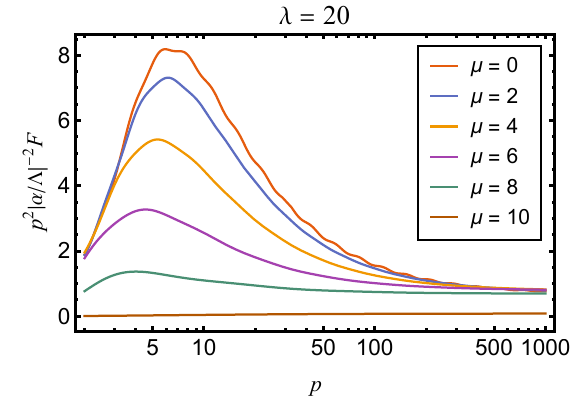} \\
        \includegraphics[scale=0.75]{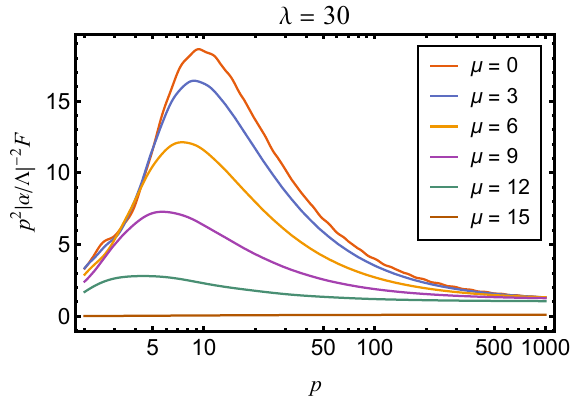} &
        \includegraphics[scale=0.75]{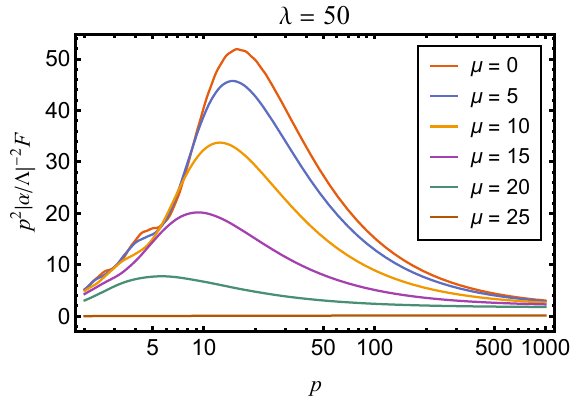}
    \end{tabular}
    \caption{The full one-loop bispectrum including the combinatorial background for various \(\lambda\) and \(\mu_1 = \mu_2 = \mu\) rescaled by \(p^2\).
    The bumps in the diagrams are artifacts of the \(p^2\) rescaling.}
    \label{fig:1-loop-full}
\end{figure}

\subsection{Combinatorial backgrounds from permuted diagrams}
\label{subsec:1-loop-perm}

Finally, there are also \emph{combinatorial backgrounds} from permuted diagrams, where the soft momentum \(k_3\) is contracted with the 3-point vertex in \fig{fig:tree-feyn}. These contributions can be obtained from \(B_{-+}\) and \(B_{++}\) by permuting \(k_3\to k_1,k_2\). In terms of \(p\) and \(\chi\):
\begin{align}\label{eq:pchi-trans}
    \begin{dcases}
        p' = \frac{p -\chi +2}{p +\chi},\\
        \chi' = -\frac{p -\chi -2}{p +\chi}
    \end{dcases} && \text{for }k_3 &\to k_1;&
    \begin{dcases}
        p' = \frac{p +\chi +2}{p -\chi},\\
        \chi' = \frac{p +\chi -2}{p -\chi}
    \end{dcases} && \text{for }k_3 &\to k_2.
\end{align}
Unfortunately, since large \(p\) corresponds to \(p'\to1\), both SPA and the analytic method described in previous sections do not apply. Instead, one needs to expand the tree-level amplitudes around \(p'\to1\).
According to the analysis in \Sec{sec:tree-SPA}, in this limit, all stationary phase points either disappear or approach fixed values. In particular, the creation and annihilation time become coincident for \(B_{-+}\). Consequently, \(F(k_2,k_1,k_3) +F(k_3,k_1,k_2)\) does not oscillate but falls off as \(p^{-2}\) in the squeezed limit as local interactions, i.e. the permuted diagrams only contribute to the analytic part of the amplitude at large \(p\). 

Nevertheless, this expansion is only of theoretical interest, and we must resort to numerically integrating the exact tree level amplitudes over the spectral density. This is because unless \(p\gg\lambda,\mu\), the number of terms needed to be included in the expansion is \(\order(\lambda/p,\mu/p)\).
For \(\lambda,\mu\sim\order(10)\), this would require either \(\order(10)\) numbers of higher order terms or \(p\gg\order(10)\) as illustrated in the right panel of \fig{fig:1-loop-breakdown}. 

It is difficult in practice to detect the NG, and one has to identify the non-analyticity using data from moderate values of \(p\), where the theory predicts a more complicated \(p\)-dependence as shown in \fig{fig:1-loop-full}.

\begin{figure}[htbp]
    \centering
    \includegraphics[scale=0.8]{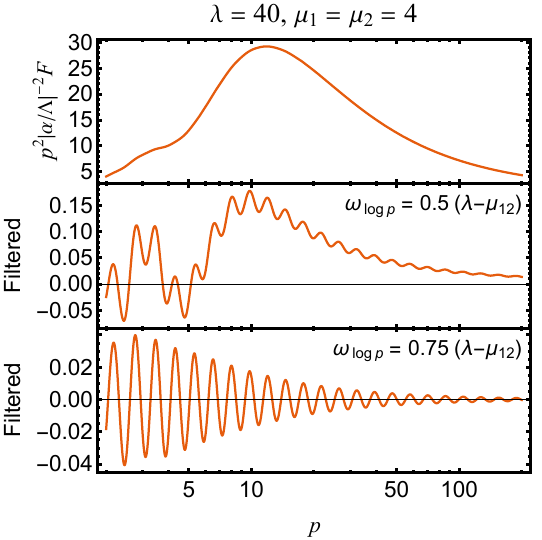}
    \caption{An example of applying high-pass filter to the full one-loop bispectrum. The Blackman-Harris window has been used in the filter. For details of how to implement the high pass filter, the reader may consult \cite{Wang:2021qez}. With a cutoff frequency at \(0.5(\lambda-\mu_{12})\), oscillatory patterns from both the signal and the permuted diagrams can pass the filter, while a slightly higher cutoff frequency will only allow the oscillatory signal to pass.}
    \label{fig:1-loop-filter}
\end{figure}

Given the large analytic backgrounds from both the resonance limit and permuted diagrams, it is natural to ask whether the non-analytic signal can be separated from them. 
This turns out not to be a serious problem as both backgrounds are mostly smooth functions in \(p\) compared to the highly-oscillatory signal. As shown in \Cite{Wang:2021qez}, one can apply a high-pass filter to extract the signal from the curve (\fig{fig:1-loop-filter}). 
To observe the signal we only require that the experimental sensitivity, \(F_{\min}\), be sufficient to resolve the oscillations, i.e.
\begin{equation}
   |\widetilde F_{\mathrm{osc}}| \gtrsim F_{\min},
\end{equation}
which is not affected by the existence of backgrounds, except in the sense that the large signal to background ratio, combined with the fact that we are yet to identify any NGs, can set an upper bound on the signal.

\subsection{Higher order corrections at one-loop}
\label{subsec:PR}

\begin{figure}[htbp]
    \centering
    \begin{tabular}{ccc}
        \includegraphics[scale=0.9]{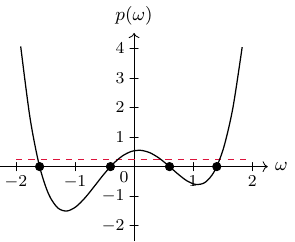} &
        \includegraphics[scale=0.9]{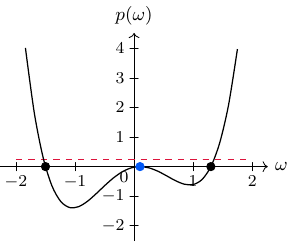} &
        \includegraphics[scale=0.9]{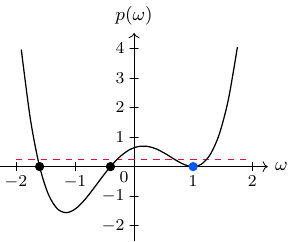}\\
        (a) & (b) & (c)
   \end{tabular}
    \caption{The condition for parametric resonance can be understood by the structure of the solutions to the quartic equation \eq{eq:param-resonance}. The solutions occur when the quartic crosses the dashed red line, and parametric resonance occurs when two of the solutions are complex. (a) For small \(\alpha\) and general values of \(\lambda\), \(\omega_1\), \(\omega_2\), there are 4 real solutions and the corrections to \(\omega_k\) are \(\order\p{\alpha^2}\); (b) when \(\lambda=\omega_1+\omega_2\), there are only two real solutions as the degenerate \(\omega_k\) (blue) receives an \(\order(\alpha)\) imaginary part (parametric resonance); (c) when \(\lambda=\abs{\omega_1-\omega_2}\) or one of \(\omega_{01}\), \(\omega_{02}\) is zero, there are 4 real solutions where the degenerate \(\omega_k\) (blue) receives an \(\order(\alpha)\) real part. }
    \label{fig:quartic}
\end{figure}

At one-loop there are also higher order contributions at each \(\order\bigl(\alpha^{2n}\bigr)\) from inserting equal numbers of \(\chi_1\chi_2\) and \(\bar\chi_1\bar\chi_2\) vertices. Resumming these contributions is equivalent to solving the full linearized equations of motion for \(\chi_{1,2}\):
\begin{align}
    \bq{-\Box +\mathbf{M}^2(\ctime) +\frac{9}{4}}  \binom{\chi_1}{\bar\chi_2} &= 0,&
    \mathbf{M}^2(\ctime)=\begin{bmatrix}
        \mu_1^2 & \bar\alpha(-\ctime)^{i\lambda}\\
        \alpha(-\ctime)^{-i\lambda} & \mu_2^2
    \end{bmatrix}.
\end{align}
In momentum space:
\begin{equation}\label{eq:exact-mfunc}
    \bq{\partial_\ctime^2 -\frac{2}{\ctime} \partial_\ctime +k^2 +\frac{\vb{M}^2(\ctime) +9/4}{\ctime^2}} \binom{\chi_1}{\bar\chi_2} = 0.
\end{equation}
It turns out that it is more intuitive to work with the proper time \(t\coloneqq-\log(-\ctime)\), so that the equation becomes
\begin{align}\label{eq:param-oscillator}
    \bq{ \partial_t^2 +k^2\ctime^2(t) +\vb{M}^2(t) } \binom{\hat\chi_1}{\hat{\bar\chi}_2} &= 0, & \hat\chi_{1,2} &\coloneqq (-\ctime)^{-3/2} \chi_{1,2}.
\end{align}
Consider the case when \(\lambda,\mu_1,\mu_2\gg1\), so that the expansion of the universe can be neglected near a pivot time \(\eta_*\). In this case, the general solution is given by
\begin{equation}\label{eq:param-oscillator-solution}
    \begin{bmatrix}
        \hat\chi_1(t)\\\hat{\bar\chi}_2(t)
    \end{bmatrix} \simeq \sum_{n=1}^4 c_n \begin{bmatrix}
        \hat\chi_1^{(n)} e^{-i\p{\omega_n+\lambda/2} t}\\
        \hat{\bar\chi}_2^{(n)} e^{-i\p{\omega_n-\lambda/2} t}
    \end{bmatrix},
\end{equation}
where \(\omega_n\) are the four roots of the quartic equation
\begin{align}\label{eq:param-resonance}
    \bq{\p{\omega +\frac{\lambda}{2}}^2 -\omega_{01}^2} \bq{\p{\omega -\frac{\lambda}{2}}^2 -\omega_{02}^2} &= \abs{\alpha}^2, &
    \omega_{01,02} \coloneqq \sqrt{k^2\eta_*^2 +\mu_{1,2}^2},
\end{align}
and \(\hat\chi_1^{(n)}\), \(\hat{\bar\chi}_2^{(n)}\) satisfy
\begin{equation}
    \begin{bmatrix}
        \p{\omega_n +\frac{\lambda}{2}}^2 -\omega_{01}^2 & -\bar\alpha\\
        -\alpha & \p{\omega_n -\frac{\lambda}{2}}^2 -\omega_{02}^2
    \end{bmatrix} \begin{bmatrix}
        \hat\chi_1^{(n)}\\\hat{\bar\chi}_2^{(n)}
    \end{bmatrix} = 0.
\end{equation}
The structure of the solutions to \eq{eq:param-resonance} for different values of \(\lambda\), \(\omega_{01}\) and \(\omega_{02}\) are illustrated in \fig{fig:quartic}.

\Eq{eq:param-resonance} implies Parametric Resonance (PR) at some time during inflation \cite{Bodas:2020yho}. This happens when any of \(\omega_k\) has an imaginary part, i.e. \(\lambda\simeq\omega_{01}+\omega_{02}\) or
\begin{equation}\label{eq:param-resonance-time}
    -k\ctime \simeq -k\ctime_{\mathrm{R}} \coloneqq \frac{\prod_{\epsilon_1,\epsilon_2=\pm} \sqrt{\lambda +\epsilon_1\mu_1 +\epsilon_2\mu_2}}{2\lambda}.
\end{equation}
Physically, this corresponds to the time when \(\lambda\) can produce a \(\chi_1\chi_2\) pair on-shell.
Around this time, \(\omega_2\) and \(\omega_3\) become nearly degenerate and are approximately given by
\begin{align}
    \omega_{2,3} &\simeq -\frac{\omega_{01}-\omega_{02}}{2} \pm \frac{1}{2} \sqrt{\p{\lambda -\omega_{01} -\omega_{02}}^2 -\frac{\abs{\alpha}^2}{\omega_{01} \omega_{02}}}.
\end{align}
Therefore, the parametric oscillator is in resonance when
\begin{equation}
    \abs{\alpha}^2 \gtrsim \omega_{01} \omega_{02} \abs{\lambda -\omega_{01} -\omega_{02}}
    \simeq \frac{\prod_{\epsilon_1,\epsilon_2=\pm} \p{\lambda +\epsilon_1\mu_1 +\epsilon_2\mu_2}^2}{4\bq{\lambda^2 -(\mu_1^2 -\mu_2^2)^2}} \delta{t}^2.
\end{equation}
where we have plugged in \(\ctime=\ctime_{\mathrm{R}} e^{-\delta{t}}\).
For \(\alpha\) not too large, PR only persists over a finite period of time. During this period, the mode function is amplified by a factor \(\mathcal{A}_{\mathrm{R}}^{1/2}\), where:
\begin{equation}
    \log\mathcal{A}_{\mathrm{R}} \simeq 
    2 \int_{t_{\min}}^{t_{\max}} \omega_2(t) \dd{t} 
    \simeq \frac{2\pi \lambda \abs{\alpha}^2}{\prod_{\epsilon_1,\epsilon_2=\pm} \p{\lambda +\epsilon_1\mu_1 +\epsilon_2\mu_2}}.
\end{equation}
Consequently, for our calculations to be valid we should require this amplification to be small:
\begin{equation}\label{eq:resonance-constraint}
    \log\mathcal{A}_{\mathrm{R}}\ll 1,\qquad\text{i.e. }
    \abs{\alpha} \ll \alpha_{\text{PR}} \coloneqq \sqrt{\frac{\prod_{\epsilon_1,\epsilon_2=\pm} \p{\lambda +\epsilon_1\mu_1 +\epsilon_2\mu_2}}{2\pi\lambda}}.
\end{equation}
For even larger \(\alpha\), PR may persist till the end of inflation, giving rise to a late-time behavior
\begin{equation}
    \abs{f_k(\ctime)} \sim (-\ctime)^{3/2 -\Im\omega_n(0)}.
\end{equation}
A simple power counting shows that, if any of \(\Im \omega_n(0)\ge1\), i.e.:
\begin{align}\label{eq:resonance-constraint-2}
    \abs{\alpha} &\gtrsim \alpha_{\max} \coloneqq \sqrt{ \max_{-\frac{\lambda}{2}<\omega<\frac{\lambda}{2}} \bq{\p{\omega +\frac{\lambda}{2}}^2 -\mu_1^2} \bq{\p{\omega -\frac{\lambda}{2}}^2 -\mu_2^2} }
    \simeq \frac{\lambda^2}{4}, & (\mu_1,\mu_2\ll\lambda),
\end{align}
there will be a late-time divergence in the perturbative calculation. This divergence will eventually be cut off by the decrease of \(\dphicl\) over time until it can no longer sustain the resonance.

\begin{figure}[htbp]
    \centering
    \includegraphics[scale=0.8]{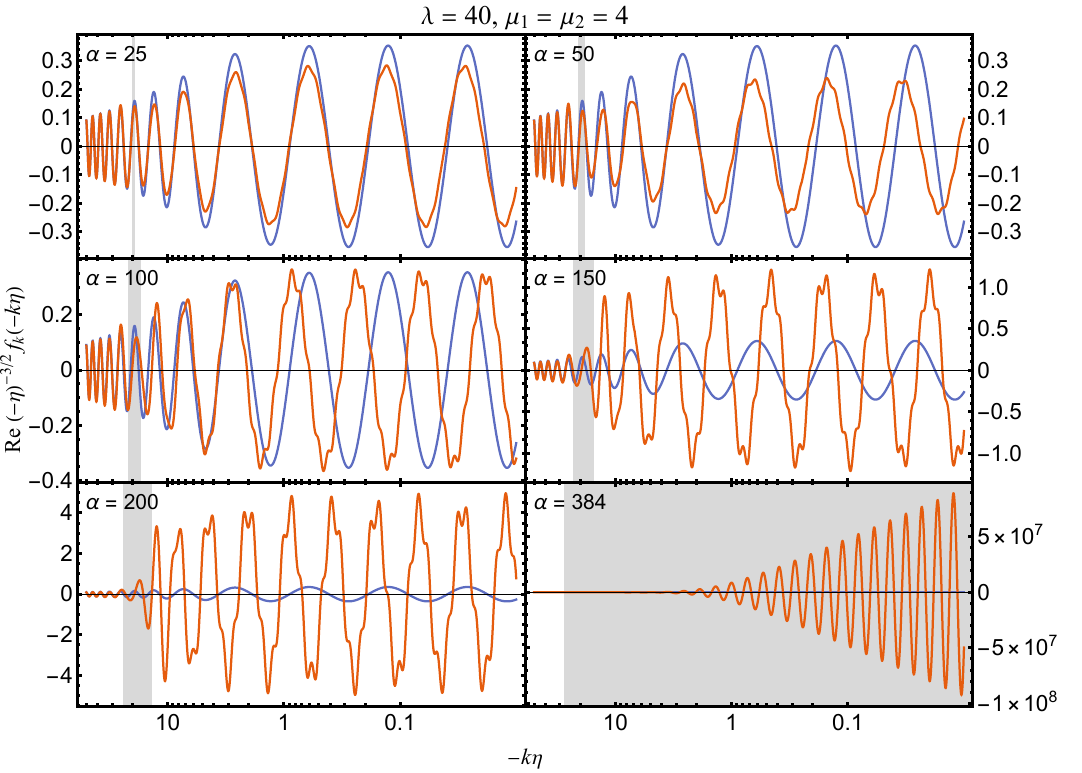}
    \caption[The numerical mode functions]{The numerical mode functions from the resummed equation of motion \eq{eq:exact-mfunc} with \(\lambda=40\) and \(\mu_1=\mu_2=4\) for various \(\alpha\) (\(\alpha_{\text{PR}}\approx99\), \(\alpha_{\max}=384\)). The blue line represents \eq{eq:heavy-mode-function} (\(\alpha=0\)) for comparison. The regions of parametric resonance where \eq{eq:param-resonance} has complex roots are marked gray. For this particular parameter choice the late-time amplitude first decreases and then increases exponentially as \(\alpha\) increases. This may not happen for other choices or parameters, e.g. \(\lambda=40\) and \(\mu_1=\mu_2=8\), where the amplitude always increases as \(\alpha\) increases.}
    \label{fig:1-loop-mfunc}
\end{figure}

So far, we have been treating the universe expansion in the adiabatic approximation. This cannot predict the solutions near the beginning and the end of PR, where \eq{eq:param-resonance} has degenerate roots and different modes evolve into each other. Here we also present the numerical solutions to \eq{eq:exact-mfunc} in \fig{fig:1-loop-mfunc} for various \(\alpha\). This shows that our previous analysis applies qualitatively even beyond the adiabatic approximation.

To see how higher order corrections at one-loop affect the bispectrum, one has to evaluate \eq{eq:loop-integral} with the corrected mode functions. Since de Sitter symmetry is badly broken by these corrections, the spectral representation method no longer applies. We can still understand its effects in the following two directions:
\begin{enumerate}
    \item The energy as a function of time is corrected according to \eq{eq:param-resonance}. For \(\mu_1=\mu_2=\mu\), the corrected energies are explicitly given by
    \begin{align}
        \omega_\pm^2 &= \frac{\lambda^2}{4} +\omega_0^2 \pm \sqrt{\lambda^2 \omega_0^2 +\abs{\alpha}^2}
        \simeq \begin{dcases}
            \p{\frac{\lambda}{2} \pm \omega_0}^2 \pm \frac{\abs{\alpha}^2}{2\lambda\omega_0}, & \abs{\alpha} \ll \lambda\omega_0,\\
            \frac{\lambda^2}{4} +\omega_0^2 \pm \abs{\alpha}, & \abs{\alpha} \gg \lambda\omega_0.
        \end{dcases}
    \end{align}
    This changes the shape of the oscillatory signal as it depends on the phase accumulation during propagation. In particular, the corrected frequency at large \(p\) can be evaluated using the late-time limit, where we can set \(k=0\) for both particles and get
    \begin{align}
        2\omega_- &= \sqrt{\lambda^2 +4\mu^2 -4\sqrt{\lambda^2\mu^2 +\abs{\alpha}^2}}\n
        &\simeq \begin{dcases}
            \lambda - 2\mu - \frac{\abs{\alpha}^2}{\lambda\mu(\lambda-2\mu)}, & \abs{\alpha} \ll \lambda\mu,\\
            \lambda - \frac{2\abs{\alpha}}{\lambda}, & \lambda\mu \ll \abs{\alpha} \ll \lambda^2.
        \end{dcases}
    \end{align}
    Such corrections can be neglected if we insist that
    \begin{equation}\label{eq:energy-constraint}
        \abs{\alpha} \ll \lambda\mu.
    \end{equation}

    \item PR increases the magnitude for both the non-analytic signal and the resonance background. 
    Comparing \(\eta_{\mathrm{R}}\) in \eq{eq:param-resonance-time} with the particle production time from \tab{tab:stationary-phases}, we have
    \begin{align}
        \frac{\lambda^2 -\mu^2}{2\lambda} < -k\eta_c &\coloneqq \frac{\lambda^2 -\mu_{12}^2}{2\lambda},&
        \frac{\ctime_c}{\ctime_{\mathrm{R}}} &= \sqrt{\frac{\lambda^2 -\mu_{12}^2}{\lambda^2 -\dmu^2}} < 1,
    \end{align}
    i.e. particle production always happens after PR. Since there are two propagators with each of them containing two mode functions evaluated at the creation and annihilation times respectively, both the signal and the background are enhanced by a factor \(\mathcal{A}_{\mathrm{R}}^2\).
\end{enumerate}

Given the smallness of the one-loop signal, an enhancement from PR may be an attractive phenomenological possibility. Unfortunately a quantitative answer cannot be obtained within the spectral decomposition framework, so we leave this direction to future work.

\section{Constraints and Signal Size}
\label{sec:signal-strength}

There are three parameters in the chemical potential model: the chemical potential, \(\lambda\), the cut-off scale, \(\Lambda\), and the symmetry breaking coupling, \(\alpha\). The chemical potential gives the maximal energy available for particle creation, and the other two are directly related to the signal strength. Note that value of \(\lambda\) is related to \(\Lambda\) by \eq{eq:lambda} with \(\dphicl\approx(60 H)^2\) fixed by measurements of the CMB power spectrum. In this section, we will discuss the constraints on \(\lambda\) and \(\alpha\) separately and combine them to predict the signal size.

\subsection{Constraints on \texorpdfstring{\(\lambda\)}{λ} and beyond de Sitter}

For \(\lambda\), A consistent effective field theory treatment requires that
\begin{equation}
    \Lambda > \sqrt{\dphicl} \approx 60 H.
\end{equation}
This is because for any operator \(\Op\), there are higher dimension operators of the form
\begin{equation}
    \frac{(\nabla\phi)^{2n}}{\Lambda^{4n}} \Op
\end{equation}
which give extra contributions suppressed by \(\dphicl^2/\Lambda^4\) after substituting \(\phi=\dphicl t\). To ensure that such higher-dimensional contributions are suppressed, we require that \(\Lambda^2>\dphicl\), or equivalently:
\begin{equation}
    \lambda \coloneqq \frac{\dphicl}{\Lambda} < \sqrt{\dphicl} \approx 60.
\end{equation}
For example, with \(\lambda\approx40\), \(\dphicl^2/\Lambda^4=\lambda^4/\dphicl^2\) is around \(0.2\).

It is worth commenting that, with even lower \(\Lambda\), it is still possible to switch to a new effective field theory where one discards the de Sitter isometry completely and writes down every interactions consistent with spatial Euclidean isometry, spacetime scale invariance and the shift symmetry of \(\phipt\), as has been shown in \cite{Cheung:2007st}. In this case, it turns out that the spectral decomposition method can still be applied provided that there is an emergent de Sitter isometry at leading order, i.e. the sound speeds of the two scalar fields \(\chi_1,\chi_2\) are equal and all \(\mathrm{U}(1)\)-breaking terms can be treated perturbatively. Nonetheless, since the energy scale of the kinetic energy, \(\sqrt{\dphicl}\approx 60H\), is outside the range of such an effective field theory, such theories are losing the power in predicting the size of the chemical potential, as can be seen already by noting that higher-order operators of the form
\begin{equation}
    \frac{\p{\nabla\phi}^{2n}}{\Lambda^{1+4n}} \nabla_\mu\phi J^\mu 
\end{equation}
changes the size of the chemical potential by \(\order\p{\dphicl^{2n}/\Lambda^{4n}}\), and becomes increasingly important when \(\Lambda<\sqrt{\dphicl}\).
Because of this, we will not consider this case in details.

\subsection{Constraints on \texorpdfstring{\(\alpha\)}{α}}

\label{subsec:alpha_constraints}

\begin{figure}[htbp]
    \centering
    \begin{tabular}{ccc}
        \includegraphics{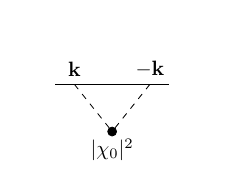} &
        \includegraphics{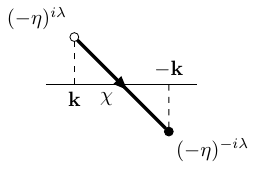} &
        \includegraphics{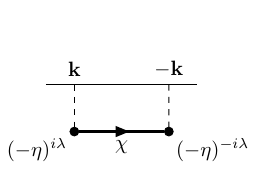}\\
        (1) & (2) & (3)
    \end{tabular}
    \caption{Corrections to the power spectrum at \(\order(\alpha^2)\). For each diagram, there is another diagram with opposite vertex types which is omitted here.}
    \label{fig:tree-ps}
\end{figure}

For the symmetry breaking coupling, \(\alpha\), the constraints are more subtle.
Firstly, there is an experimental bound on the equilateral NG from CMB data. Depending on the shape, it is given by \(\fNL < \order\p{10}\) \cite{Planck:2019kim}. Given that the non-analytic signal is small compared to the big analytic background, the \(\fNL\) constraint on the background will also effectively constrain the size of the signal. Besides this, there are also several constraints from the theory side:

\begin{enumerate}
    \item \textbf{Higher order corrections at one-loop:} These corrections are similar to \fig{fig:1-loop-feyn} but with more insertions of \(\exp\p{\pm i\phipt/\Lambda}\). As has been discussed in \Sec{subsec:PR}, the smallness of these corrections require both \eqs{eq:energy-constraint}{eq:resonance-constraint}. In later analysis, we will assume that these two conditions are satisfied so that the de Sitter symmetry is not seriously broken.
    \item \textbf{Corrections to the power spectrum:} Another constraint comes from the correction to the power spectrum:
\begin{equation}
    \bk{\phipt_{\vb{k}} \phipt_{\vb{k}}'}=\Ps_\varphi(k)\cdot (2\pi)^3\delta^3(\vb{k}-\vb{k}').
\end{equation}
Such corrections can also be evaluated using the spectral representation, with the tree-level contributions shown in \fig{fig:tree-ps}. 
\end{enumerate}

Here we quote the crude estimation \eq{eq:power-spectru-1-loop} for the corrections to the power spectrum, and leave their calculation to \app{app:power-spectrum}:
\begin{align}\begin{aligned}
    2k^3\delta{\Ps_\phi^{(1,\wloop)}}(\mu_1,\mu_2) &\simeq -\frac{\abs{\alpha}^2}{\Lambda^2} \cdot \frac{\pi}{3\lambda^2} \rho^{\text{Mink}}_{\mu_1\mu_2} (\lambda),\\
    2k^3\delta{\Ps_\phi^{(2,\wloop)}}(\mu_1,\mu_2) &\simeq \frac{\abs{\alpha}^2}{\Lambda^2} \cdot \frac{\pi^2}{3\lambda} \rho^{\text{Mink}}_{\mu_1\mu_2} (\lambda),\\
    2k^3\delta{\Ps_\phi^{(3,\wloop)}}(\mu_1,\mu_2) &\simeq \frac{\abs{\alpha}^2}{\Lambda^2} \cdot \frac{1}{\lambda^2} \Bq{ \frac{\pi}{12} \rho^{\text{Mink}}_{\mu_1\mu_2} (\lambda) +0.482 \bq{\rho^{\text{Mink}}_{\mu_1\mu_2} (\lambda) -\lambda \rho'^{\text{Mink}}_{\mu_1\mu_2} (\lambda)} }.
\end{aligned}\end{align}
Since these corrections are proportional to the zeroth-order \(1/2k^3\), as required by scale invariance, they can be regarded as a pure renormalization from an observational point of view.
However, there are some other physical effects associated with these corrections:
\begin{itemize}
    \item \Fig{fig:tree-ps} (2) produces \(\chi_1\chi_2\)-pairs that survive until the end of inflation. If it dominates over the power spectrum from the free field theory, the number density of \(\chi_1\), \(\chi_2\) will be as large as the one of \(\phipt\), while the energy density will be \(\order(\mu/k\ctime)\) larger for each \(k\) mode.
    \item \Fig{fig:tree-ps} (3) corrects the \emph{sound speed} of the inflaton fluctuations. This follows from the observation that \(\chipt\) is coupled only to \(\phipt'\) at the 2-point vertex in \eq{eq:tree-int-basis}.
\end{itemize}
To avoid these complications we require that
\begin{equation}\label{eq:constraint-ps}
    2k^3\delta{\Ps_\phi^{(\wloop)}} \ll 1.
\end{equation}

\subsection{Signal size}

\begin{figure}[htbp]
    \centering
    \begin{tabular}{cc}
        \includegraphics[scale=0.75]{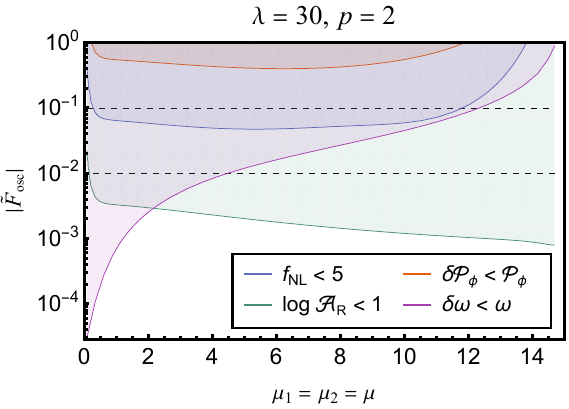}&
        \includegraphics[scale=0.75]{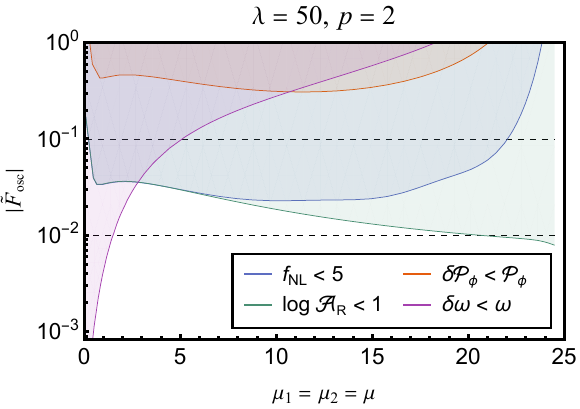}
    \end{tabular}
    \caption{The maximal signal size at \(p=2\) with various constraints. The 4 constraints shown by the shaded regions are (i) \(\fNL<5\) from CMB data \cite{Planck:2019kim}, (ii) the power spectrum constraint \eq{eq:constraint-ps}, (iii) the parametric resonance constraint \eq{eq:resonance-constraint} and (iv) the frequency correction constraint \eq{eq:energy-constraint}. Note that the last 3 constraints are only to ensure the validity of our calculations and do not imply the breakdown of the theory on their own.}
    \label{fig:1-loop-signal-constraints}
\end{figure}

\begin{figure}[htbp]
    \centering
    \begin{tabular}{c}
        \includegraphics[scale=0.8]{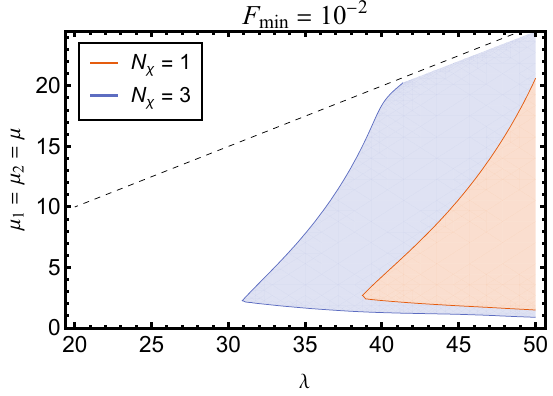}
    \end{tabular}
    \caption{The parameter space for \(\lambda\) and \(\mu_1=\mu_2=\mu\) assuming the largest \(\alpha\). The shaded regions are allowed by all the 4 constraints. The dashed line is \(\lambda=2\mu\), above which the signal is exponentially suppressed. \(N_\chi\) is the multiplicity of \(\chi_1\) and \(\chi_2\).}
    \label{fig:1-loop-param-space}
\end{figure}

In figures \ref{fig:1-loop-signal-constraints} and \ref{fig:1-loop-param-space} we take into account all the theoretical and experimental constraints discussed above. In these plots we compare \(F_{\min}\) at \(p=2\) with the maximal signal size from various constraints. We did not consider the signal at larger \(p\) since for most values of \(\lambda\) and \(\mu_1,\mu_2\) we consider, the oscillation is at frequency \(\lambda-\mu_{12}\sim \order(10)\) and only requires data from a small range of \(p\) to be identified. 
From the diagrams, we see that with \(\lambda\gtrsim40\), a signal observable with \(F_{\min}=0.01\) is possible within all the constraints. This parameter space could be probed in future \qty{21}{cm} experiments.

Some further features of these constraints worth noting:
\begin{itemize}
    \item It turns out that the most stringent constraint comes from the parametric resonance constraint \eq{eq:resonance-constraint}, which combined with \(\Lambda=\dot\phi/\lambda\) gives
    \begin{equation}
        \frac{\abs{\alpha}^2}{\Lambda^2} \lesssim \frac{\lambda^5}{2\pi (60)^2} \approx \num{4.4e-5}\:\lambda^5.
    \end{equation}
    The maximum allowed value of the interaction strength, \(\alpha\), is thus proportional to \(\lambda^5\). This explains why the signal size is small for small \(\lambda\).

    \item Another important constraint comes from \(\fNL\), which becomes increasingly important for large \(\lambda\). Since the background is larger than the signal, the \(\fNL\) constraint curve is roughly proportional to the signal-to-background ratio
    \begin{equation}
        r_{\mathrm{sbr}} \coloneqq \lim_{p\to\infty} \frac{p \lvert\widetilde{F}_{\mathrm{osc}}\rvert}{F_{\mathrm{bkg}}} \simeq \sqrt{\frac{\pi}{2}} \frac{\lambda^{5/2} (\mu_1\mu_2)^{1/2}}{\mu_{12}^{5/2} \p{\lambda-\mu_{12}}^{3/2} \p{\lambda^2 -\dmu^2}^{1/2}},
    \end{equation}
    following from \eqs{eq:1-loop-na-F}{eq:1-loop-a}. For \(\mu_1,\mu_2\sim\order(\lambda)\), \(r_{\mathrm{sbr}}\sim\order\p{\lambda^{-1}}\). This explains why the \(\fNL\) constraint is more stringent at large \(\lambda\). Note that for a typical point in the parameter region, \(r_{\mathrm{sbr}}\sim\order(0.1)\).
\end{itemize}

Up to this point, we have been considering \(\chi_1\), \(\chi_2\) as single complex scalars charged under some \(\mathrm{U}(1)_V\) group. This can easily be promoted to the case where \(\chi_1\), \(\chi_2\) are \(\vb{N}_\chi\) and \(\vb{\overline{N}}_\chi\) multiplets under some non-abelian group. In this case, the bispectrum will be larger by a factor of \(N_\chi\), i.e. the constraint curves from parametric resonance \eq{eq:resonance-constraint} will be multiplied by \(N_\chi\), while the curves from \(\fNL\) and the power spectrum constraint \eq{eq:constraint-ps} will stay unchanged as both the smooth background and the contribution to the power spectrum are also enhanced by \(N_\chi\). In figure \ref{fig:1-loop-param-space} we plot the allowed parameter space for \(N_\chi = 1,3\) to illustrate how the added multiplicity can open up new regions of parameter space.

\section{Application: Colored Higgs in SUSY Orbifold GUTs
}
\label{sec:SUSY_GUT}

Supersymmetric (SUSY) Grand Unified Theories (GUTs) (for a review, see \cite{Mohapatra:1999vv}) provide natural candidates for the pair of charged scalars appearing in the loop.
It is well known that SUSY requires the existence of a pair of Higgs doublet chiral superfields (\(H_u,H_d\)) with conjugate quantum numbers to give masses to up-type and down-type quarks respectively.
At energies above the unification scale, \(\Mgut\), these Higgses must be embedded into a larger multiplet (\(\Higgs_u,\Higgs_d\)) to form a representation of the GUT gauge group.

For example, in the simplest \(\mathrm{SU}(5)\) GUTs, the \(\Higgs_{u,d}\) multiplets are in the \(\vb{5}\) and \(\bar{\vb{5}}\) representation respectively, which include both the electroweak Higgs and the \emph{colored Higgs} bosons:
\begin{equation}
    \Higgs_u=\begin{pmatrix}
        \chi_u\\
        h_u
    \end{pmatrix},\qquad
    \Higgs_d=\begin{pmatrix}
        \chi_d & h_d
    \end{pmatrix}.
\end{equation}
While being far too heavy for direct detection at colliders, these particles may have been produced on-shell during inflation through the chemical potential mechanism and manifest themselves in non-Gaussianities. In particular, since they carry color charge and hypercharge,  they can only be pair-produced by the singlet inflaton, and contribute to non-Gaussianities at the one-loop level.

In the minimal supersymmetric \(\mathrm{SU}(5)\) GUT the unification of gauge couplings predicts \(\Mgut\sim\qty{e16}{GeV}\), and the masses of the colored Higgs masses are around this scale, which is somewhat higher than the reach of the chemical potential mechanism.
Another problem with SUSY GUTs emerges once proton decay is considered: if the colored Higgs bosons obtain their masses through a \(\Higgs_u\Higgs_d\) term in the superpotential, they will lead to too rapid proton decay unless their masses are \(\order(10)\) times larger than \(\Mgut\) \cite{Mohapatra:1999vv}.

There is however a well-known variant called \emph{orbifold} SUSY GUTs \cite{Kawamura:1999nj,Kawamura:2000ev,Hall1} that can naturally avoid the proton decay problem. In these models, the GUT partners are Kaluza-Kelin (KK) excitations from a compact extra dimension. The colored Higgs bosons then appear at the lower compactification scale \(\Mcom\sim10^{14-16}\:\unit{GeV}\) \cite{Hall1,Hall2,Hall3,Hall4} and become suitable targets for the cosmological collider.

\subsection{SUSY inflation and the fifth dimension}

In this section we will briefly review the basic setup of SUSY inflation, as well as the extra dimension. For our purposes we will follow the construction in \Cite{Kawasaki:2000yn,Kallosh:2010xz} for realizing the inflaton potential in SUSY. As a phenomenological discussion, we will not attempt to address any deeper origin behind the inflaton potential.

Inflation necessarily breaks SUSY as it is incompatible with the de Sitter isometries. Meanwhile, since the electroweak hierarchy problem suggests a SUSY breaking scale \(\Msusy\gtrsim\unit{TeV}\) today, much lower than the Hubble constant \(H\sim\qty{e14}{GeV}\) during a high-scale inflation, we can think of SUSY as approximately restored at the end of the inflation, i.e. SUSY is \emph{spontaneously broken} by the classical background of the inflaton. 

To realize such a scenario, consider an inflaton chiral superfield, \(\Phi\), with its scalar component given by
\begin{equation}
    \Phi = \frac{\sigma +i\phi}{\sqrt{2}},
\end{equation}
where \(\phi\) will be the inflaton scalar field, and the other components of \(\Phi\) are its various superpartners. To main the approximate shift symmetry in \(\phi\),  \(\Phi\) must appear in the K\"{a}hler potential in the combination \(\Phi +\bar\Phi\). This ensures that \(\phi\) has no potential, even with (spontaneously) broken SUSY. In comparison, the real component \(\sigma\), called the \emph{sinflaton}, is only massless under when SUSY is unbroken and acquires a mass typically of \(\order\p{H}\) after spontaneous SUSY breaking \cite{Kallosh:2010xz}.

We introduce the inflaton slow-roll potential by adding an extra chiral superfield, \(S\), and coupling it to the inflaton through the following superpotential:
\begin{align}
    \mathcal{L} &\supset -\int \dd[2]{\theta} Sf\p{-i\sqrt{2}\Phi} +\cc = -F_S f(\phi) +\cdots +\cc,
\end{align}
where \(f(\phi)\) is some model-dependent holomorphic function. After integrating out \(F_S\), the scalar potential, the expectation value of \(F_S\),  and the Hubble constant are related by
\begin{equation}
    V_{\mathrm{inf}}(\phicl) = \abs{\bk{F_S}}^2 = \abs{f(\phicl)}^2 \simeq 3\Mpl^2 H^2,
\end{equation}
with \(\Mpl\gtrsim\num{4e4}H\) from \eq{eq:mplanck}.
Finally, to avoid extra complications due to a light sinflaton, or the scalar component of \(S\) (denoted as \(s\)), one can make them heavy by including the following term in the K\"ahler potential:
\begin{align}
    \mathcal{L} &\supset -\int\dd[4]{\theta} \bq{ \frac{\kappa_\sigma}{2\Lambda^2} \abs{S}^2 (\Phi +\bar\Phi)^2 +\frac{\kappa_S}{4\Lambda^2} \abs{S}^4}\n
    &\supset -\frac{\kappa_\sigma}{\Lambda^2} \abs{F_S}^2 \sigma^2 -\frac{\kappa_S}{\Lambda^2} \abs{F_S}^2 \abs{s}^2.
\end{align}
This can gives them masses as large as \(\order\p{\Mpl H/\Lambda}\gtrsim\lambda=(60H)^2/\Lambda\).

Note that for simplification we have considered global supersymmetry only and ignored supergravity corrections. This can be justified because supergravity effects are suppressed by \(1/\Mpl^2\), so they can only shift particle masses by \cite{Kallosh:2010xz}
\begin{equation}
    \Delta m^2_{\mathrm{SUGRA}} \sim \frac{V_{\mathrm{inf}}(\phicl)}{\Mpl^2} = \order\p{H^2}.
\end{equation}
This is negligible if all particles we considered are assumed to be heavy compared to \(H\), except for the inflaton, for which the approximate shift symmetry eliminates such contributions.

\begin{table}[htbp]
    \centering
    \begin{tabular}{llll}\toprule
        Particles &  Boundary conditions & KK spectrum & \(\mathrm{U}(1)_R\)\\\midrule
        \(h_u,h_d\) & \((+,+)\) & \(n\Mcom\) & 0\\
        \(\chi_u,\chi_d\) & \((-,+)\) & \((n+1/2)\Mcom\) & 0\\
        \(h_u^c,h_d^c\) & \((+,-)\) & \((n+1/2)\Mcom\) & 2\\
        \(\chi_u^c,\chi_d^c\) & \((-,-)\) & \((n+1)\Mcom\) & 2\\\bottomrule
    \end{tabular}
    \caption{The boundary conditions, masses and \(\mathrm{U}(1)_R\) charges for the Higgs bosons in the orbifold SUSY GUT (\(n=0,1,2,\cdots\)). For boundary conditions, plus signs represent essentially Neumann boundary conditions while negative signs represent essentially Dirichlet boundary conditions. The KK spectrum is obtained with a flat 5D geometry.}
    \label{tab:5D-Higgs}
\end{table}

\begin{figure}[htbp]
    \centering
    \includegraphics{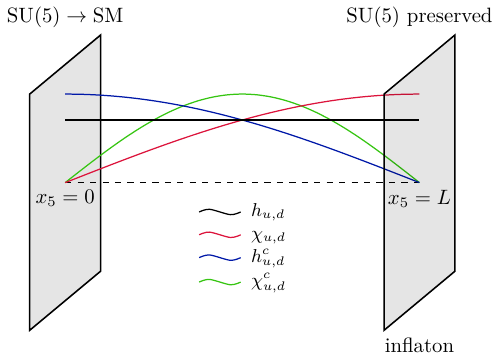}
    \caption{The 5D spacetime with two boundaries in orbifold SUSY GUTs and the 5D profiles of the lightest KK excitations of the Higgs bosons. The left boundary at \(x_5=0\) breaks \(\mathrm{SU}(5)\) down to the standard model gauge group, while the right boundary at \(x_5=L\) preserves \(\mathrm{SU}(5)\) but reduces the 5D \(\mathcal{N}=1\) SUSY down to 4D \(\mathcal{N}=1\) SUSY. The inflaton is chosen to live on the second boundary where \(\chi_{u,d}\) have the largest 5D profiles.}
    \label{fig:5D-Higgs}
\end{figure}

In order to couple the inflaton to the colored Higgs bosons in orbifold SUSY GUTs, the extra dimension should also be taken into account. We consider the simplest model in (4+1)D with the extra dimension being an interval of length \(L=\pi/\Mcom\), suggested by \cite{Hall1,Hall2,Hall3,Hall4}. In this model, \(\Higgs_u\) and \(\Higgs_d\) live in the bulk and, together with \(\Higgs_u^c\) and \(\Higgs_u^c\), form two hypermultiplets of the 5D \(\mathcal{N}=1\) SUSY. The boundary conditions at the left endpoint are chosen to break \(\mathrm{SU}(5)\) down to the standard model gauge symmetry, while the boundary conditions at the right endpoint are chosen to respect the full GUT symmetry, but reduce the 5D \(\mathcal{N}=1\) SUSY down to 4D \(\mathcal{N}=1\) SUSY. The boundary conditions are tabulated in \tab{tab:5D-Higgs}, and illustrated in \fig{fig:5D-Higgs}. Note that we assume an approximately flat 5D geometry for simplicity, which can easily be generalized to warped 5D geometries if necessary.

While the profile of the inflaton in the extra dimension is unclear, here we discuss a simple possibility that it lives on one of the 4D boundaries as in \Cite{Kumar:2018jxz}. To maximize the couplings, we put the inflaton on the right boundary where the 5D profiles of \(\chi_u\), and \(\chi_d\) are their largest. 
As has been discussed in \cite{Kumar:2018jxz}, the inflationary 4D boundary has significant gravitational back-reaction on the 5D geometry if \(\Mcom\lesssim H\), the case one would consider without a chemical potential. Fortunately, the chemical potential allows us to work in the \(\Mcom\gg H\) region, where the back-reaction can be negligible.

\subsection{Realizing the chemical potential}

In the setup described above it is more natural to write down the chemical potential model in the basis in \eq{eq:1-loop-exp-basis}. In particular, the \(\mathrm{U}(1)\) breaking \(\chi_u\chi_d\)-term can be implemented via an interaction localized on the boundary
\begin{equation}\label{eq:5D-chem}
    S_{\text{\cancel{$\mathrm{U}(1)$}}} = -\int_{x_5=L} \dd[4]{x} \p{ \alpha_{\mathrm{5D}} \chi_u\chi_d e^{i\phi/\Lambda} +\cc },
\end{equation}
where \(\alpha_{\mathrm{5D}}\) is a dimension one coupling. After plugging in the KK decomposition of \(\chi_u\) and \(\chi_d\):
\begin{equation}
    \chi_{u,d}(x,x_5) = \sqrt{\frac{2}{L}} \sum_{n=0}^\infty \chi_{u,d}^{(n)}(x) \sin \bq{\p{n+\frac{1}{2}}\frac{\pi x_5}{L}},
\end{equation}
we have
\begin{equation}
    \mathcal{L}_{\text{\cancel{$\mathrm{U}(1)$}}} = -\frac{2\alpha_{\mathrm{5D}} \Mcom}{\pi} \chi_u^{(0)} \chi_d^{(0)} e^{i\phi/\Lambda} +\cc +\cdots.
\end{equation}

When SUSY is taken into account, we consider the supermultiplet \(X_{u,d}\) containing \(\chi_{u,d}\) as scalar components. It turns out that a term like \eq{eq:5D-chem} cannot be obtained without breaking SUSY. Combined with the requirement that SUSY breaking be spontaneous during inflation, the naive way to implement the coupling between \(\chi_u\chi_d\) and \(F_S\) would be to include the following term in the superpotential:
\begin{align}\label{eq:F-term-mass}
    \mathcal{L}_{\text{\cancel{$\mathrm{U}(1)$}}} &= -\int \dd[2]{\theta} \frac{\kappa_0}{\Lambda_{\mathrm{5D}}} X_uX_d S \exp \p{\frac{\sqrt{2}\Phi}{\Lambda}} +\cc\n
    &= -\frac{2\kappa_0}{\pi} \frac{\Mcom}{\Lambda_{\mathrm{5D}}} F_S \chi_u^{(0)}\chi_d^{(0)} e^{i\phi/\Lambda} +\cdots +\cc
\end{align}
This implementation is flawed as it breaks the \(\mathrm{U}(1)_R\) symmetry of the SUSY GUT theory explicitly, under which \(\theta\) has \(+1\) charge and the charges of Higgs bosons are listed in \tab{tab:5D-Higgs}. This places the model in danger as the \(\mathrm{U}(1)_R\) symmetry has been utilized in \cite{Hall2,Hall3,Hall4} to eliminate dangerous dimension-5 operators that can mediate proton decay.

To fix this, one can consider the Giudice-Masiero mechanism \cite{Giudice:1988yz}, where the interaction now comes from the following term in the K\"ahler potential:
\begin{align}\label{eq:SUSYGUT-F-term-origin}
    \mathcal{L}_{\text{\cancel{$\mathrm{U}(1)$}}} &= -\int \dd[4]{\theta} \bq{ \frac{\kappa_1}{\Lambda^3_{\mathrm{5D}}} X_uX_d \lbar{S} S \exp\p{\frac{\sqrt{2}\Phi}{\Lambda}} +\cc}\n
    &= -\frac{2\kappa_1}{\pi} \frac{\Mcom}{\Lambda^3_{\mathrm{5D}}} \abs{F_S}^2 \chi_u^{(0)} \chi_d^{(0)} e^{i\phi/\Lambda} +\cdots +\cc
\end{align}
This new Lagrangian renders the \(\mathrm{U}(1)_R\) breaking spontaneous by \(\bk{F_S}\) and restores it at the end of the inflation. We now have the following expression for \(\alpha\):
\begin{equation}
    \alpha = \frac{2\kappa_1}{\pi} \frac{\Mcom\abs{\bk{F_S}}^2}{\Lambda^3_{\mathrm{5D}}} \simeq \frac{6\kappa_1}{\pi} \frac{\Mcom \Mpl^2}{\Lambda_{\mathrm{5D}}^3} H^2,
\end{equation}
where \(\Mpl\gtrsim\num{4e4}H\).

There is another way to obtain the chemical potential coupling. We can couple \(\chi_u\chi_d\) to \(-(\nabla_\mu\phi)^2\supset\dphicl^2=(60H)^4\) through the following K\"ahler potential term:
\begin{align}\label{eq:SUSYGUT-kinetic-origin}
    \mathcal{L}_{\text{\cancel{$\mathrm{U}(1)$}}} &= -\int \dd[4]{\theta} \bq{ \frac{\kappa_2}{\Lambda_{\mathrm{5D}}^3} X_uX_d (\Phi +\bar{\Phi})^2 \exp \p{\frac{\sqrt{2}\Phi}{\Lambda}} +\cc}\n
    &= \frac{2\kappa_2}{\pi} \frac{\Mcom}{\Lambda_{\mathrm{5D}}^3} (\nabla_\mu\phi)^2 \chi_u^{(0)} \chi_d^{(0)} e^{i\phi/\Lambda}
    +\cdots +\cc
\end{align}
After plugging in \(\sigma=0\) and \(\phi = \dphicl t\), we have the following relation
\begin{equation}
    \alpha = \frac{2\kappa_2}{\pi} \frac{M_C \dphicl^2}{\Lambda^3_{\mathrm{5D}}} \simeq \frac{2\kappa_2}{\pi} \frac{\Mcom}{\Lambda_{\mathrm{5D}}^3} (60H)^4.
\end{equation}
Note that in this realization, there are extra couplings between \(\chi_u\chi_d\) (or the effective \(\chi_\mu\)) and \(\phipt\) in addition to those described by \eq{eq:tree-int-basis}:
\begin{align}
    \delta\mathcal{L}_{\chipt}^{(\text{int})} &= -\frac{\alpha}{\lambda^2} (-\ctime)^{-i\lambda} \bq{2(-\ctime)\frac{\lambda}{\Lambda} \phipt' -\frac{1}{\Lambda^2} (\nabla\phipt)^2} e^{i\phipt/\Lambda} \chipt_\mu +\cc
\end{align}
This will make the parametric dependence on \(\lambda\) and \(\mu\) somewhat different in the bispectrum.

If we set \(\Lambda_{\mathrm{5D}}\simeq\Lambda\), \(\Mcom=2\mu H\), the upper-bounds on \(\alpha\) in both scenarios are higher than the two constraints from \eqs{eq:resonance-constraint}{eq:energy-constraint}, sufficient to cover all the parameter region in \fig{fig:1-loop-param-space} with \(N_\chi=3\).
Consequently, a non-analytic signal of order \(\fNL\gtrsim\order(0.01)\) can be used to probe these colored Higgs bosons with \(M_u=M_d=\Mcom/2\) lighter than \(30H\lesssim\qty{1.5e15}{GeV}\).

\begin{figure}[htbp]
    \centering
    \includegraphics[scale=0.8]{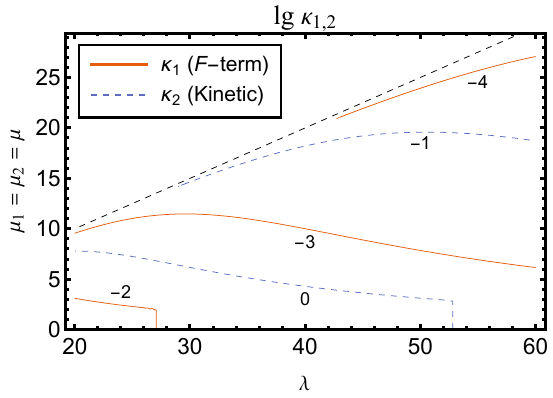}
    \caption{The maximal sizes (\(\log_{10}\)) of the Wilson coefficients of the \(F\)-term-induced symmetry-breaking operator in \eq{eq:SUSYGUT-F-term-origin} and the kinetic-induced symmetry-breaking operator in \eq{eq:SUSYGUT-kinetic-origin} under all the constraints on \(\alpha\) discussed in \Sec{subsec:alpha_constraints}, assuming \(\Lambda_{\mathrm{5D}}=\Lambda\) and maximal \(H\) from \eq{eq:mplanck}. Compared with \fig{fig:1-loop-param-space}, a typical choice of \(\alpha\) with \(F_{\min}>\num{e-2}\) while also compatible with all constraints requires \(\kappa_1\sim\num{e-3}\) or \(\kappa_2\sim\num{e-1}\). Note that with different choice of the ratio \(\Lambda_{\mathrm{5D}}/\Lambda\), \(\kappa_{1,2}\) scale as \((\Lambda_{\mathrm{5D}}/\Lambda)^3\).}
    \label{fig:1-loop-param-space-SUSYGUT}
\end{figure}

In the generic model \eq{eq:1-loop-chem-basis}, it is hard to estimate the natural size of the \(\alpha\) parameter as it has positive mass dimensions, while the system has enhanced symmetry when it vanishes, just like the fermion mass. With supersymmetry, we have seen that the size \(\alpha\) is controlled by the SUSY-breaking effect during the inflation. In this case, we can further measure the naturalness of the parameter \(\alpha\) by assuming that both terms contributing to \(\alpha\) arises and see whether extremely small Wilson coefficients are required to satisfy the constraints in \Sec{subsec:alpha_constraints} and obtain an observable signal. Under the assumption \(\Lambda_{\mathrm{5D}}=\Lambda\) and maximal \(H\) from \eq{eq:mplanck}, the result is summarized in \fig{fig:1-loop-param-space-SUSYGUT}. Combined with \fig{fig:1-loop-param-space}, we see that a typical parameter set that can satisfy all the constraints with an observable signal requires \(\kappa_1\sim\num{e-3}\) or \(\kappa_2\sim\num{e-1}\). While the required value for \(\kappa_1\) may seem to be too small to be natural, this can be solved by introducing a small hierarchy of \(\order(10)\) between \(\Lambda_{\mathrm{5D}}\) and \(\Lambda\), which will then predict \(\kappa_1\sim 1\) with contribution from \(\kappa_2\) being suppressed.

\section{Discussion and Conclusions}
\label{sec:discussion}

In this paper we have studied the chemical potential mechanism for a pair of charged scalars at the one-loop level. We have shown that the leading-order contribution to the bispectrum can be computed using the de Sitter spectral representation, where one integrates a fictitious tree-level amplitude against the spectral density function.
We have demonstrated that the non-analytic signal comes from
the threshold limit of the integral with the invariant mass \(\mu\simeq\mu_1+\mu_2\), while there is also a smooth background coming from the resonance limit when \(\mu\simeq\lambda\) .

In the limit of a large chemical potential, \(\lambda\), and masses \(\mu_1,\mu_2\), both the non-analytic signal and the analytic background have simple forms given by \eq{eq:1-loop-na-F} and \eq{eq:1-loop-a} respectively.
We have that a model with only the lowest dimensional symmetry-breaking operator can satisfy all constraints, and generate an oscillatory signal in the bispectrum of order \(\fNL\gtrsim\order(0.01)\), which could be within the sensitivity of future \qty{21}{cm} experiments. This mechanism may apply to the colored Higgs bosons in orbifold SUSY GUTs up to \(M_u=M_d\lesssim\qty{e15}{GeV}\).

In this work, we have been considering the exchange of two scalars as a spin-0 multi-particle state. It may be interesting to considering the generalization to nonzero spin exchange. One can consider two heavy particles with spin \(s_1\), \(s_2\) exchanged as a spin-\(s\) multi-particle state, where:
\begin{align}
    \vb{s} &\in \vb{s}_1 \otimes \vb{s}_2 \otimes \vb{j}
\end{align}
for some \(j=0,1,2,\cdots\) (the orbital angular momentum). 
The amplitude is then given by integrating the tree-level amplitude of a spin-\(s\) particle times the corresponding spectral density.

Another direction for future work could be relaxing the parametric resonance constraints. The spectral decomposition approach requires treating the breaking of de Sitter isometry down to scale invariance perturbatively. This put very stringent constraints on the coupling \(\alpha\) in the scalar model. However, parametric resonance might be desirable as it can induce a provide a large enhancement of the non-analytic signal. Thus, it will be helpful to compute the bispectrum in this regime, and study its phenomenology.

Finally, while our model predicts a striking oscillatory signal, it is accompanied by a large smooth background. It would be interesting to see if some variant of the minimal model can improve the signal-to-background ratio.

\acknowledgments
AB is supported by DOE grant DE-SC-0013642 at the University of Chicago and the Fermi Forward Discovery Group, LLC, under Contract No. 89243024CSC000002 with the DOE Office of Science, Office of High Energy Physics. EB, RS, and ZX are supported by the NSF grant PHY-2210361 and the Maryland Center for Fundamental Physics.

\appendix

\section{Integrals and Identities Involving Special Functions}

In this appendix, we collect some integrals involving special functions used in the main text and subsequent sections. For \(\Re\rho>0\):
\begin{align}
    &\quad~e^{-\pi\mu/2}\sqrt{\frac{\pi}{2}} \int_0^{\infty_+} x^{\rho-1} e^{ipx} H_{i\mu}^{(1)}(x) \dd{x} = \frac{e^{(\rho-1)\pi i/2}}{2^{\rho-1/2}} \vb{F}^{\rho}_{i\mu} (p),  
    \hspace{3.05cm}p>-1,\label{eq:hankel-int}\\
    &\quad~e^{-\pi\mu/2}\sqrt{\frac{\pi}{2}} \int_0^{\infty_-} x^{\rho-1} e^{ipx} H_{i\mu}^{(1)}(x) \dd{x} = \frac{e^{(\rho-1)\pi i/2}}{2^{\rho-1/2}} \vb{F}^{\rho}_{i\mu} (p+i\epsilon),
    \hspace{2.25cm}p<-1,\label{eq:hankel-intt}\\ 
    &\quad~e^{-\pi\mu/2}\sqrt{\frac{\pi}{2}} \int_0^x x^{\rho-1} e^{\pm ix} H_{i\mu}^{(1)}(x) \dd{x} = \frac{e^{(\rho-1)\pi i/2}}{2^{\rho-1/2}} \bq{\mathbf{G}_{i\mu}^{\rho,\pm}(x) +\mathbf{G}_{-i\mu}^{\rho,\pm}(x)}.\label{eq:hankel-int-2}
\end{align}
where the function on the RHS of the first two equations is given by
\begin{equation}\label{eq:f-func'}
    \mathbf{F}_{i\mu}^\rho(p) \coloneqq \mGamma{\rho+i\mu,\rho-i\mu}{\frac{1}{2}+\rho} \hF{\rho+i\mu}{\rho-i\mu}{\frac{1}{2}+\rho}{\frac{1-p}{2}},
\end{equation}
with \(\Gamma,F\) defined as in footnote \ref{foot:defs}. In the final equation
\begin{align}\label{eq:f-func-2}
    \mathbf{G}_{i\mu}^{\rho,\pm}(x) &\coloneqq \frac{1}{\rho +i\mu} \mGamma{-2i\mu}{\frac{1}{2}-i\mu} (-2ix)^{\rho+i\mu} \hFPQ22{\frac{1}{2}+i\mu,\rho+i\mu}{1+2i\mu,1+\rho+i\mu}{\pm2ix}\n
    &= \frac{(-2ix)^{\rho+i\mu}}{4\pi\sinh\pi\mu}
    \int_L \frac{(\mp 2ix)^t}{\rho+i\mu+t} \mGamma{-t,\frac{1}{2}+i\mu+t}{1+2i\mu+t} \dd{t},
\end{align}
where \(L:-i\infty\to i\infty\) is some path separating poles of \(\Gamma(-t)\) from other poles of the integrand. The first two equations \eqref{eq:hankel-int} and \eqref{eq:hankel-intt} can be derived from eq. (10.43.22) in \cite{NIST} after Wick rotations (also see the appendices of \Cite{Bodas:2020yho} and \cite{Sleight:2019mgd}). For the second equation, we can first use the following power expansion of \(H_{i\mu}^{(1)}(x)\): 
\begin{align}
    e^{-\pi\mu/2 \pm ix} H_{i\mu}^{(1)}(x) &= -\frac{2i}{\sqrt{\pi}} \bq{ \mGamma{-2i\mu}{\frac{1}{2}-i\mu} (-2ix)^{i\mu} \sum_{n=0}^\infty \frac{\p{\frac{1}{2}+i\mu}_n}{(1+2i\mu)_n} \frac{(\pm2ix)^n}{n!} +(\mu\to-\mu)}
\end{align}
and then integrate term-by-term.

The \(\mathbf{F}_{i\mu}^{\rho}(p)\) in \eq{eq:f-func'} also satisfies the following recurrence relation \NIST{15.5.19}:
\begin{align}\label{eq:f-func-recur}
 \frac{1}{4} \p{p^2-1} \mathbf{F}_{i\mu}^{2+\rho}(p)
 -\p{\frac{1}{2} +\rho} p \mathbf{F}_{i\mu}^{1+\rho}(p) +\p{\rho^2 +\mu^2} \mathbf{F}_{i\mu}^{\rho}(p) = 0. 
\end{align}
This has been used to prove the equivalence of the two forms in \eq{eq:tree-MP-exact}, where the latter form is equivalent to the one in \Cite{Bodas:2020yho}.

Finally, to extract the non-analytic piece of the integral over the spectral density, we made use of the identity \NIST{5.13.1}
\begin{equation}
\label{eq:GammaIntegral}
\int_L \Gamma(a+s)\Gamma(b-s)z^{-s} \dd{s} = \Gamma(a+b) \cdot\frac{z^a}{(1+z)^{a+b}},
\end{equation}
where \(L:-i\infty\to i\infty\) is some path separating the poles of \(\Gamma(a+s)\) from the poles of \(\Gamma(b-s)\).

\section{Asymptotic Expansions of \texorpdfstring{\(\mathbf{\widetilde F}_{i\mu}^s (p)\)}{Fiμ s}}
\label{app:asymp-bounds}

In \eq{eq:Fseparation} we split the integrand into two pieces to facilitate use of the residue theorem. In this appendix, we collect identities needed to justify that step. Firstly, the \(\mathbf{\widetilde F}_{i\mu}^s (p)\) defined in \eq{eq:tilde-f-func} has the following integral representation \NIST{15.6.1}:
\begin{align}
    \mathbf{\widetilde F}_{i\mu}^{s}(p) &= \frac{2^{-2i\mu}}{\sqrt{\pi}} \mGamma{1-i\mu}{\frac{1}{2}-i\mu} \int_0^1 [t(1-t)]^{-1/2-i\mu} \p{\frac{p-1}{2}+t}^{-s} \dd{t}.
\end{align}
In this appendix, we will use this to derive the asymptotic approximation of \(\mathbf{\widetilde F}_{i\mu}^s\) in three limits assuming \(p>1\).

The first limit is when \(s=\rho -i\mu\) and \(\mu\to\infty\) with \(\Im\mu\ge0\):
\begin{align}
    2^{2i\mu} \mathbf{\widetilde F}_{i\mu}^{\rho -i\mu}(p) &= \frac{1}{\sqrt{\pi}} \mGamma{1-i\mu}{\frac{1}{2}-i\mu} \int_0^1 [t(1-t)]^{-1/2} \p{\frac{p-1}{2}+t}^{-\rho} e^{-i\mu g(t;p)} \dd{t},
\end{align}
where
\begin{equation}
    g(t;p) = \log t(1-t) -\log\p{\frac{p-1}{2}+t}.
\end{equation}
For \(p>1\), \(g'(t;p) = 0\) has a unique solution between \((0,1)\) at
\begin{align}
    t_* &= \frac{1}{2} \p{1-p +\sqrt{p^2-1}}, &g''(t_*;p) = -\frac{4}{\sqrt{p^2-1} \p{p-\sqrt{p^2-1}}}<0.
\end{align}
Therefore, as \(\mu\to\infty\), the integral above can be approximated as
\begin{align}\label{eq:tilde-F-asymp-3}
    2^{2i\mu} \mathbf{\widetilde F}_{i\mu}^{\rho-i\mu}(p) &\simeq \frac{e^{\pi i/4}}{\sqrt{\pi}} \mGamma{1-i\mu}{\frac{1}{2}-i\mu} \sqrt{\frac{2\pi}{-\mu g''(t_*;p)}} [t_*(1-t_*)]^{-1/2-i\mu} \p{\frac{p-1}{2}+t_*}^{-\rho+i\mu}\n
    &\simeq \frac{1}{\sqrt{2}} \p{\frac{\sqrt{p^2-1}}{2}}^{-\rho} \p{p-\sqrt{p^2-1}}^{-i\mu}.
\end{align}
In the main text we use the asymptotic expansion of \eqref{eq:tilde-F-asymp-3} to demonstrate that we can close the contour in the upper half plane.

There are other two limits that will be used to calculate \(B_{++}\) in \app{app:exact-B++}. The second limit is when \(s\to\infty\), with \(\Re s>0\). In this limit, the integration is dominated by the \(t=0\) boundary:
\begin{align}\label{eq:tilde-F-asymp-1}
    2^{2i\mu} \mathbf{\widetilde F}_{i\mu}^{s}(p) &\simeq \frac{1}{\sqrt{\pi}} \mGamma{1-i\mu}{\frac{1}{2}-i\mu} \int_0^\infty t^{-1/2-i\mu} \p{\frac{p-1}{2}+t}^{-s} \dd{t}\n
    &= \frac{1}{\sqrt{\pi}} \mGamma{1-i\mu,-\frac{1}{2}+i\mu+s}{s} \p{\frac{p-1}{2}}^{1/2-i\mu-s}\n
    &\simeq \frac{\Gamma\p{1 -i\mu}}{\sqrt{\pi}} s^{-1/2 +i\mu} \p{\frac{p-1}{2}}^{1/2-i\mu-s}.
\end{align}
The third limit is when \(\abs{\mu}\to\infty\) with \(\Im\mu\ge0\):
\begin{align}
    \mathbf{\widetilde F}_{i\mu}^{s}(p) &= \frac{2^{-2i\mu}}{\sqrt{\pi}} \mGamma{1-i\mu}{\frac{1}{2}-i\mu} \int_0^1 [t(1-t)]^{-1/2} \p{\frac{p-1}{2}+t}^{-s} e^{-i\mu h(t)} \dd{t},
\end{align}
where \(h(t)\coloneqq\log t(1-t)\). In this case:
\begin{align}\label{eq:tilde-F-asymp-2}
    \mathbf{\widetilde F}_{i\mu}^{s}(p) 
    &\simeq \frac{2^{-2i\mu} e^{\pi i/4}}{\sqrt{\pi}} \mGamma{1-i\mu}{\frac{1}{2}-i\mu} \eval{ \sqrt{\frac{2\pi}{-\mu h''(t)}} \bq{t(1-t)}^{-1/2-i\mu} \p{\frac{p-1}{2}+t}^{-s} }_{t=1/2}\n
    &\simeq \p{\frac{p}{2}}^{-s}.
\end{align}

\section{Spectral Density in Non-relativistic Limit of de Sitter Spacetime}

\label{app:NRdS}

In this section, we will derive \eq{eq:ds-spectral-threshold} directly using the non-relativistic limit of de Sitter spacetime without referring to \eq{eq:ds-spectral}. To begin with, we first show that the non-relativistic limit of de Sitter spacetime is described by the \emph{quantum inverted harmonic oscillator} with Hamiltonian
\begin{equation}\label{eq:QIHO}
    \mathscr{H} = \frac{\vb{P}^2}{2M} -\frac{1}{2} M \vb{X}^2.
\end{equation}
The equivalence can either be established at the level of isometry groups or metrics. For a derivation using group contractions readers may refer to \Cite{Bacry:1968zf}. We will take the second approach. We start from the expression for the de Sitter metric in the \emph{static coordinates} with the speed-of-light \(c\) restored\footnote{The coordinate transformation from the static coordinate to the conformal coordinates \eq{eq:dS-metric} we have been using so far is given by
\begin{equation*}\begin{aligned}
    X^i &= -\frac{x^i}{\ctime},&\qquad
    e^{-T} &= \sqrt{\ctime^2 -\frac{x^2}{c^2}}.
\end{aligned}\end{equation*}}
\begin{equation}
    \dd{s}^2 = -\p{1 -\frac{R^2}{c^2}} c^2 \dd{T}^2 {}+\p{1 -\frac{R^2}{c^2}}^{-1} \dd{R}^2 {}+R^2 \dd{\mathrm{\Omega}}^2.
\end{equation}
The nontrivial Christoffel symbols are as follows:
\begin{align}
    \Gamma^R_{RR} &= -\Gamma^T_{RT} = \frac{R}{c^2} \p{1 -\frac{R^2}{c^2}}^{-1}, & \Gamma^R_{TT} &= -R \p{1 -\frac{R^2}{c^2}}.
\end{align}
In the limit \(c\to\infty\), only \(\Gamma^R_{TT}\) survives and the geodesic equation becomes
\begin{equation}
    \frac{\dd[2]{X^i}}{\dd{T}^2} +\Gamma^R_{TT} \hat{R}^i = \frac{\dd[2]{X^i}}{\dd{T}^2} -X^i = 0,
\end{equation}
which is exactly the Newtonian equation of motion from \eq{eq:QIHO}.

In the non-relativistic limit, the scalar propagator can be approximated as
\begin{equation}\label{eq:NR-propagator}
    G_{-+}(X_1, X_2) \simeq \frac{1}{2M} G_{\mathrm{NR}}(\vb{X}_1,\vb{X}_2, T_1-T_2) e^{-iMc^2(T_1-T_2)},
\end{equation}
where
\begin{align}
    G_{\mathrm{NR}}(\vb{X}_1,\vb{X}_2, T) &\coloneqq \p{\frac{M}{2\pi i\sinh T}}^{3/2} \exp\Bq{\frac{iM}{2\sinh T} \bq{\p{\vb{X}_1^2 +\vb{X}_2^2} \cosh T -2\vb{X}_1\cdot\vb{X}_2}}
\end{align}
is the propagator of the inverted harmonic oscillator \eq{eq:QIHO}\footnote{The readers may be more familiar with the propagator of the quantum harmonic oscillator given by
\[
    G(\vb{X}_1,\vb{X}_2, T) \coloneqq \p{\frac{M\omega}{2\pi i\sin\omega T}}^{3/2}
    \exp\Bq{\frac{iM\omega}{2\sin\omega T} \bq{\p{\vb{X}_1^2 +\vb{X}_2^2} \cos \omega T -2\vb{X}_1\cdot\vb{X}_2}},
\]
following from path integral. The propagator of the inverted harmonic oscillator can be obtained in the same manner, or by analytic continuation \(\omega\to iH\), where we have then set \(H=1\).}. Now, we rewrite the spectral decomposition \eq{eq:kallen} using \(M\simeq\mu\) as
\begin{align}
    G_{-+}(X_1, X_2; M_1) G_{-+}(X_1, X_2; M_2)
    &\simeq \int_0^\infty \rho^{\mathrm{dS}}_{M_1M_2}(M) G_{-+}(X_1, X_2; M) \dd{M}.
\end{align}
Given that \(G_{-+}(X_1,X_2;M)\) contains the highly oscillatory \(e^{-iMc^2(T_1-T_2)}\) term and \(\rho^{\mathrm{dS}}_{M_1M_2}(M)\) changes rapidly around \(M=M_{12}\coloneqq M_1+M_2\), we can approximate the integration as
\begin{align}
    \rhs &\simeq G_{-+}(X_1, X_2; M_{12}) \int_{-M_{12}}^\infty \rho^{\mathrm{dS}}_{M_1M_2}(M_{12} +\varepsilon) e^{-i\varepsilon c^2(T_1-T_2)} \dd{\varepsilon},
\end{align}
i.e.
\begin{align}
    \int_{-M_{12}}^\infty \rho^{\mathrm{dS}}_{M_1M_2}(M_{12} +\varepsilon) e^{-i\varepsilon c^2T} \dd{\varepsilon}
    &\simeq \eval{\frac{G_{-+}(X_1, X_2; M_1) G_{-+}(X_1, X_2; M_2)}{G_{-+}(X_1, X_2; M_{12})}}_{T_1-T_2=T}\n
    &= \frac{(2M_{12})}{(2M_1)(2M_2)} \frac{G_{-+}^{\mathrm{NR}}(\vb{X}_1,\vb{X}_2, T;M_1) G_{-+}^{\mathrm{NR}}(\vb{X}_1,\vb{X}_2, T;M_2)}{G_{-+}^{\mathrm{NR}}(\vb{X}_1,\vb{X}_2, T;M_{12})}\n
    &= \frac{e^{-3\pi i/4}}{2^{5/2}\pi^{3/2}} \sqrt{\frac{M_1M_2}{M_{12}}} \p{\sinh T}^{-3/2}.
\end{align}
The spectral density near \(M=M_{12}\) can thus be recovered using the inverse Fourier transformation (\(c=1\)):
\begin{align}
    \rho^{\mathrm{dS}}_{M_1M_2}(M_{12} +\varepsilon) &\simeq \frac{1}{2^{5/2}\pi^{3/2}} \sqrt{\frac{M_1M_2}{M_{12}}} \biggl[ e^{-3\pi i/4} \int_0^\infty \frac{\dd{T}}{2\pi} \p{\sinh T}^{-3/2} e^{i\varepsilon T}\n
    &\phantom{=\frac{1}{2^{5/2}\pi^{3/2}} \sqrt{\frac{M_1M_2}{M_{12}}} \biggl[}
    +e^{3\pi i/4} \int_{-\infty}^0 \frac{\dd{T}}{2\pi} \p{-\sinh T}^{-3/2} e^{i\varepsilon T} \biggr]\n
    &= \frac{1}{4\pi^{5/2}} \sqrt{\frac{M_1M_2}{M_{12}}} \bq{ e^{-3\pi i/4} \int_0^1 \frac{t^{1/2-i\varepsilon} \dd{t}}{\p{1-t^2}^{3/2}} +\cc }\n
    %
    %
    &= \frac{1}{4\pi^3} \sqrt{\frac{M_1M_2}{M_{12}}} \Gamma\p{\frac{3}{4} +\frac{i\varepsilon}{2}} \Gamma\p{\frac{3}{4} -\frac{i\varepsilon}{2}} e^{\pi\varepsilon/2},
\end{align}
where the divergence at \(T=0\) cancels between the \(T>0\) and \(T<0\) integral. This agrees with \eq{eq:ds-spectral-threshold}.

The quantum inverted harmonic oscillator naturally explains why \(\rho^{\text{dS}}_{\mu_1\mu_2}\) does not drop to zero, but is only exponentially suppressed, when \(M<M_1+M_2\). This is because a particle can decay into two particles with a higher total mass by extracting energy from the expanding universe and going through a quantum tunneling process.

\section{Exact Expression of \texorpdfstring{\(B_{++,\An}\)}{B++,A}}
\label{app:exact-B++}

\begin{figure}[htbp]
    \centering
    \includegraphics{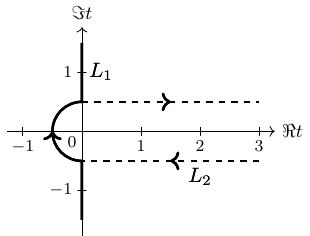}
    \caption{Two contours used to perform the integration in \eq{eq:h-func}.}
    \label{fig:complex-t-PP}
\end{figure}

In this subsection, we provide an analytic evaluation of the integral \eq{eq:tree-PP-A} and derive some of its properties. We first complete the \(x_1\)-integral using \eq{eq:hankel-int-2}:
\begin{equation}\begin{aligned}
    \int^{x_2}_0 \dd{x_1} \Int_{1,++}^{(\pm)}(\lambda;x_1)
    &= \frac{e^{-\pi i/4} e^{\mp\pi\lambda/2}}{2^{\pm i\lambda}} \bq{\mathbf{G}_{i\mu}^{\frac{1}{2}\pm i\lambda,-}(x_2) +\mathbf{G}_{-i\mu}^{\frac{1}{2}\pm i\lambda,-}(x_2)},\\
    \int_0^{x_2} \dd{x_1} \Int_{1,-+}^{(\mp)*} (\lambda;x_1)
    &= \frac{e^{\pi i/4} e^{\pm\pi\lambda/2}}{2^{\pm i\lambda}} \bq{\mathbf{G}_{i\mu}^{\frac{1}{2}\mp i\lambda,+*}(x_2) +\mathbf{G}_{-i\mu}^{\frac{1}{2}\mp i\lambda,+*}(x_2)},
\end{aligned}\end{equation}
where \(\mathbf{G}_{i\mu}^{\rho,\pm}(x)\) is defined in \eq{eq:f-func-2} with the contour \(L\) shown by the contour \(L_1\) in \fig{fig:complex-t-PP}. Using the integral representation from \eq{eq:f-func-2}, the \(x_2\)-integration can be evaluated first using \eqs{eq:hankel-int}{eq:hankel-intt}, leaving
\begin{align}\label{eq:tree-PP-A-exact}
    B_{++,\An}^{(\pm)}(k_1,k_2,k_3) &=
    \frac{i}{8k_1^3k_2^3} \frac{\abs{\alpha}^2}{\Lambda^3} \frac{\lambda}{\mu \bq{\p{\lambda^2 -M^2}^2 +9\lambda^2}} \sum_{n=0}^2 \frac{c_n^\pm}{2^n} \bq{ \mathbf{H}_{n,i\mu}^{\pm i\lambda} (p) -\mathbf{H}^{\pm i\lambda}_{n,-i\mu}(p) },
\end{align}
where we have used \eq{eq:Fseparation} and the function \(\mathbf{\widetilde{F}}_{i\mu}^s (p)\) defined in \eq{eq:tilde-f-func} to write
\begin{align}\label{eq:h-func}
    \mathbf{H}_{n,i\mu}^{i\lambda} (p) &\coloneqq
    \frac{1}{2\pi i} \mGamma{1+2i\mu}{\frac{1}{2}+i\mu} \int_L \frac{\mathbf{\widetilde{F}}_{i\mu}^{1+n+t} (p) \dd{t}}{\frac{1}{2} +i\lambda +i\mu +t} \mGamma{-t, 1 +n +t, \frac{1}{2}+i\mu +t}{1+2i\mu +t}.
\end{align}

Although the \(t\)-integral in \eq{eq:h-func} cannot be evaluated analytically, it can be represented as an infinite sum over the residues of \(\Gamma(-t)\) if the path \(L\) can be deformed to start and end at \(+\infty\), as shown by the contour \(L_2\) in \fig{fig:complex-t-PP}. To see when such a deformation is allowed, we need the following asymptotic bound on the integrand as \(t\to \infty\) with \(\Re t\ge0\):\footnote{For the gamma functions, the asymptotic property \(\Gamma(a+t)\sim t^a \Gamma(t)\) and the identity \(\Gamma(-t) \Gamma(1+t) = -\pi \csc \pi t\) are used. For \(\mathbf{\widetilde F}_{i\mu}^{1+n+t}(p)\), the asymptotic expansion is given by \eq{eq:tilde-F-asymp-1} in \app{app:asymp-bounds}.}
\begin{align}\label{eq:h-integrand-asymp}
    \frac{\mathbf{\widetilde{F}}_{i\mu}^{1+n+t} (p)}{\frac{1}{2} +i\lambda +i\mu +t} \mGamma{-t, 1 +n +t, \frac{1}{2}+i\mu +t}{1+2i\mu +t}
    &\sim t^{-2 +n} \p{\frac{p-1}{2}}^{-t} \csc \pi t\n
    &\lesssim \begin{dcases}
        t^{-2 +n} \p{\frac{p-1}{2}}^{-t}, & t\to \pm i\tau +\infty,\\
        t^{-2 +n} e^{-\pi\abs{t}}, & t\to \pm i\infty.
    \end{dcases}
\end{align}
From this it follows that, for \(p>3\), the integral converges exponentially for \(L_2\) in \fig{fig:complex-t-PP} and can be expressed as a sum over the residues of \(\Gamma(-t)\):
\begin{align}
    \mathbf{H}_{n,i\mu}^{i\lambda} (p) &= \sum_{m=0}^\infty \frac{(-)^m (m+1)_n \p{\frac{1}{2}+i\mu}_m}{\p{\frac{1}{2} +m +i\lambda +i\mu} (1+2i\mu)_m} \mathbf{\widetilde{F}}_{i\mu}^{1+n+m} (p),&(p>3).
\end{align}
In particular, this shows that \(B_{++,\An}^{(\pm)}\) are analytic functions at \(p\to\infty\), and in the squeezed limit:
\begin{align}
    B_{++,\An}^{(\pm)}(k_1,k_2,k_3) &\sqlimit
    \frac{3}{8k_1^3k_2^3} \frac{\abs{\alpha}^2}{\Lambda^3} \frac{\lambda(1 \pm i\lambda)}{\bq{\p{\frac{1}{2} \pm i\lambda}^2 +\mu^2} \bq{\p{\lambda^2 -\mu^2 -\frac{9}{4}}^2 +9\lambda^2}} p,\\[1em]
    B_{++,\An}(k_1,k_2,k_3) &\coloneqq B_{++,\An}^{(+)}(k_1,k_2,k_3) +B_{++,\An}^{(-)}(k_1,k_2,k_3)\n
    &\sqlimit \frac{3}{4k_1^3k_2^3} \frac{\abs{\alpha}^2}{\Lambda^3} \frac{\lambda \p{\mu^2 +\frac{1}{4}}}{\bq{\p{\lambda^2 -\mu^2 -\frac{1}{4}}^2 +\lambda^2} \bq{\p{\lambda^2 -\mu^2 -\frac{9}{4}}^2 +9\lambda^2}} p.
\end{align}
On the other hand, \eq{eq:h-integrand-asymp} also shows that the integral always converges exponentially for \(L_1\) in \fig{fig:complex-t-PP}. Therefore, \(\mathbf{H}_{n,i\mu}^{i\lambda} (p)\) can always be evaluated numerically along \(L_1\). In practice, the contour should be far from all poles of the integrand for numerical stability. We choose the radius of the semicircle in \fig{fig:complex-t-PP} to be \(1/4\) in our numerical evaluation.

Returning to \eq{eq:tree-PP-A-exact}, in the limit when both \(\abs{\lambda\pm\mu},\mu\gg1\), we can approximate \(\mathbf{H}_{n,i\mu}^{i\lambda} (p)\) using \eq{eq:tilde-F-asymp-2} as
\begin{align}
    \mathbf{H}_{n,i\mu}^{i\lambda} (p) 
    &\simeq \frac{1}{\frac{1}{2} +i\lambda +i\mu} \cdot \frac{1}{2\pi i} \int_L \p{1 -\frac{t}{\frac{1}{2} +i\lambda +i\mu}} \Gamma(-t) \Gamma(1 +n +t) 2^{-t} \p{\frac{p}{2}}^{-1-n-t} \dd{t}\n
    &= \frac{1}{\frac{1}{2} +i\lambda +i\mu} \p{\frac{p}{2}}^{-1-n} \sum_{k=0}^\infty \p{1 -\frac{k}{\frac{1}{2} +i\lambda +i\mu}} \frac{(n+k)!}{k!} (-p)^{-k}\n
    &= \frac{1}{\frac{1}{2} +i\lambda +i\mu} \p{\frac{p+1}{2}}^{-1-n} \bq{n! +\frac{1}{\frac{1}{2} +i\lambda +i\mu} \frac{(n+1)!}{p+1} } .
\end{align}
Therefore:
\begin{align}
    B_{++,\An}^{(\pm)}(k_1,k_2,k_3) &\simeq
    \frac{1}{2k_1^3k_2^3} \frac{\abs{\alpha}^2}{\Lambda^3} \frac{\lambda}{\p{\lambda^2 -\mu^2}^2 \bq{\p{\frac{1}{2}\pm i\lambda}^2 +\mu^2}}\n
    &\peq \times\sum_{n=0}^2 \bq{ n!
    \mp \frac{2i\lambda}{\lambda^2 -\mu^2} \frac{(n+1)!}{p+1} } c_n^\pm (p+1)^{-1-n}\n
    &\simeq
    \frac{1}{2k_1^3k_2^3} \frac{\abs{\alpha}^2}{\Lambda^3} \frac{\lambda}{\p{\lambda^2 -\mu^2}^2 \bq{\p{\frac{1}{2}\pm i\lambda}^2 +\mu^2}} \biggl[ \frac{(1 \pm i\lambda) p^2}{p+1}\n
    &\peq -\frac{\p{1 \pm i\lambda} (p +2)Q}{(p+1)^2} -\frac{1}{2} +\frac{2\lambda^2}{\lambda^2 -\mu^2} \frac{1}{(p+1)^2} \p{ p^2 -\frac{2Q}{p+1} } \biggr],\\[1em]
    B_{++,\An}(k_1,k_2,k_3) &\coloneqq B_{++,\An}^{(+)}(k_1,k_2,k_3) +B_{++,\An}^{(-)}(k_1,k_2,k_3)\n
    &\simeq \frac{1}{k_1^3k_2^3} \frac{\abs{\alpha}^2}{\Lambda^3} \frac{\lambda}{\p{\lambda^2-\mu^2}^4} \biggl\{ \frac{\mu^2}{p+1} \bq{ p^2 -\frac{(p +2)Q}{p+1}} +\frac{1}{2} \p{\lambda^2 -\mu^2}\n
    &\peq -\frac{2\lambda^2}{(p+1)^2} \p{ p^2 -\frac{2Q}{p+1} } \biggr\}.
\end{align}
This does not agree with \eq{eq:tree-PP-loc} in its na\"ive form. However, after summing over permutations of external momenta and using \(\mu\coloneqq\sqrt{M^2 -9/4}\simeq M\), both expressions are equal to
\begin{align}
    &\peq B_{++,\mathrm{A}}(k_1,k_2,k_3) +(k_3\to k_1,k_2)\n
    &\simeq \frac{1}{k_1^3k_2^3k_3^3} \frac{\abs{\alpha}^2}{\Lambda^3} \frac{\lambda}{\p{\lambda^2 -\mu^2}^3} 
    \Biggl[ k_1k_2k_3 -\frac{k_1^3 +k_2^3 +k_3^3}{2}\n
    &\peq +\frac{2\p{k_1^2k_2^2 +k_2^2k_3^2 +k_1^2k_3^2}}{k_{123}} -\frac{k_1 k_2 k_3 \p{k_1^2 +k_2^2 +k_3^2}}{k_{123}^2} -\frac{12\lambda^2}{\lambda^2-\mu^2} \frac{k_1^2 k_2^2 k_3^2}{k_{123}^3} \Biggr].
\end{align}
This confirms \eq{eq:tree-PP-loc} as the leading-order approximation of \eq{eq:tree-PP-A-exact}.

Finally, even for \(\abs{\lambda\pm\mu}\sim\order(1)\) with \(\mu\gg1\), \eqs{eq:tilde-F-asymp-2}{eq:h-func} still imply that
\begin{align}\label{eq:h-func-spa}
    \mathbf{H}_{n,i\mu}^{i\lambda} (p) &\simeq
    \frac{1}{2\pi i} \int_L \frac{\Gamma(-t) \Gamma(1+n+t)}{\frac{1}{2} +i\lambda +i\mu +t} 2^{-t} \p{\frac{p}{2}}^{-1-n-t}  \dd{t}\n
    &= \frac{n!}{\frac{1}{2}+i\lambda +i\mu} \p{\frac{p}{2}}^{-1-n} \hF{1+n}{\frac{1}{2}+i\lambda +i\mu}{\frac{3}{2}+i\lambda +i\mu}{-\frac{1}{p}},
\end{align}
this allows the corresponding one-loop amplitude being calculated approximately.

\section{Full Evaluation of the One-loop Model}
\label{app:full-results}

In this section, we present the full evaluation of the one-loop bispectrum. Section \ref{subsec:1-loop-exact} has shown that all contributions to the bispectrum can be written as integrations of the following form:
\begin{align}
    F^{(\pm)}_{X,\wloop}(k_1,k_2,k_3) &= \int_{-\infty}^{\infty} \rho_{\mu_1\mu_2}^{\mathrm{dS}}(\mu) \widetilde F^{(\pm)}_X(k_1,k_2,k_3;\mu) \dd{\mu}{} +\cc,
\end{align}
where \(F_X\in\Bq{F_{-+},F_{++,\NAn},F_{++,\An}}\) and
\begin{align}
    \widetilde{F}^{(\pm)}_{-+}(k_1,k_2,k_3;\mu) &\coloneqq \frac{\mathcal{F}}{\sqrt{\pi}} \frac{2^{2i\mu} e^{\mp\pi\lambda}}{\p{\frac{3}{2}\mp i\lambda}^2 +\mu^2} \sum_{n=0}^2 \frac{d_n}{2^n} \mathbf{\widetilde F}_{i\mu}^{-\frac{3}{2}+n\mp i\lambda-i\mu} (p)\n
    &\peq \times \mGamma{i\mu,-\frac{3}{2}+n\mp i\lambda -i\mu,\frac{1}{2}\pm i\lambda+i\mu,\frac{1}{2}\pm i\lambda-i\mu}{1\pm i\lambda},\\
    \widetilde{F}^{(\pm)}_{++,\NAn}(k_1,k_2,k_3;\mu) &\coloneqq ie^{\pi(\pm\lambda-\mu)} \widetilde{F}^{(\pm)}_{-+}(k_1,k_2,k_3;\mu),\\\label{eq:tree-PP-A-exact-F}
    \widetilde{F}^{(\pm)}_{++,\An}(k_1,k_2,k_3;\mu)  &\coloneqq -\frac{i\mathcal{F}}{2} \frac{1}{\mu \bq{\p{\lambda^2 -\mu^2 -\frac{9}{4}}^2 +9\lambda^2}} \sum_{n=0}^2 \frac{c_n^\pm}{2^n} \mathbf{H}_{n,i\mu}^{\pm i\lambda} (p)\n
    &= -\frac{i\mathcal{F}}{4} \frac{1}{\mu \bq{\p{\lambda^2 -\mu^2 -\frac{9}{4}}^2 +9\lambda^2}} \sum_{n=0}^2 \frac{c_n^\pm}{2^n}\n
    &\peq \times\sum_{m=0}^\infty \frac{(-)^m (m+1)_n \p{\frac{1}{2}+i\mu}_m}{\p{\frac{1}{2} +m \pm i\lambda +i\mu} (1+2i\mu)_m} \mathbf{\widetilde{F}}_{i\mu}^{1+n+m} (p). &(p>3)
\end{align}
Here we have extracted the common factor:
\begin{align}
 \mathcal{F} &\coloneqq \frac{5\lambda^2}{6} \abs{\frac{\alpha}{\Lambda}}^2 \frac{4}{p^3 +3\chi^2 p +4}
\end{align}
and introduce the following notations:
\begin{align}
 \left\{\begin{aligned}
  d_0 &\coloneqq 1,\\
  d_1 &\coloneqq p,\\
  d_2 &\coloneqq Q,
 \end{aligned}\right. &&
 \left\{\begin{aligned}
  c_0^\pm &\coloneqq \frac{1}{2} \bq{(1\pm2i\lambda) \p{p^2 -2Q}-1},\\
  c_1^\pm &\coloneqq \frac{1}{2} p \bq{p^2-2(1\mp i\lambda)Q-1},\\
  c_2^\pm &\coloneqq \frac{1}{2} Q \p{p^2-1},\\
 \end{aligned}\right.
\end{align}
where \(Q\coloneqq \frac{1}{4}\p{p^2-\chi^2}\).
All of these integrations can be expressed as sums over residues at poles with \(\Im\mu>0\). Such poles are summarized in \tab{tab:pole-structure-MP} and \tab{tab:1-loop-poles}. Consequently,
\begin{align}
 F_{\wloop} &= \sum_{\mu_*,s_1=\pm} \bq{\tF_{-+,\wloop}^{(s_1,\mu_*)} + \tF_{++,\NAn,\wloop}^{(s_1,\mu_*)} +\tF_{++,\An,\wloop}^{(s_1,\mu_*)}} +(k_1\to k_2,k_3) +\cc,
\end{align}
where \(\mu_* = \pm\lambda,\pm\mu_{12},\pm\dmu\) are the real parts of the poles and \(\tF_{X,\wloop}^{(s_1,\mu_*)}\) are the some of residues at these poles.

\begin{table}[htbp]
    \centering
    \begin{tabular}{l|ll|ll}\hline
        \multirow{2}{*}{\(\tF\)} & \multicolumn{2}{c|}{\(\Re\mu=\lambda\)} & \multicolumn{2}{c}{\(\Re\mu=-\lambda\)}\\\cline{2-5}
        & source & \(\mu\) & source & \(\mu\)\\\hline
        \(\tF_{-+}^{(-)}\) & \(\begin{aligned} &\Gamma\p{\tfrac{1}{2} -i\lambda +i\mu},\\
        &\Gamma\p{-\tfrac{3}{2}+n+i\lambda -i\mu}\end{aligned}\)\rule[-1.5em]{0pt}{3.5em} & \(\lambda +\p{\frac{1}{2}+k}i\) & \(\frac{1}{(3/2 +i\lambda)^2 +\mu^2}\) & \(-\lambda +\frac{3}{2}i\)\\\hline
        \(\tF_{++,\NAn}^{(+)}\) & \(\frac{1}{(3/2 -i\lambda)^2 +\mu^2}\) & \(\lambda +\frac{3}{2}i\) & \(\begin{aligned} &\Gamma\p{\tfrac{1}{2} +i\lambda +i\mu},\\
        &\Gamma\p{-\tfrac{3}{2}+n -i\lambda -i\mu}\end{aligned}\)\rule[-1.5em]{0pt}{3.5em} & \(-\lambda +\p{\frac{1}{2}+k}i\)\\\hline
        \(\tF_{++,\An}^{(+)}\) & \(\frac{1}{(\lambda^2 -\mu^2 -9/4)^2 +9\lambda^2}\) & \(\lambda +\frac{3}{2}i\) & \(\begin{aligned}
        &\tfrac{1}{\frac{1}{2}+m+i\lambda+i\mu},\\
        &\tfrac{1}{(\lambda^2 -\mu^2 -9/4)^2 +9\lambda^2}
        \end{aligned}\)\rule[-1.65em]{0pt}{3.9em} & \(-\lambda +\p{\frac{1}{2}+k}i\)\\\hline
    \end{tabular}
    \caption{The position of poles of the integrand in \eqs{eq:kallen-bispectrum-MP-contour}{eq:kallen-bispectrum-PP-contour} over the upper complex-\(\mu\) plane and their sources (\(n=0,1,2\) and \(m,\:k\in\N\)). The positions of poles at \(\Re\mu=\pm\mu_{12},\:\dmu\) are from the spectral density and are the same for all amplitudes, given in \tab{tab:pole-structure-MP}. For other amplitudes, e.g. \(\tF_{++,\An}^{(-)}\), the position of poles can be obtained using \eq{eq:relation}.}
    \label{tab:1-loop-poles}
\end{table}

\begin{table}[htbp]
 \centering
 \begin{tabular}{l|ll|lll}\hline
  \(\tF\) & \(s_1\) & \(\Re\mu\) & \(\lambda>\mu_{12}\) & \(\dmu<\lambda<\mu_{12}\) & \(\lambda<\dmu\)\\\hline
  \multirow{7}{*}{\(-+\)} & \multirow{6}{*}{\(-\)} & \(\lambda\) & \(\mathbf{0}\) & \(-\pi(\mu_{12}-\lambda)\) & \(-2\pi(\mu_1-\lambda)\)\\
  & & \(\mu_{12}\) & \(\mathbf{0}\) & \multicolumn{2}{l}{\(-2\pi(\mu_{12}-\lambda)\)}\\
  & & \(\dmu\) & \multicolumn{2}{l}{\(-2\pi\mu_2\)} & \(-2\pi(\mu_1-\lambda)\)\\
  & & \(-\dmu\) & \multicolumn{2}{l}{\(-\pi\mu_{12}\)} & \(-\pi(2\mu_1-\lambda)\)\\
  & & \(-\mu_{12}\) & \(-\pi\mu_{12}\) & \multicolumn{2}{l}{\(-\pi(2\mu_{12}-\lambda)\)}
  \\
  & & \(-\lambda\) & \(-\pi\lambda\) & \(-\pi\mu_{12}\) & \(-\pi(2\mu_1-\lambda)\)\\\cline{2-6}
  & \(+\) & any \(\mu_*\) & \multicolumn{3}{l}{\(-2\pi\lambda + {}\)(the one with \(-\mu_*\) above)}\\\hline
  \multirow{7}{*}{\(++,\NAn\)} & \multirow{6}{*}{\(+\)} & \(-\lambda\) & \(\mathbf{0}\) & \(-\pi(\mu_{12}-\lambda)\) & \(-2\pi(\mu_1-\lambda)\)\\
  & & \(-\mu_{12}\) & \(-\pi(\lambda-\mu_{12})\) & \multicolumn{2}{l}{\(-\pi(\mu_{12}-\lambda)\)}\\
  & & \(-\dmu\) & \multicolumn{2}{l}{\(-\pi(\lambda+3\mu_2-\mu_1)\)} & \(-\pi(\mu_{12}-\lambda)\)\\
  & & \(\dmu\) & \multicolumn{2}{l}{\(-\pi(\lambda+2\mu_1)\)} & \(-\pi(3\mu_1-\mu_2)\)\\
  & & \(\mu_{12}\) & \(-\pi(\lambda+2\mu_{12})\) & \multicolumn{2}{l}{\(-3\pi\mu_{12}\)}\\
  & & \(\lambda\) & \(-3\pi\lambda\) & \(-\pi(2\lambda+\mu_{12})\) & \(-\pi(\lambda+2\mu_1)\)\\\cline{2-6}
  & \(-\) & any \(\mu_*\) & \multicolumn{3}{l}{\(2\pi\mu_* + {}\)(the one with \(-\mu_*\) above)}\\\hline
  \multirow{3}{*}{\(++,\An\)} & \multirow{3}{*}{both} & \(\pm\lambda\) & \multicolumn{3}{l}{\(\mathbf{0}\)}\\
  & & \(\pm\mu_{12}\) & \multicolumn{3}{l}{\(\mathbf{0}^*\)}\\
  & & \(\pm\dmu\) & \multicolumn{3}{l}{\(-2\mu_2\)}\\\hline
 \end{tabular}
 \caption{Estimates of the exponential suppression factors for all contributions to \(\tF\), assuming \(p\gg1\) and \(\mu_1\ge\mu_2\). For brevity, we only show the exponents of the suppression factors. Note that \(\tF^{(s_1,s_2\mu_{12})}_{++,\text{A}}\) is polynomially suppressed and is indicated by the * sign.}
 \label{tab:1-loop-order-estimation}
\end{table}

Assuming \(\mu_1\ge\mu_2\) w.l.o.g., we have the estimation of orders for all 36 terms in \tab{tab:1-loop-order-estimation} with the Stirling approximation. For \(\lambda,\lambda-\mu_{12},\mu_1,\mu_2\) all much larger than 1, only the following terms are not exponentially suppressed (\(s_1,s_2=\pm\)):
\begin{align}
    \tF_{-+,\wloop}^{(-,\mu_{12})},\qquad
    \tF_{-+,\wloop}^{(-,\lambda)},\qquad
    \tF_{++,\NAn,\wloop}^{(+,-\lambda)},\qquad
    \tF_{++,\An,\wloop}^{(s_1,s_2\lambda)},\qquad \tF_{++,\An,\wloop}^{(s_1,s_2\mu_{12})}.
\end{align}
Among these terms, the non-analytic signal is given by
\begin{align}
 \tF_{-+,\wloop}^{(-,\mu_{12})} &= \sum_{k=0}^\infty \tF_{-+,\wloop}^{(-,\mu_{12},k)},&
 \tF_{-+,\wloop}^{(-,\mu_{12},k)} &\coloneqq \tF_{-+}^{(-)}\p{\mu_{12} +2ik +\frac{3i}{2}} \hat\rho_k(\mu_1,\mu_2),
\end{align}
where
\begin{align}
 &\hat\rho_k(\mu_1,\mu_2)
 \coloneqq 2\pi i\Res \bq{ \rho_{\sigma_1\sigma_2}(\mu), \mu_{12} +2ik +\frac{3i}{2}}\n
 &= \frac{(-)^k (2k+1)!!}{2^{k+2} \pi^{5/2} k!}
 \frac{\prod_{i=1,2,12} \Gamma(-k +i\mu_i) \Gamma\p{\frac{3}{2} +k -i \mu_i}}{\Gamma\p{-\frac{3}{2}-2k+i\mu_{12}} \Gamma\p{\frac{3}{2}+2k-i\mu_{12}} \Gamma(-2k+i\mu_{12}) \Gamma(3+2k-i\mu_{12})}.
\end{align}
The remaining four terms are all resonance backgrounds. The background terms from \(\tF_{-+}\) are given by
\begin{align}
 \tF_{-+,\wloop}^{(-,\lambda)} &= \sum_{k=0}^\infty \tF_{-+,\wloop}^{(-,\lambda,k)} \coloneqq \sum_{k=0}^\infty \tF_{-+}^{(-,\lambda,k)} \rho^{\text{dS}}_{\mu_1\mu_2} \p{\lambda +ik +\frac{i}{2}},\\
 \tF_{-+}^{(-,\lambda,k)} &\coloneqq 2\pi i\Res\bq{\tF_{-+}^{(-,\lambda,k)}(\mu), \lambda +ik +\frac{i}{2}}.
\end{align}
where the residues are explicitly given by\footnote{\(\psi(z)\eqqcolon \frac{\Gamma'(z)}{\Gamma(z)}\) is the digamma function.
The derivative of hypergeometric function in \(\tF_{-+}^{(-,\lambda,1)}\) can be expressed in terms of Meijer-G functions as
\begin{align}
    \eval{\frac{\partial}{\partial s} \mathbf{\widetilde F}_{-\frac{3}{2}+i\lambda -s}^s(p)}_{s=0}
    &= -\frac{\Gamma\p{4-2i\lambda}}{\Gamma\p{2 -i\lambda}} \MeijerG{22}{33}{0,-1+i\lambda}{1}{0,0}{-3+2i\lambda}{\frac{2}{p-1}} +\psi\p{4 -2i\lambda} -\psi\p{2 -i\lambda}\n
    &= -\log \frac{p}{2} +\frac{1}{2\p{5 -2i\lambda} p^2} +\frac{3}{4 (5 -2i\lambda) (7 -2i\lambda) p^4}
    +\cdots.
\end{align}
To prove this, one uses the eq. (16.17.2) in \cite{NIST} to convert
\begin{equation}
    \MeijerG{22}{33}{0,-1+i\lambda}{1}{0,z}{-3+2i\lambda}{\frac{2}{p-1}}
\end{equation}
into hypergeometric functions, and then take the limit \(z\to0\).
}
\begin{align}
 \tF_{-+}^{(-,\lambda,0)} &\coloneqq -\frac{\sqrt{\pi} \mathcal{F}}{8} \mGamma{1-2i\lambda,-\frac{1}{2}+i\lambda}{2-i\lambda} \frac{2^{2i\lambda} e^{\pi\lambda}}{1 +2i\lambda}
 \bq{ \p{1-2i\lambda} p +\frac{c_1^+}{p}  \mathbf{\widetilde{F}}_{-\frac{1}{2}+i\lambda}^1(p)},\\[1em]
 \tF_{-+}^{(-,\lambda,1)} &\coloneqq -\frac{\sqrt{\pi} \mathcal{F}}{12} \mGamma{2-2i\lambda,-\frac{3}{2}+i\lambda}{1-i\lambda} \frac{2^{2i\lambda} e^{\pi\lambda}}{2i\lambda}
 \Biggl\{ \eval{\frac{\partial}{\partial s} \mathbf{\widetilde F}_{-\frac{3}{2}+i\lambda -s}^s(p)}_{s=0}\n
 &\peq\phantom{\times \Biggl[}
 +\sum_{n=1}^2 \frac{d_n}{2^n} \mathbf{\widetilde{F}}_{-\frac{3}{2}+i\lambda}^n(p) -\frac{1}{2} \sum_{\epsilon,\epsilon_1,\epsilon_2=\pm1} \epsilon \psi\bq{\frac{3(1-\epsilon)}{4} +\frac{i}{2} \epsilon\p{\lambda +\epsilon_1\mu_1 +\epsilon_2\mu_2}}\n
 &\peq\phantom{\times \Biggl[}
 +\psi(i\lambda) -\psi(2-2i\lambda) -\frac{1}{1-i\lambda} -\frac{1}{2-i\lambda} +\frac{1}{2i\lambda} -\frac{4}{3} \Biggr\},\\[1em]
 \tF_{-+}^{(-,\lambda,k\ge2)} &\coloneqq \frac{(-)^k \sqrt{\pi} \mathcal{F}}{2^{2k} k!(k+2)} \mGamma{1+k-2i\lambda,-\frac{1}{2}-k+i\lambda}{1-i\lambda}\n
 &\peq \times \frac{2^{2i\lambda} e^{\pi\lambda}}
 {1-k+2i\lambda} \sum_{n=0}^2 \frac{(n+k-2)! d_n}{2^n} \mathbf{\widetilde F}_{-\frac{1}{2} -k +i\lambda}^{-1+n+k}(p).
\end{align}
The background terms from \(\tF_{++,\NAn}\) can be expressed in term of those from \(\tF_{-+}\) as
\begin{align}
 \tF_{++,\NAn,\wloop}^{(+,-\lambda)} &= \sum_{k=0}^\infty \tF_{++,\NAn,\wloop}^{(+,-\lambda,k)}
 \coloneqq \sum_{k=0}^\infty \tF_{++,\NAn}^{(+,-\lambda,k)} \rho^{\text{dS}}_{\mu_1\mu_2} \p{-\lambda +ik +\frac{i}{2}},\\
 \tF_{++,\NAn}^{(+,-\lambda,k)} &= (-)^k e^{2\pi\lambda} \tF_{-+}^{(+,-\lambda,k)} +\delta_{k1} \cdot\frac{i\pi^{3/2} \mathcal{F}}{12} \mGamma{2+2i\lambda,-\frac{3}{2}-i\lambda}{1+i\lambda} \frac{2^{-2i\lambda} e^{\pi\lambda}}{2i\lambda},
\end{align}
where \(\tF_{-+}^{(+,-\lambda,k)}\) is equal to \(\tF_{-+}^{(-,\lambda,k)}\) with \(\lambda\) replaced by \(-\lambda\). Finally, the background terms from \(\tF_{++,\An}\) are given by
\begin{align}
 \tF_{++,\An,\wloop}^{(s_1,-s_1\lambda)} &= \sum_{k=0,k\ne1}^\infty \tF_{++,\An}^{(s_1,-s_1\lambda,k)} \tF_{++,\An,\wloop}^{(s_1,-s_1\lambda,k)} +\sum_{m=0}^\infty \tF_{++,\An,\wloop}^{(s_1,-s_1\lambda,1,m)}\n
 &\coloneqq \sum_{k=0,k\ne1}^\infty \tF_{++,\An}^{(s_1,-s_1\lambda,k)} \rho_{\mu_1\mu_2}^{\mathrm{dS}}\p{-s_1\lambda +ik +\frac{i}{2}}\n
 &\peq +\sum_{m=0}^\infty \tF_{++,\An}^{(s_1,-s_1\lambda,1,m)} \rho_{\mu_1\mu_2}^{\mathrm{dS}}\p{-s_1\lambda +\frac{3i}{2}},\\
 \tF_{++,\An,\wloop}^{(s_1,s_1\lambda)} &= \sum_{m=0}^\infty \tF_{++,\An,\wloop}^{(s_1,s_1\lambda,m)}
 \coloneqq \sum_{m=0}^\infty \tF_{++,\An}^{(s_1,s_1\lambda,m)} \rho_{\mu_1\mu_2}^{\mathrm{dS}}\p{s_1\lambda +\frac{3i}{2}},\\
 \tF_{++,\An,\wloop}^{(s_1,s_2\mu_{12})} &= \sum_{k=0}^\infty \tF_{++,\An,\wloop}^{(s_1,s_2\mu_{12},k)}
 \coloneqq \sum_{k=0}^\infty \tF_{++,\An}^{(s_1)} \p{s_2\mu_{12} +2ik +\frac{3i}{2}} \hat\rho_k(s_2\mu_1,s_2\mu_2),
\end{align}
where, for \(s_1=+\),
\begin{align}
 \tF_{++,\An}^{(+,-\lambda,k\ne1)} &\coloneqq \frac{2\pi \mathcal{F}}{(k-1)(k+2)} \frac{(-)^k}{(1-k-2i\lambda) (1 +2k +2i\lambda) (2+k+2i\lambda)}\n
 &\peq \times \frac{(-k -i\lambda)_k}{(-2k-2i\lambda)_k} \sum_{n=0}^2 \frac{(k+1)_n c_n^+}{2^n} \mathbf{\widetilde{F}}_{-\frac{1}{2}-k-i\lambda}^{1+n+k} (p),\\[1em]
 \tF_{++,\An}^{(+,-\lambda,1,m\ne1)} &\coloneqq \frac{2\pi \mathcal{F}}{3(m-1)} \frac{(-)^m}{2i\lambda (3 +2i\lambda)^2} \frac{(-1-i\lambda)_m}{(-2-2i\lambda)_m} \sum_{n=0}^2 \frac{(m+1)_n c_n^+}{2^n}  \mathbf{\widetilde{F}}_{-\frac{3}{2}-i\lambda}^{1+n+m} (p),\\[1em]
 \tF_{++,\An}^{(+,-\lambda,1,1)} &\coloneqq \frac{2\pi \mathcal{F}}{3} \frac{1}{2i\lambda (3 +2i\lambda)}
 \Biggl\{\frac{-1}{3+2i\lambda} \sum_{n=0}^1 \frac{f_n}{2^n(p-1)} \mathbf{\widetilde{F}}_{-\frac{3}{2}-i\lambda}^{1+n} (p)\n
 &\peq\phantom{\times\Biggl\{}
 -\frac{1}{2} \sum_{\epsilon,\epsilon_1,\epsilon_2=\pm1} \epsilon \psi\bq{\frac{3(1-\epsilon)}{4} +\frac{i}{2} \epsilon\p{-\lambda +\epsilon_1\mu_1 +\epsilon_2\mu_2}}\n
 &\peq\phantom{\times\Biggl\{} -\frac{2\pi i}{\tanh2\pi\lambda} +\frac{1}{3+2i\lambda} -\frac{1}{2+i\lambda} -\frac{1}{1+i\lambda} -\frac{1}{2i\lambda} -\frac{1}{3} \Biggr\},\\[1em]
 \tF_{++,\An}^{(+,\lambda,m)} &\coloneqq \frac{2\pi \mathcal{F}}{3}  \frac{(-)^{1+m}}{2i\lambda (3 -2i\lambda)^2}
 \frac{(-1+i\lambda)_m}{(-2+2i\lambda)_m} \sum_{n=0}^2 \frac{(m+1)_n c_n^+}{2^n (-1+m+2i\lambda)} \mathbf{\widetilde{F}}_{-\frac{3}{2}+i\lambda}^{1+n+m} (p),
\end{align}
and
\begin{align}
 \left\{\begin{aligned}
  f_0 &\coloneqq p^2-2Q,\\
  f_1 &\coloneqq 2pQ.
 \end{aligned}\right.
\end{align}
For \(s_1=-\), one simply replaces all \(\lambda\) with \(-\lambda\).

The asymptotic estimation \eq{eq:asymp-bound} implies that the infinite sum over residues is also an series expansion over \(1/p\) with the following power-counting rules:
\begin{align}
    \tF_{-+,\wloop}^{(-,\mu_{12},k)} &\sim \order\p{p^{-3-2k}},\\
    \tF_{-+,\wloop}^{(-,\lambda,k)} \sim \tF_{++,\NAn,\wloop}^{(+,-\lambda,k)} &\sim \begin{dcases}
        \order\p{p^{-2-k}}, & k\ne1,\\
        \order\p{p^{-3}\log p}, & k=1,
    \end{dcases}\\
    \tF_{++,\An,\wloop}^{(s_1,-s_1\lambda,k)} &\sim \order\p{p^{-2-k}},\\
    \tF_{++,\An,\wloop}^{(s_1,-s_1\lambda,1,m)} \sim \tF_{++,\An,\wloop}^{(s_1,s_1\lambda,m)} &\sim \order\p{p^{-2-m}},\\
    \tF_{++,\An,\wloop}^{(s_1,s_2\mu_{12},k)} &\sim \order\p{p^{-2}}.
\end{align}
For \(F_{-+}\) and \(F_{-+,\NAn}\), the series is convergent for any \(p>1\), while for \(F_{++,\An}\), the series is at least convergent for \(p>3\) as indicated by \eq{eq:tree-PP-A-exact-F}. Note that \(\tF_{++,\An,\wloop}^{(s_1,s_2\mu_{12},k)}\) are \(\order\p{p^{-2}}\) for all \(k\). This explains why the difference between the numerical result and the sum over poles does not approach zero in \fig{fig:loop-bkg} at large \(p\).

\section{Corrections to the Power Spectrum}

\label{app:power-spectrum}

In this appendix, we evaluate the corrections to the power spectrum by our model. We first evaluate the correction for the tree-level model given by \eqref{eq:tree-int-basis}. At \(\order\bigl(\abs{\alpha}^2\bigr)\), there are three diagrams contributing to the power spectrum as shown in \fig{fig:tree-ps}. The first diagram is simply given by
\begin{equation}
    2k^3 \delta\Ps^{(1)}_\phi = -\frac{2\abs{\chicl}^2}{\Lambda^2}.
\end{equation}
The second diagram is given by
\begin{align}
    2k^3 \delta\Ps^{(2)}_\phi &= 2k^3 \abs{\chicl}^2 \abs{e^{-\pi\mu/2} \sqrt{\frac{\pi}{2}} \int^0_{-\infty_+} \frac{\dd{\ctime}}{\ctime^{5/2}} \cdot 2(-\ctime)^{1-i\lambda} \frac{\lambda}{\Lambda} \partial_{\ctime} D_-(k,\ctime) H_{i\mu}^{(1)}(-k\ctime)}^2\n
    &\peq +(\lambda\to -\lambda) +\cc\n
    &= \frac{2\lambda^2\abs{\chicl}^2}{\Lambda} \abs{\int_0^{\infty_+} \Int^{(-)}_{1,-+}(\lambda;x) \dd{x}}^2 +(\lambda\to -\lambda)\n
    &= \frac{4\lambda^2 \abs{\chicl}^2 \cosh\pi\lambda}{\Lambda^2} \abs{\mathbf{F}_{i\mu}^{\frac{1}{2}-i\lambda} (1)}^2\n
    &= \frac{2\abs{\chicl}^2}{\Lambda^2} \frac{2\pi\lambda \sinh 2\pi\lambda}{\cosh 2\pi\lambda +\cosh 2\pi\mu},
\end{align}
where \(\Int_{1,\pm +}^{(\pm)}\) is defined by \eq{eq:int}.
Finally, the third diagram can be evaluated in the same manner as in \Sec{subsec:tree-subdom}, except now we can utilize the symmetry between the two vertices and write
\begin{align}
    2k^3 \delta\Ps^{(3)}_{\phi} &= -2k^3 \abs{\chicl}^2 \int_{-\infty_+}^0 \frac{\dd{\ctime_2}}{\ctime_2^4} \int_{-\infty_+}^0 \frac{\dd{\ctime_1}}{\ctime_1^4} \cdot 2(-\ctime_1)^{1-i\lambda} \frac{\lambda}{\Lambda} \partial_{\ctime_1} D_+(k,\ctime_1)\n
    &\peq \times 2(-\ctime_2)^{1+i\lambda} \frac{\lambda}{\Lambda} \partial_{\ctime_2} D_+(k,\ctime_2) G_{++}(k,\ctime_1,\ctime_2) +(\lambda\to-\lambda) +\cc\n
    &= -\frac{\lambda^2\abs{\chicl}^2}{\Lambda^2} \int^{\infty_+}_0 \dd{x_2} \int^{x_2}_0 \dd{x_1} \bq{ \Int^{(-)}_{1,++}(\lambda;x_1) \Int^{(-)*}_{1,-+}(\lambda;x_2) +(\lambda\to -\lambda)} +\cc
\end{align}
Then, by applying \eqs{eq:hankel-int}{eq:hankel-int-2}, we obtain
\begin{align}
    2k^3 \delta\Ps^{(3)}_{\phi} &= \frac{i \lambda^2\abs{\chicl}^2}{4\pi \Lambda^2} \frac{e^{\pi\mu}}{\sinh\pi\mu} \int_L \frac{\dd{t}}{\frac{1}{2}-i\lambda+i\mu+t} \mGamma{-t,1+t,\frac{1}{2}+i\mu+t}{\frac{3}{2}+i\mu+t}\n
    &\peq +(\lambda\to -\lambda,\mu\to -\mu) +\cc\n
    %
    %
    &= -\frac{\lambda^2\abs{\chicl}^2}{2\Lambda^2} \frac{e^{\pi\mu}}{\sinh\pi\mu} \sum_{n=0}^\infty \frac{(-)^n}{\p{\frac{1}{2} +n +i\mu} \p{\frac{1}{2} +n -i\lambda +i\mu}}\n
    &\peq +(\lambda\to -\lambda,\mu\to -\mu) +\cc\n
    %
    %
    %
    &= \frac{i\abs{\chicl}^2}{\Lambda^2} \frac{\lambda e^{\pi\mu}}{4\sinh\pi\mu} \bq{ \psi\p{\frac{1}{4} -\frac{i\lambda -i\mu}{2}} -\psi\p{\frac{3}{4} -\frac{i\lambda -i\mu}{2}} }\n
    &\peq +(\lambda\to -\lambda,\mu\to -\mu) +\cc
\end{align}
Note that a similar calculation without chemical potential has been carried out in \Cite{CCP2}. Our results for the second and the third diagrams agree with their expressions in the limit \(\lambda\to0\).

\begin{figure}[htbp]
    \centering
    \begin{tabular}{cc}
        \includegraphics[scale=0.75]{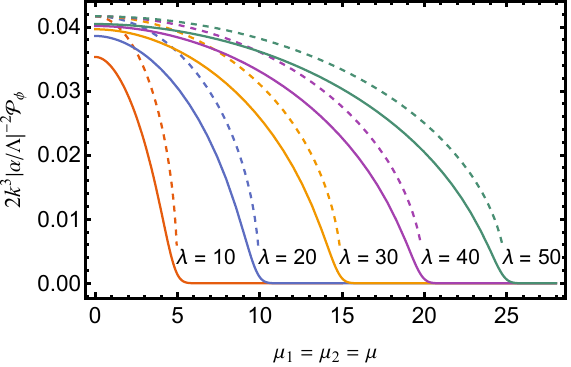}
        \includegraphics[scale=0.75]{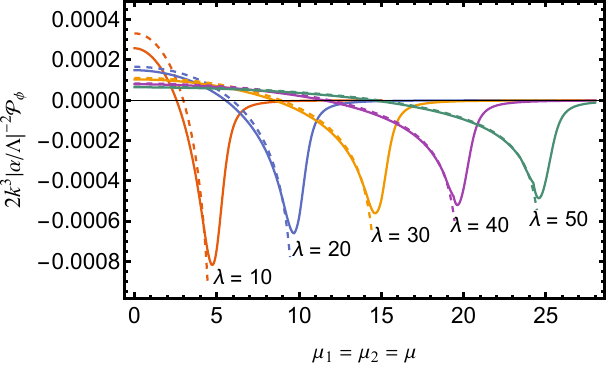}
    \end{tabular}
    \caption{The corrections to the power spectrum in the one-loop model for various \(\lambda\) and \(\mu_1=\mu_2=\mu\) from \(\delta{\Ps_\phi^{(2)}}\) (\emph{left}) and \(\delta{\Ps_\phi^{(3)}}\) (\emph{right}). The dashed lines are estimations from \eq{eq:power-spectru-1-loop}.}
    \label{fig:power-spectrum}
\end{figure}

For the one-loop model, the correction can also be derived from the tree-level one using the K\"all\'en–Lehmann representation:
\begin{align}
    \delta{\Ps_\phi^{(\wloop)}}(\mu_1,\mu_2) &= \int_0^\infty \delta{\Ps_\phi^{(\tree)}} (\mu) \rho^{\text{dS}}_{\mu_1\mu_2} (\mu) \dd{\mu}.
\end{align}
For \(\lambda,\mu_1,\mu_2,\lambda-\mu_{12}\gg1\), all three contributions peak around \(\mu\simeq\lambda\) and we can expand the integrand around this point, giving:
\begin{align}\begin{aligned}\label{eq:power-spectru-1-loop}
    2k^3\delta{\Ps_\phi^{(1,\wloop)}}(\mu_1,\mu_2) &\simeq -\frac{\abs{\alpha}^2}{\Lambda^2} \cdot \frac{\pi}{3\lambda^2} \rho^{\text{Mink}}_{\mu_1\mu_2} (\lambda),\\
    2k^3\delta{\Ps_\phi^{(2,\wloop)}}(\mu_1,\mu_2) &\simeq \frac{\abs{\alpha}^2}{\Lambda^2} \cdot \frac{\pi^2}{3\lambda} \rho^{\text{Mink}}_{\mu_1\mu_2} (\lambda),\\
    2k^3\delta{\Ps_\phi^{(3,\wloop)}}(\mu_1,\mu_2) &\simeq \frac{\abs{\alpha}^2}{\Lambda^2} \cdot \frac{1}{\lambda^2} \Bq{ \frac{\pi}{12} \rho^{\text{Mink}}_{\mu_1\mu_2} (\lambda) +\zeta \bq{\rho^{\text{Mink}}_{\mu_1\mu_2} (\lambda) -\lambda \rho'^{\text{Mink}}_{\mu_1\mu_2} (\lambda)} },
\end{aligned}\end{align}
where
\begin{equation}
    \zeta \coloneqq \frac{1}{4} \int_{-\infty}^{\infty} \frac{-ix \bq{\psi\p{\frac{1}{4}+\frac{ix}{2}} -\psi\p{\frac{3}{4}+\frac{ix}{2}}}}{\frac{9}{4} +x^2} \dd{x} = 0.482003\cdots.
\end{equation}
For a more accurate result, one has to carry out numerical calculations and the results are presented in \fig{fig:power-spectrum} for \(\delta{\Ps_\phi^{(2)}}\) and \(\delta{\Ps_\phi^{(3)}}\).

\bibliographystyle{JHEP}
\bibliography{main.bib}

\end{document}